\documentclass[3p,review,authoryear]{elsarticle}

\usepackage{commonm}
\usepackage{hyperref}
\usepackage{lineno}
\modulolinenumbers[1]

\journal{ }

\begin{document}

\begin{frontmatter}

\title{Capturing thin structures in VOF simulations with two-plane reconstruction}

\author[cornell]{Austin Han\corref{mycorrespondingauthor}}
\ead{ah2262@cornell.edu}
\author[lanl]{Robert Chiodi}
\author[cornell]{Olivier Desjardins}

\cortext[mycorrespondingauthor]{Corresponding author}

\address[cornell]{Sibley School of Mechanical and Aerospace Engineering, Cornell University, Ithaca, NY 14853, United States}
\address[lanl]{Computer, Computational, and Statistical Sciences Division, Los Alamos National Laboratory,
Los Alamos, NM 87545, United States}

\begin{abstract}
    A novel interface reconstruction strategy for volume of fluid (VOF) methods is introduced that represents the liquid-gas interface as two planes that co-exist within a single computational cell. In comparison to the piecewise linear interface calculation (PLIC), this new algorithm greatly improves the accuracy of the reconstruction, in particular when dealing with thin structures such as films. The placement of the two planes requires the solution of a non-linear optimization problem in six dimensions, which has the potential to be overly expensive. An efficient solution to this optimization problem is presented here that exploits two key ideas: an algorithm for extracting multiple plane orientations from transported surface data, and an efficient and mass-conserving distance-finding algorithm that accounts for two planes with arbitrary orientation. Additionally, a simple and robust strategy is presented to accurately represent the surface tension forces produced at the interface of subgrid-thickness films. The performance of this new VOF reconstruction is demonstrated on several test cases that illustrate the capability to handle arbitrarily thin films.
\end{abstract}

\begin{keyword}
Volume of fluid \sep interface reconstruction \sep multiphase flow \sep thin films \sep subgrid scale
\end{keyword}

\end{frontmatter}


\section{Introduction}\label{introduction}
Volume of fluid (VOF) methods are commonly used to track moving material interfaces in multiphase flow simulations.
VOF methods implicitly represent the interface between two immiscible fluids $a$ and $b$ in each computational cell $\Omega_{i}$ as a local fraction of volume
\begin{equation}
    \alpha_i=\frac{1}{V_i}\int_{\Omega_{i}}\chi(\bm{x})dV,
\end{equation}
where $V_i$ is the cell volume, $\bm{x}\in\bbR^3$, and $\chi$ is an indicator function that follows
\begin{equation}
    \chi(\bm{x})=
    \begin{cases}
    1 & \text{for } \bm{x}\in \text{fluid } a \\
    0 & \text{for } \bm{x}\in \text{fluid } b.
    \end{cases}
    \label{eq:vof}
\end{equation}
From these volume fractions, typical geometric VOF methods utilize a piecewise-linear interface calculation (PLIC) \citep{Debar1974,youngsTimedependentMultimaterialFlow1982} to model the interface as a single plane
\begin{equation}
    P(\bm{x})=\bm{n}\cdot\bm{x}-d=0,
    \label{eq:plane}
\end{equation}
where $\bm{n}$ is the interface normal vector that points from fluid $a$ to fluid $b$, and $d$ is the normal-distance to the plane, where $d$ is chosen such that the volume fractions on either side of the plane are $\alpha$ and $1-\alpha$. The normal vector can be computed using several methods including Youngs' method \citep{youngsTimedependentMultimaterialFlow1982}, the least-squares volume of fluid interface reconstruction algorithm (LVIRA) and its ``efficient'' variant for structured meshes (ELVIRA), both proposed by \cite{Pilliod2004}, and the moment-of-fluid method (MOF) \citep{Dyadechko2008}. However, regardless of normal computation method, PLIC methods are subject to an inherent mesh resolution limit where the minimum thickness for a fluid (e.g., liquid or gas) structure is approximately the size of a single computational cell. This leads to the erroneous breakup of poorly-resolved fluid structures due to numerical error, or ``numerical breakup."

The discrepancy between numerical breakup and physical breakup depends on the shape of the subgrid structure. For liquid ligaments, capillary instabilities can develop at diameters on the order of a few cell lengths, and therefore, numerical breakup can still predict most of the resulting droplet sizes \citep{Kim2020}. In constrast, the characteristic instability and breakup scales for liquid films can be on the order of a few microns and below, which is several orders of magnitude lower than the characteristic sizes of the liquid structures, such as droplets and liquid jets, from which these films are formed \citep{villermauxSingledropFragmentationDetermines2009}. The mesh resolution requirement to resolve this breakup is therefore immense. For instance, the numerical study of \cite{Ling2023} examined the aerodynamic breakup of a millimeter-scale drop in the bag breakup regime utilizing adaptive mesh refinement with a minimum cell size corresponding to 2048 cells across the initial diameter. The ability to represent thin films on coarse meshes may enable the prediction of the evolution and breakup of thin films at minimal computational cost. Several methods have been proposed to represent subgrid films within a VOF framework by approximating the interface with multiple planes per cell \citep{Lopez2005,Jemison2015,Shashkov2023,hergiboMomentoffluidMethodResolving2023,hergiboQuadtreebasedAdaptiveMomentoffluid2024,chiodiMomentoffluidInterfaceReconstruction}. However, these methods are limited to two-dimensional interfaces.

This paper presents an interface reconstruction in three dimensions that uses up to two planes in each cell to represent thin structures. This approach is hereinafter referred to as R2P (reconstruction with 2 planes). First, spherical k-means clustering \citep{Dhillon2001} is utilized on advected interfacial segments to determine the number of planes in the new reconstruction and form an initial guess of the plane orientation. Then, nonlinear optimization is performed to improve the orientation estimate of the planes by minimizing the error between the advected volume moments and the reconstructed moments, integrated over a neighborhood of cells. For each iteration of the optimization step, local volume conservation is enforced with an efficient routine. Section \ref{method} provides the mathematical formulation and implementation of the method, while Section \ref{tests} compares the method to standard single-plane reconstruction methods through canonical test cases. Finally, Section \ref{surface_tension} discusses and demonstrates the modeling of surface tension for subgrid liquid structures in various scenarios. Note that although this work refers to the fluids $a$ and $b$ in Eq.\ \eqref{eq:vof} as liquid and gas, respectively, the proposed method is not constrained to liquid-gas interface reconstruction, but could be useful for any immiscible interface. An implementation of the proposed method is accessible via the open-source Interface Reconstruction Library (IRL) \citep{Chiodi2022,chiodiInterfaceReconstructionLibrary2023}.

\section{Reconstruction with 2 Planes}\label{method}
Here, the nomenclature used for the method description and the assumptions necessary for the operation of the algorithm are introduced.
At a given time $t^n$, where $n$ is the time step, each computational cell volume $\Omega_i$ is associated with an interface reconstruction $\Gamma^n_i$ composed of $N^n_i$ planes, where $0\le N^n_i \le 2$. The set of polygons formed from the intersection $\Omega_i \cap \Gamma^n_i$ is subdivided into a set of triangles $\mathcal{T}^n_i$ within each cell, where $\mathcal{T}^n_i=\emptyset$ for non-interfacial cells, and where a particular triangulation is not prescribed for the purposes of this work. Each triangle is advected forward in time using a Lagrangian advection scheme, subdivided at its intersection with the cell faces, and deposited onto the Eulerian mesh as a set of polygons, such that each cell contains a set of polygons
\begin{equation}
    \mathcal{P}^{n+1}_i=\bigcup_{j=1}^{N_{\text{cell}}}\left\{\Omega_i \cap \mathscr{T}(\mathcal{T}^n_j)\right\},
\end{equation}
where $N_{\text{cell}}$ is the number of cells in the domain.
Here, $\mathscr{T}$ is an operator representing the Lagrangian transport from $t^n$ to $t^{n+1}$ of a triangle wherein the velocity of each vertex is computed by trilinear interpolation of the face-centered velocities of a staggered flow solver, and the time integration is performed with a second order implicit Runge-Kutta method. 
These polygons collectively form a second-order estimate of the interface in the cell at time $t^{n+1}$. The choice of triangulation of $\Gamma^n_i$ should not affect the order of accuracy of the method but will alter the collection $\mathcal{P}^{n+1}_i$. Figure \ref{subfig:advectednormal} shows a 2D example of the advected polygons within a cell and their precursor triangles. Note that in 2D, the polygons appear as line segments in the figure. The goal of the algorithm proposed herein is to construct the phase interface at the new time $t^{n+1}$, i.e., to build $\Gamma_i^{n+1}$.
Since the following steps comprise cell-local operations at a single timestep, the cell index $i$ and time step $n$ are omitted, except for the time variable $t$, in the remainder of this section.

\subsection{Formation of Initial Guess with Spherical K-Means}
\begin{figure}[tbp]
    \begin{subfigure}[b]{0.4\textwidth}
        \centering\includegraphics[width=\textwidth]{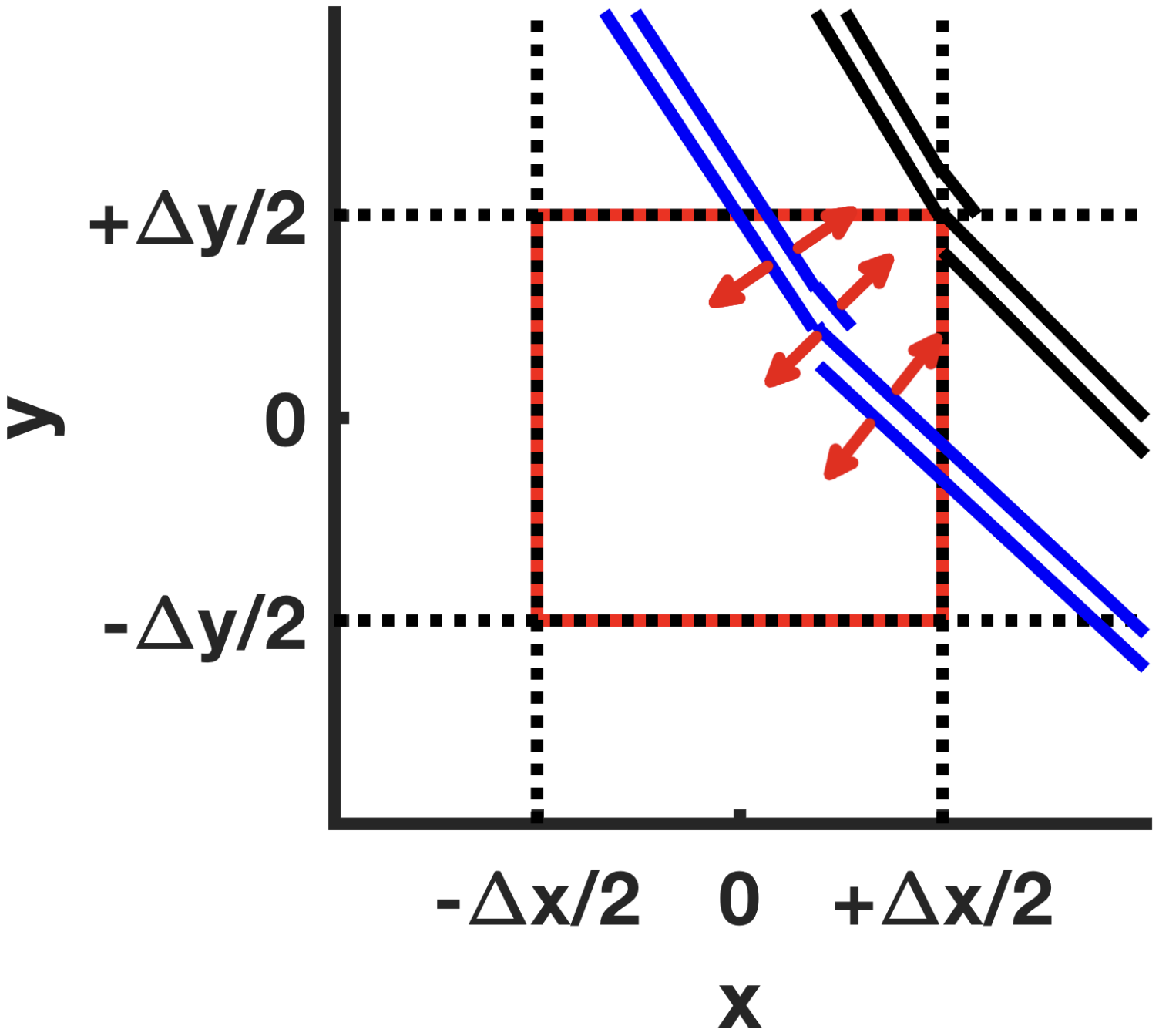}
        \caption{2D example of surface elements $\mathscr{T}(\mathcal{T}^n)$ (\fullblue) advected from the reconstruction $\Gamma^{n}$ (\fullblack). The segment of surface element contained within the red box constitutes $\mathcal{P}^{n+1}$. Dashed lines represent cell boundaries.}
        \label{subfig:advectednormal}
    \end{subfigure}
    \hfill
    \begin{subfigure}[b]{0.4\textwidth}
        \centering\includegraphics[width=\textwidth]{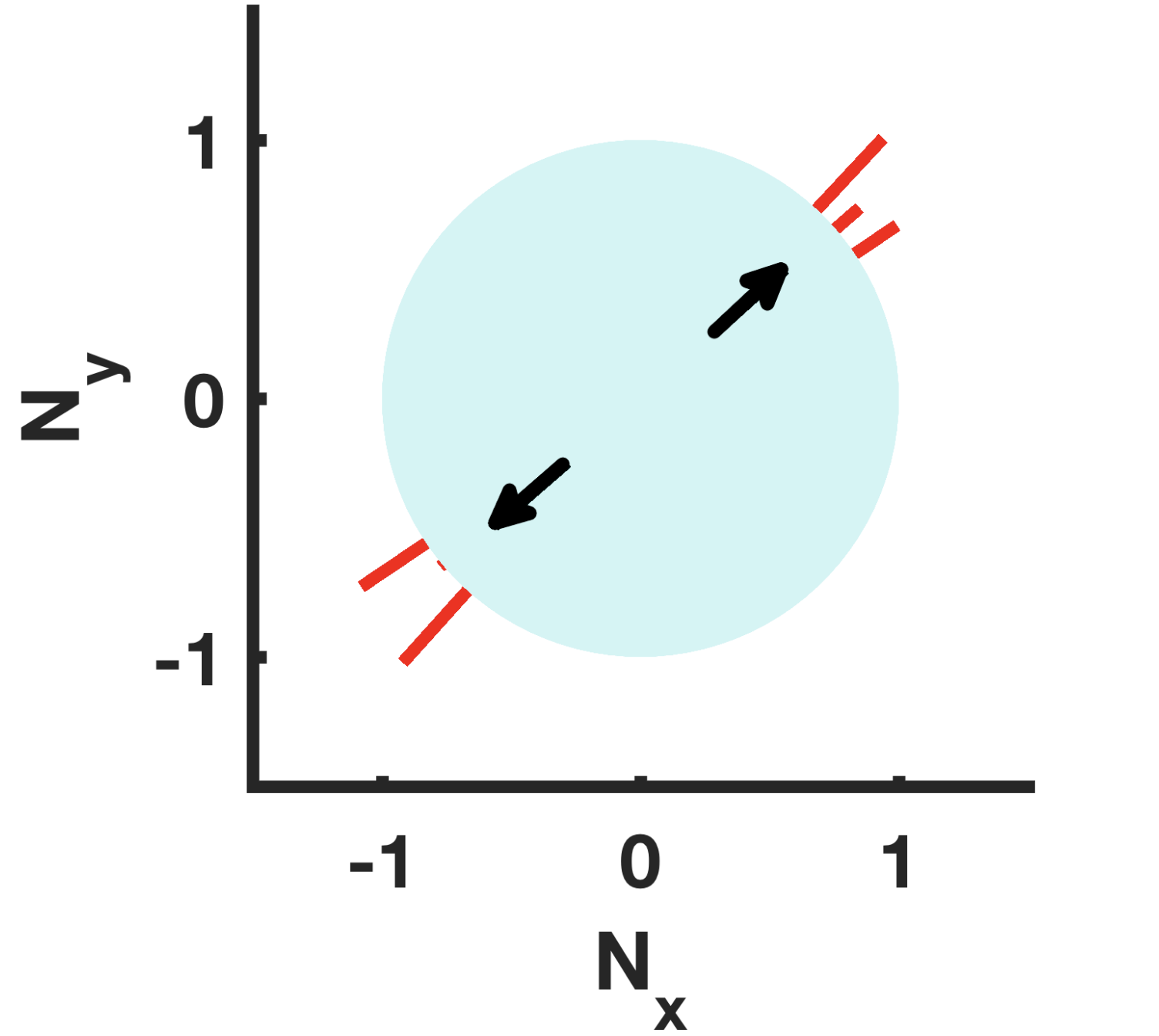}
        \caption{Normal vectors of the surface elements of $\mathcal{P}^{n+1}$ in (\subref{subfig:advectednormal}) shown on the unit circle. Normals are represented by a red bar with bar length indicating relative surface area. Black arrows show the two mean norms found from spherical k-means clustering.}
        \label{subfig:kmeans}
    \end{subfigure}
    \caption{Example of surface interface elements advected between $t^n$ and $t^{n+1}$ and the resulting normal vector distribution for the elements in the red box.}
    \label{fig:advected-normals}
\end{figure}

The first step of the algorithm is to form an initial guess of the interface normal vectors in each cell $\Omega$.
Each advected polygon in $\mathcal{P}$ has a corresponding normal in $\mathcal{N}=\{\bm{N}_1,\bm{N}_2,...,\bm{N}_Q\}$ along with a surface area in $\mathcal{W}=\{w_1,w_2,...,w_Q\}$ that is used as a weight, where $Q=|\mathcal{P}|$. Within each cell, these data form a weighted directional distribution deﬁned on a unit sphere in 3D (unit circle in 2D), i.e., $\mathcal{N}\subseteq S = \{ \bm{x}\in \bbR^3 | \, \|\bm{x}\| =1 \}$. For reference, the directional distribution resulting from the surface in the red box of Figure \ref{subfig:advectednormal} is shown in Figure \ref{subfig:kmeans}, where the location of the red bar indicates the normal of an advected polygon, and the length of the red bar represents its normalized surface area. To determine the number of planes that should constitute the reconstruction at time $t^{n+1}$, an average normal vector $\bm{m}$ is computed from the normals of the advected polygons in each cell, weighted by their respective surface areas:
\begin{equation}
    \bm{m}=\frac{\sum^Q_{j=1}w_j\bm{N}_j}{\sum^Q_{j=1}w_j}.
\end{equation}
As $0 \le \|\bm{m}\| \le 1 $, the magnitude of $\bm{m}$ measures the variance of the surface normal distribution. A magnitude $\|\bm{m}\|$ close to 1 indicates that the distribution is strongly unimodal and that the interface at $t^{n+1}$ in the cell is adequately represented with a single plane ($N=1$) with a normal given by $\bm{n}=\bm{m}/\|\bm{m}\|$, while lower values of $\|\bm{m}\|$ indicate that the interface should be represented by two planes ($N=2$). 
A threshold value $\lambda$ is defined such that
\begin{equation}
    N=
    \begin{cases}
        1, & \text{if $\|\bm{m}\|\ge \lambda$}, \\
        2, & \text{if $\|\bm{m}\|  < \lambda$},
    \end{cases}
\end{equation}
where $\lambda=0.99$ is chosen in this work.
For two advected polygons with equal surface area, $\lambda=0.99$ corresponds to an angle between the two corresponding normals of $\ang{16.22}$. In the case that $\|\bm{m}\| < \lambda$, two representative average normals are chosen through the spherical k-means algorithm \citep{Dhillon2001}. While it is possible to utilize a threshold of $\lambda=1$, thereby enforcing a two-plane reconstruction for all interfacial cells, lowering $\lambda$ decreases the cost of the reconstruction by allowing it to use a single plane in well-resolved regions of the interface.

The spherical k-means algorithm is an iterative algorithm where at each iteration $k$, $\mathcal{N}$ is partitioned into two disjoint subsets $\mathcal{N}_1^k$ and $\mathcal{N}_2^k$. Defining a normal vector for each subset as $\bm{n}_1^k$ and $\bm{n}_2^k$, the subsets are defined as $\mathcal{N}_1^k=\{ \bm{N} \in \mathcal{N} | \bm{N}\cdot\bm{n}_1^k \ge \bm{N}\cdot\bm{n}_2^k \}$ and $\mathcal{N}_2^k=\{ \bm{N} \in \mathcal{N} | \bm{N}\cdot\bm{n}_1^k < \bm{N}\cdot\bm{n}_2^k \}$, sorting $\mathcal{N}$ based on colinearity with $
\bm{n}_1^k$ and $\bm{n}_2^k$. A mean vector is computed for each subset as 
\begin{equation}
    \bm{m}_1^k=\frac{\sum_{l=1}^{|\mathcal{N}_1^k|}w_l\bm{N}_l}{\sum_{l=1}^{|\mathcal{N}_1^k|}w_l}\quad \textrm{and} \quad\bm{m}_2^k=\frac{\sum_{l=1}^{|\mathcal{N}_2^k|}w_l\bm{N}_l}{\sum_{l=1}^{|\mathcal{N}_2^k|}w_l},
\end{equation} and normal vectors computed from those as $\bm{n}_1^{k+1}=\bm{m}_1^k/\|\bm{m}_1^k\|$ and $\bm{n}_2^{k+1}=\bm{m}_2^k/\|\bm{m}_2^k\|$. The algorithm terminates when $\mathcal{N}_1^k=\mathcal{N}_1^{k+1}$ and $\mathcal{N}_2^k=\mathcal{N}_2^{k+1}$, at which point the two plane normal vectors are taken as $\bm{n}_1^k$ and $\bm{n}_2^k$.
The initial guess for $\bm{n}_1$ is chosen to be the normal vector with the largest corresponding surface area, i.e., $\bm{n}_1^0=\bm{N}_l$ where $l=\argmax_{j\in[1,Q]}(w_j)$. The initial normal vector $\bm{n}_2^0$ is then chosen to be the normal in $\mathcal{N}$ most dissimilar to $\bm{n}_1^0$, i.e., $\bm{n}_2^0=\bm{N}_l$, where $l=\argmin_{j\in[1,Q]}(\bm{n}_1^0\cdot\bm{N}_j)$.

The two representative normals identified for the example distribution are given by the black arrows in Figure \ref{subfig:kmeans}. Note that while this work limits the reconstructions to $N=2$, this process is theoretically extendable to an $N$-plane interface representation (for $N<Q$), where the number of subsets for $\mathcal{N}$ is increased until $\min(\|\bm{m}\|=\{\|\bm{m}_1\|,\|\bm{m}_2\|,...,\|\bm{m}_N\|\})>\lambda$.

\subsection{Volume Conserving Placement of Planes}\label{distance-solver}
The spherical k-means clustering of the weighted directional distribution from the previous section determines the number of planes in the initial guess and their normal vectors. To fully define the planes as in Eq.\ \eqref{eq:plane}, the normal-distance $d$ to each plane must be determined. Furthermore, the distances must result in an interface representation that exactly recovers the phase volumes in the cell to machine precision in order for the interface reconstruction to be conservative.

Many methods have been developed to determine a volume-conservative distance for a single plane, as required by PLIC. For specific cell geometries, such as rectangular prisms and tetrahedra, analytical solutions exist \citep{Scardovelli2000,Yang2006}. For the more general case, $d$ is typically found through a bracketing and solution approach, where possible values of $d$ are bracketed and the exact value is subsequently found analytically or iteratively \citep{Lopez2019,Ahn2007,Ivey2015,Lopez2016,Diot2016}. To find the distance for a set of planes that co-exist in a cell, R2P utilizes a similar approach in the form of non-linear root finding.

Let $d_1$ and $d_2$ be the two distances corresponding to the two planes. An initial guess for $d_1$ and $d_2$ is obtained from the spherical k-means clustering, and the guess is iteratively improved to conserve volume. From spherical k-means, the initial distances are computed using the surface centroids $\bm{s}$ of the corresponding surface elements in $\mathcal{P}$ whose normal vectors belong to $\mathcal{N}_1$ and $\mathcal{N}_2$. This initial distance guess can be computed as
\begin{equation}
    d^0_1=\bm{n}_1\cdot\bm{s}_1\quad\text{and}\quad d^0_2=\bm{n}_2\cdot\bm{s}_2.
\end{equation}
In all other cases, such as during the non-linear optimization algorithm presented in the next section, the initial guess for distances is directly supplied.

These two distance values, along with the normal vectors $\bm{n}_1$ and $\bm{n}_2$, define the multi-plane reconstruction consisting of the two planes $P_1$ and $P_2$. From this reconstruction, the liquid volume fraction for the cell, $\Tilde{\alpha}$, is computed using the Interface Reconstruction Library \citep{Chiodi2022,chiodiInterfaceReconstructionLibrary2023}. The error in liquid volume fraction can then be calculated as
\begin{equation}
    e_{\alpha}=\alpha-\Tilde{\alpha},
    \label{eq:vf-error}
\end{equation}
\noindent where $\alpha$ is the liquid volume fraction in the cell at $t^{n+1}$. As a matter of convention, quantities derived from a planar reconstruction are denoted with $\Tilde{} \;$.

To drive Eq.\ \eqref{eq:vf-error} to zero, thereby finding a volume-conservative placement of the two planes, the distances at the $k^{th}$ iteration are written as
\begin{equation}
    d^k_1=d^0_1+\delta^k\quad\text{and}\quad d^k_2=d^0_2+\delta^k,
\end{equation}
and non-linear root-finding is used to find the value of $\delta$ such that $e_{\alpha}=0$. Importantly, there is guaranteed to be a unique value of $\delta$ to achieve this since $\delta$ can be made sufficiently small or large such that $\Tilde{\alpha}=0$ or $\Tilde{\alpha}=1$. Note that there is the potential for the volume-conservative solution to force one of the planes outside of the cell, effectively reducing the reconstruction to a single plane. This has been found to be desirable in practice.

Numerically, $\delta$ is found using the Illinois method \citep{Dowell1971}, which is a modification to the Regula-Falsi method for finding the zero of a function. The Illinois method requires an initial bracket of values for $\delta$, between which the root lies. These bracketing values are computed as
\begin{equation}
    \delta_{\text{lower}}=\min_n(\min_m(\bm{x}_m\cdot \bm{n}_n)-d^0_n)\quad\text{and}\quad \delta_{\text{upper}}=\max_n(\max_m(\bm{x}_m\cdot \bm{n}_n)-d^0_n),
\end{equation}
where $\bm{x}_m$ are the vertices of the cell, and $n\in[1,2]$. At both values, the planes in the reconstruction exist outside the cell, with $\Tilde{\alpha}=0$ for $\delta=\delta_{\text{lower}}$ and $\Tilde{\alpha}=1$ for $\delta=\delta_{\text{upper}}$. Although R2P only necessitates two planes for the bracketing, this approach works for any number of planes.
\subsection{Optimization of the Reconstruction}
The initial guess is further improved using non-linear optimization towards a final reconstruction. When the single-plane case is identified through spherical k-means, the initial guess is overwritten by the estimate
\begin{equation}
    \bm{n}=\frac{\bm{c}^b-\bm{c}^a}{\|\bm{c}^b-\bm{c}^a\|},
\end{equation}
where $\bm{c}^b$ and $\bm{c}^a$ are the gas and liquid phase centroids in the cell $\Omega$ for which the interface is being reconstructed, respectively. This estimate was found to be more accurate than the initial guess from k-means for single-plane reconstructions. Ordinarily, calculating $\bm{n}$ in this manner is unreliable due to the potential for structures of the same phase to enter the cell from different directions. However, such situations are associated with highly dissimilar interface normal vectors in the cell and consequently are treated with multiple planes. In the case where a two-plane reconstruction is deemed necessary, the spherical k-means initial guess, $\bm{n}_{1}$ and $\bm{n}_{2}$, is kept.

For a cell $\Omega$ containing an interface, the goal of the optimization is to best approximate geometric information in a stencil of cells $\mathscr{N}$ (e.g., a $3\times 3\times 3$ stencil in a Cartesian mesh) by minimizing the error
\begin{equation}
    e_{R}=\left\Vert\sum_{\Omega_{j}\in \mathscr{N}}w_{\alpha,j}(\alpha_j-\Tilde{\alpha}_j)\right\Vert^2+\left\Vert\sum_{\Omega_j\in \mathscr{N}}w_{a,j}(\bm{c}^a_j-\Tilde{\bm{c}}^a_j)\right\Vert^2+\left\Vert\sum_{\Omega_j\in \mathscr{N}}w_{b,j}(\bm{c}^b_j-\Tilde{\bm{c}}^b_j)\right\Vert^2+\left|w_{\Sigma}\biggl(\sqrt{\Sigma}-\sqrt{\Tilde{\Sigma}}\biggr)\right|^2
    \label{eq:cost-fcn}
\end{equation}
where the reference quantities, written without the $\Tilde{}\;$, are the geometric quantities computed during the Lagrangian advection of the phase volumes and interface surface.

As can be seen, $e_{R}$ is constructed from four separate types of error, which are, from left to right: liquid volume fractions in the stencil ($\alpha$), liquid centroids in the stencil ($\bm{c}^a$), gas centroids in the stencil ($\bm{c}^b$), and the interface surface area ($\Sigma$) for the cell being reconstructed. Each of these types of error also have an associated weighting vector, $\bm{w}$, with the requirement that 
\begin{equation}\label{eq:normalizedweights}
    \|\bm{w}_{\alpha}\|+\|\bm{w}_a\|+\|\bm{w}_b\|+|w_{\Sigma}|=1.
\end{equation}
As $e_{R}$ only depends on the surface area error of the reconstructed cell and not the surrounding stencil cells, $w_{\Sigma}$ is a scalar.

These weights serve two purposes. First, they are functions of space, which allows the relative importance of each cell in the stencil to be modified for each individual type of error through the use of spatially dependent weighting coefficients. Second, the relative importance of each type of error to one another can be changed through modification of the weighting vector magnitude. It is interesting to note that both LVIRA \citep{Pilliod2004} and MOF \citep{Dyadechko2008} are contained as subsets of this optimization problem solely through the selection of proper weighting vectors.

For all results presented herein, $\|\bm{w}_{\alpha}\|=0$, removing the contribution from errors in representing local volume fractions. The centroid weightings for a given cell $\Omega_j$ in the stencil depend on their proximity to the phase centroids of the cell being reconstructed:
\begin{equation}
    \begin{split}
        w_{a,j} &= e^{-\|\bm{c}^a-\bm{c}^a_j\|^2} \\
        w_{b,j} &= e^{-\|\bm{c}^b-\bm{c}^b_j\|^2}.
    \end{split}
\end{equation}

For two-plane reconstructions, the relative importance of the surface area error and centroid error is computed based on the volume fraction in the stencil as
\begin{equation}
    \begin{split}
        |w_{\Sigma}| &= \frac{3}{4}(1-2\Bar{\alpha})^2 \\
        \|\bm{w}_a \| &= \|\bm{w}_b\|=0.5,
    \end{split}
\end{equation}
where $\Bar{\alpha}$ is the average liquid volume fraction computed over the entire stencil, and the weights are subsequently normalized to satisfy Eq.\ \eqref{eq:normalizedweights}. In this way, the creation or destruction of large amounts of surface area is prevented for small liquid and gas structures. For single-plane reconstructions, the centroid error is solely considered by setting $|w_{\Sigma}|=0$. These weightings have been developed heuristically but have delivered good results in practice, as will be shown. There is potential for further optimization of the weighting by using techniques such as machine learning.

During the optimization, the two normal vectors are parameterized as an equal but opposite rotation of a normal vector about a rotation axis. This can be written as
\begin{equation}
    \bm{n}_1=R_{\bm{t}}(\beta)\bm{n}_s \quad \text{and} \quad \bm{n}_2=R_{\bm{t}}(-\beta)\bm{n}_s,
    \label{eq:quaternion}
\end{equation}
where $R_{\bm{t}}(\beta)$ is a quaternion performing a rotation about an axis $\bm{t}$ by $\beta$ degrees, and $\bm{n}_s$ is a normal vector. Here, a shorthand notation of a quaternion-normal vector multiplication resulting in a vector is used, neglecting the transformations to and from quaternions. The single-plane reconstruction is encompassed in this parameterization by enforcing a value of $\beta=0$, leading to $\bm{n}_1=\bm{n}_2=\bm{n}_s$.

\begin{figure}[tbp]
    \centering\includegraphics[width=0.3\textwidth]{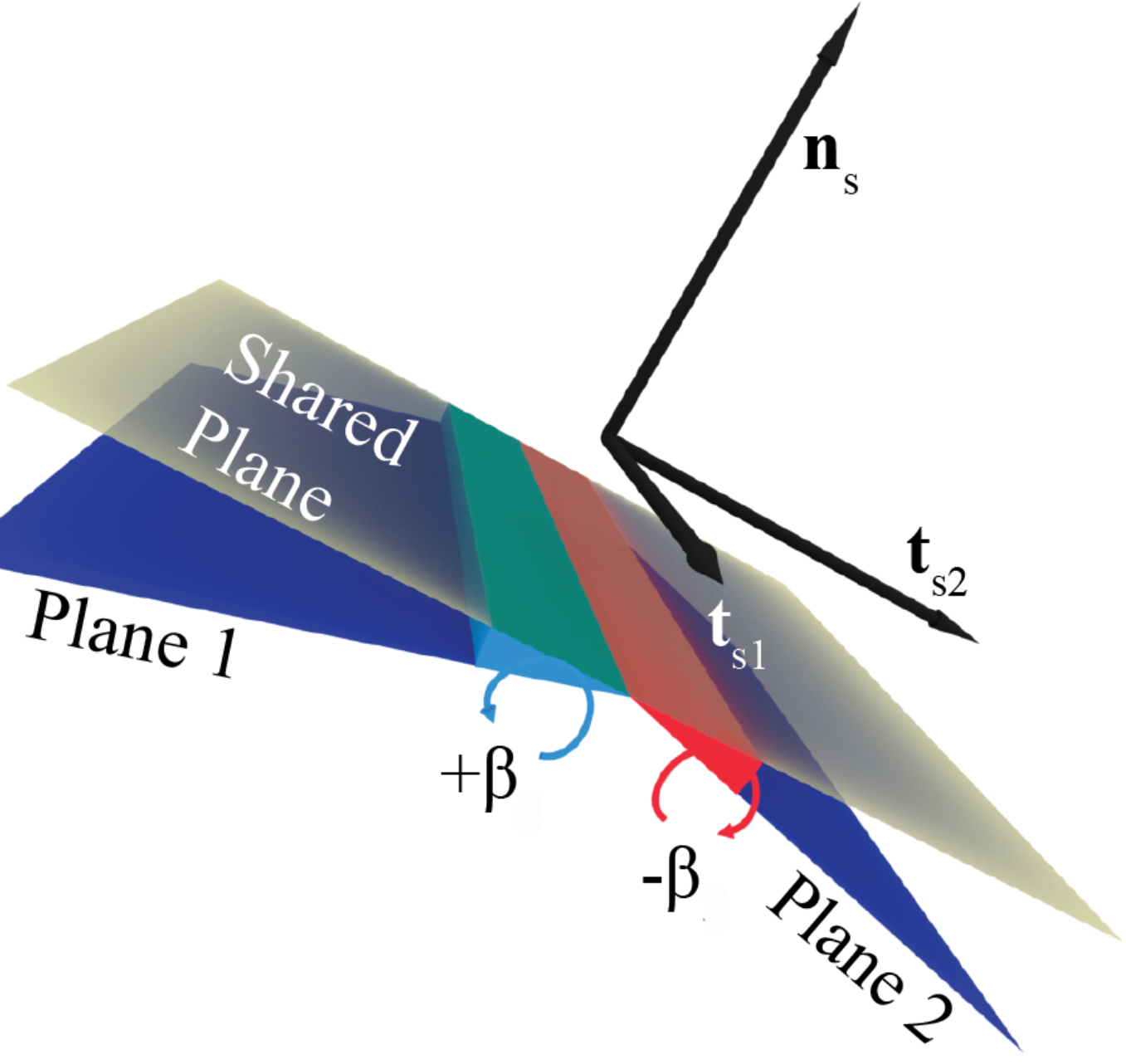}
    \caption{Diagram of the local orthonormal reference frame created and the relationship between the shared plane given by $\bm{n}_s$ and Plane 1 and Plane 2 defined by $\bm{n}_1$ and $\bm{n}_2$.}
    \label{fig:optimization}
\end{figure}

The optimization is performed inside a local orthonormal reference frame constructed as shown in Figure \ref{fig:optimization}, where $\bm{t}_{s1}$ and $\bm{t}_{s2}$ are two tangent vectors orthogonal to $\bm{n}_s$. The tangent vector $\bm{t}_{s1}$ is also used as the rotation vector $\bm{t}$ in the parameterization of $\bm{n}_{1}$ and $\bm{n}_{2}$ given in Eq.\ \eqref{eq:quaternion}. The error function given by Eq.\ \eqref{eq:cost-fcn} is then minimized using the Levenberg-Marquardt algorithm \citep{Levenberg1944} to rotate the reference frame of $[\bm{t}_{s1},\bm{t}_{s2},\bm{n}_s]$ to find the optimal solution. In a two-plane reconstruction, the value of $\beta$ also needs to be determined, along with optimal values of $d_1$ and $d_2$. To ensure that values of $d_1$ and $d_2$ lead to a volume-conservative reconstruction, they are projected onto a volume-conserving solution space using the one-dimensional minimization problem described in Section \ref{distance-solver}. The inclusion of $d_1$ and $d_2$ in the optimization allows for the translation of the interface planes, which is important to accurately capture thin films spanning multiple cells. For single-plane reconstructions, $d$ is computed, either using an analytical method for hexahedral and tetrahedral cells \citep{Scardovelli2000,Yang2006} or an iterative method for other cell geometries.

\section{Verification Tests}\label{tests}
The proposed reconstruction method is evaluted with a series of canonical test cases, in each case comparing the accuracy, error mesh convergence, and cost of the reconstruction with those of the single-plane ELVIRA reconstruction \citep{Pilliod2004}. The test cases use velocity fields that return the initial interface to its original position. Errors in the volume fraction transport and reconstruction steps, however, accumulate and produce errors in the final interface shape. In each case, two error metrics are computed: the shape error
\begin{equation}
    E_{\text{shape}}=\frac{\sum_{i=1}^{N_{\text{cell}}}V_i\bigl|\alpha_i(T)-\alpha_i(0)\bigr|}{\sum_{i=1}^{N_{\text{cell}}}V_i\alpha_i(0)},
\end{equation}
which measures the deviation of the transported and reconstructed interface from the initial condition, and the conservation error 
\begin{equation}
    E_{c}=\frac{\Biggl|\sum_{i=1}^{N_{\text{cell}}}V_i\alpha_i(T)-\sum_{i=1}^{N_{\text{cell}}}V_i\alpha_i(0)\Biggr|}{\sum_{i=1}^{N_{\text{cell}}}V_i}, 
\end{equation}
which measures the change in liquid volume integrated over the domain. Since the test cases utilize uniform meshes, $V_i$ is constant for a given mesh resolution. In addition, the computational time necessary to perform the volume transport, interface transport, and interface reconstruction steps during each time step , $T_c$, is measured.

All simulations are performed on a Beowulf cluster with nodes containing dual six-core Intel X5670 processors. All test cases are performed with the NGA2 flow solver \citep{Desjardins2008,desjardinsNGA22023}, in which the advection of the volume moments is performed with the method of \cite{Owkes2014}. The Interface Reconstruction Library (IRL) \citep{Chiodi2022,chiodiInterfaceReconstructionLibrary2023} is used for the necessary computational geometry and reconstruction algorithms. The tolerance in volume fraction error (Eq.\ \eqref{eq:vf-error}) used in the root-finding step in Section \ref{distance-solver} is $10^{-12}$. 

\subsection{Zalesak's Disk}
The case proposed by \cite{Zalesak1979} tests the ability of a VOF scheme to transport a two-dimensional shape with sharp corners. In this test, a notched disk is subjected to rigid-body rotation by the velocity field
\begin{equation}
    (u,v)=(-2\pi y,2\pi x).
\end{equation}
The disk has a diameter 0.3, a notch width of 0.05, a notch height of 0.25, and is initially centered at $(x,y)=(0,0.25)$ within a square domain $[-0.5,0.5]^2$. The velocity field returns the disk to its initial position after $t=1$ time units. The simulations are performed on computational domains of $N=[16,32,64,128,256]$ cells in each direction. For each mesh resolution, the time step size is $\Delta t = 32/(100N)$. The two-dimensional simulations are performed on a three-dimensional Cartesian mesh that is one cell in depth with cell thickness $1/N$. For all mesh sizes, the simulations are performed on nine cores within a single node of the Beowulf cluster.

Figure \ref{fig:zalesak-interface} compares the initial and final interface shapes after one rotation for the mesh resolutions $N=[32,64,128]$. Across all mesh resolutions, R2P produces a final interface shape that better retains the sharp features of the initial shape than ELVIRA. With the default parameters, including the two-plane threshold $\lambda=0.99$, R2P does not perfectly retain the sharp corners due to the switching between one- and two-plane reconstructions. If $\lambda$ is set to 1, thereby enforcing two-plane reconstructions at all instances, then the sharp corners are retained after one rotation.
Figure \ref{fig:zalesak-scatter} shows the mesh dependence of the shape error from and computational cost of the compared reconstruction methods. Both the ELVIRA and R2P methods exhibit first-order convergence in shape error, but the R2P method produces errors that are lower than those from the ELVIRA method by a factor of two or greater. The increased accuracy comes with an increased computational cost $T_c$, where the cost difference is greatest in the lowest resolution simulations where two-plane reconstruction is utilized more often. Finally, the conservation error is shown in Table \ref{tab:zalesak-econs}, where for all mesh resolutions, R2P preserves liquid volume to machine precision.

\begin{figure}
    \begin{subfigure}[b]{.33\textwidth}
        \centering\includegraphics[width=\textwidth]{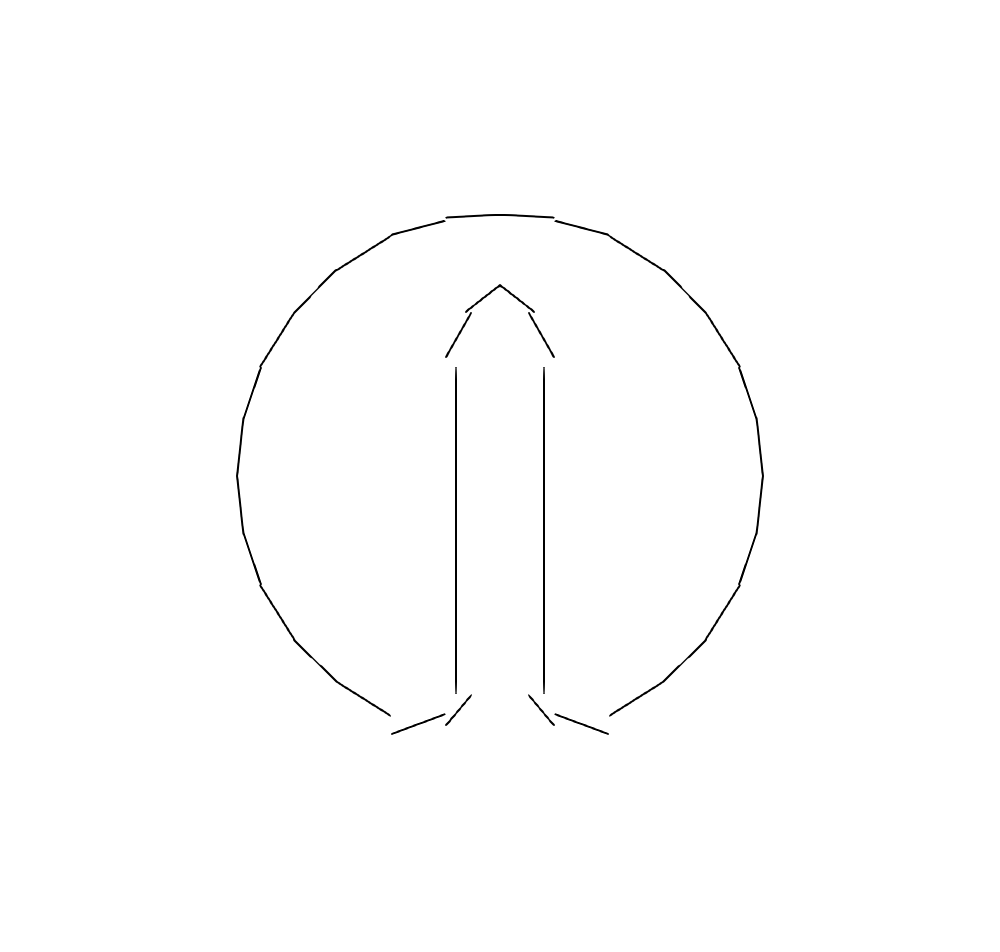}
        \caption*{$N=32,t=0$}
    \end{subfigure}\hfill
    \begin{subfigure}[b]{.33\textwidth}
        \centering\includegraphics[width=\textwidth]{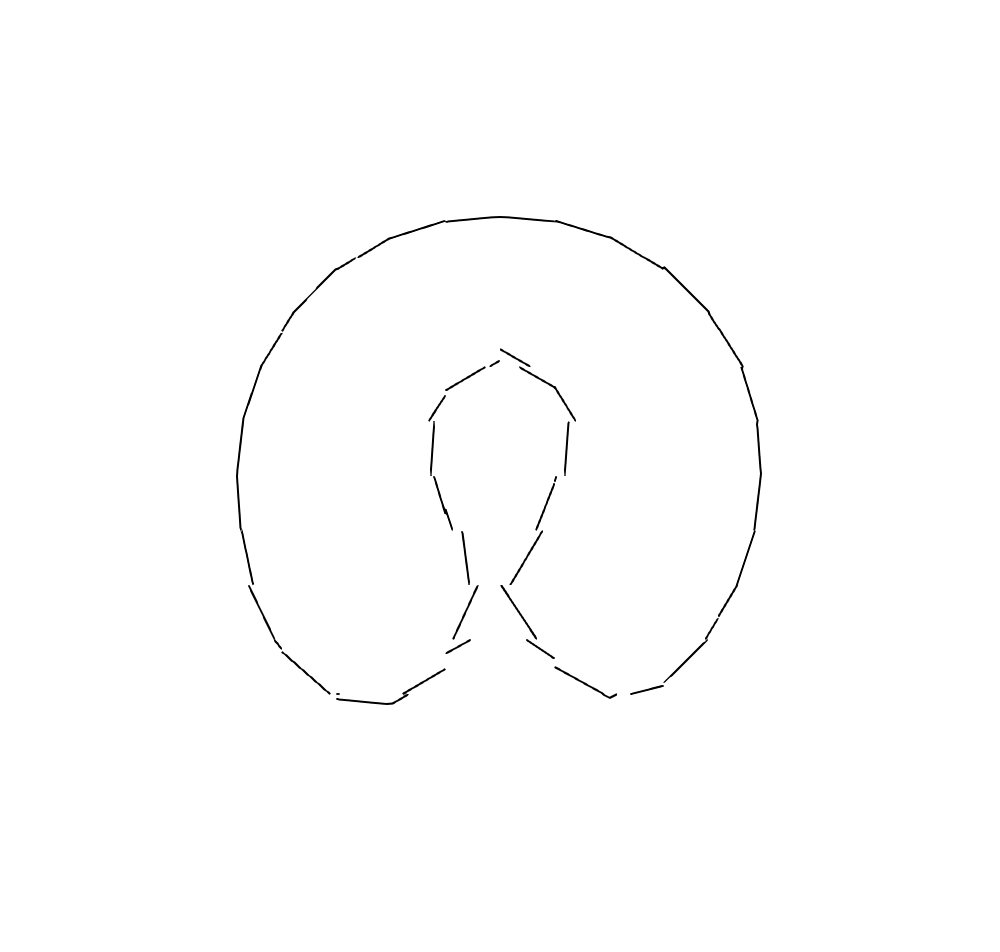}
        \caption*{$N=32,t=1,\text{ELVIRA}$}
    \end{subfigure}\hfill
    \begin{subfigure}[b]{.33\textwidth}
        \centering\includegraphics[width=\textwidth]{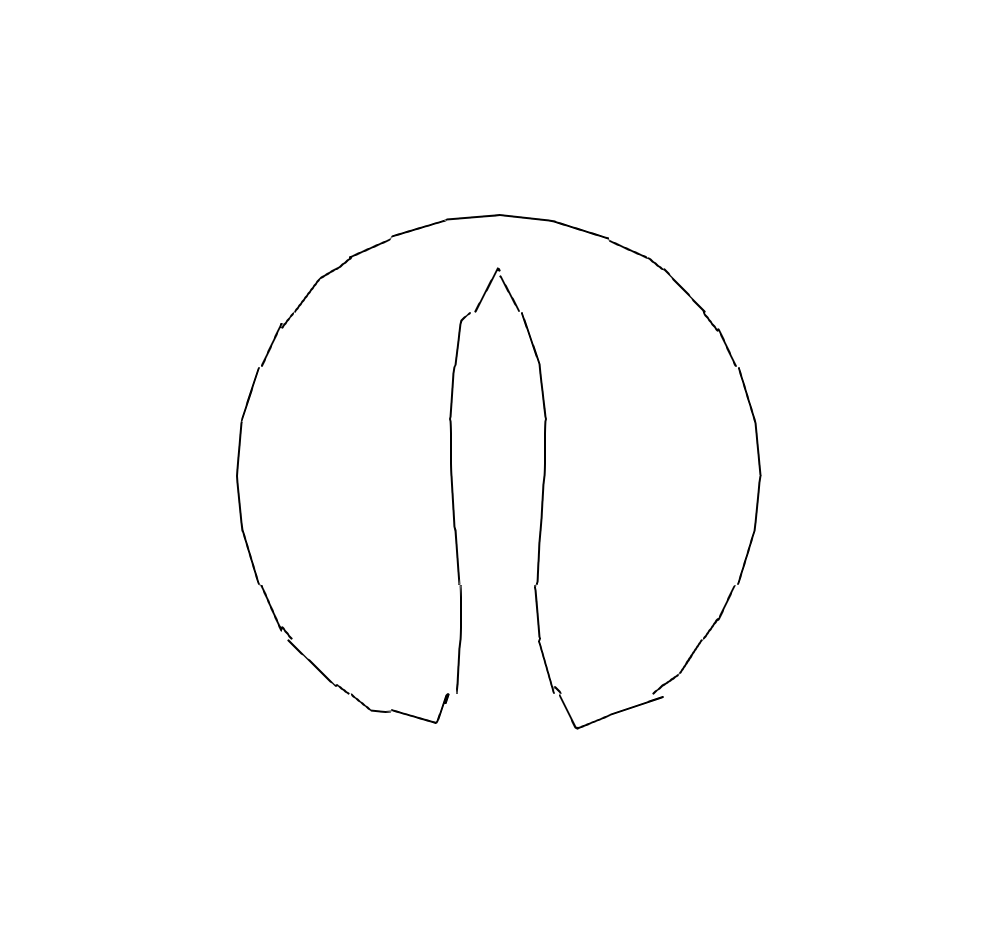}
        \caption*{$N=32,t=1,\text{R2P}$}
    \end{subfigure}\hfill
    \begin{subfigure}[b]{.33\textwidth}
        \centering\includegraphics[width=\textwidth]{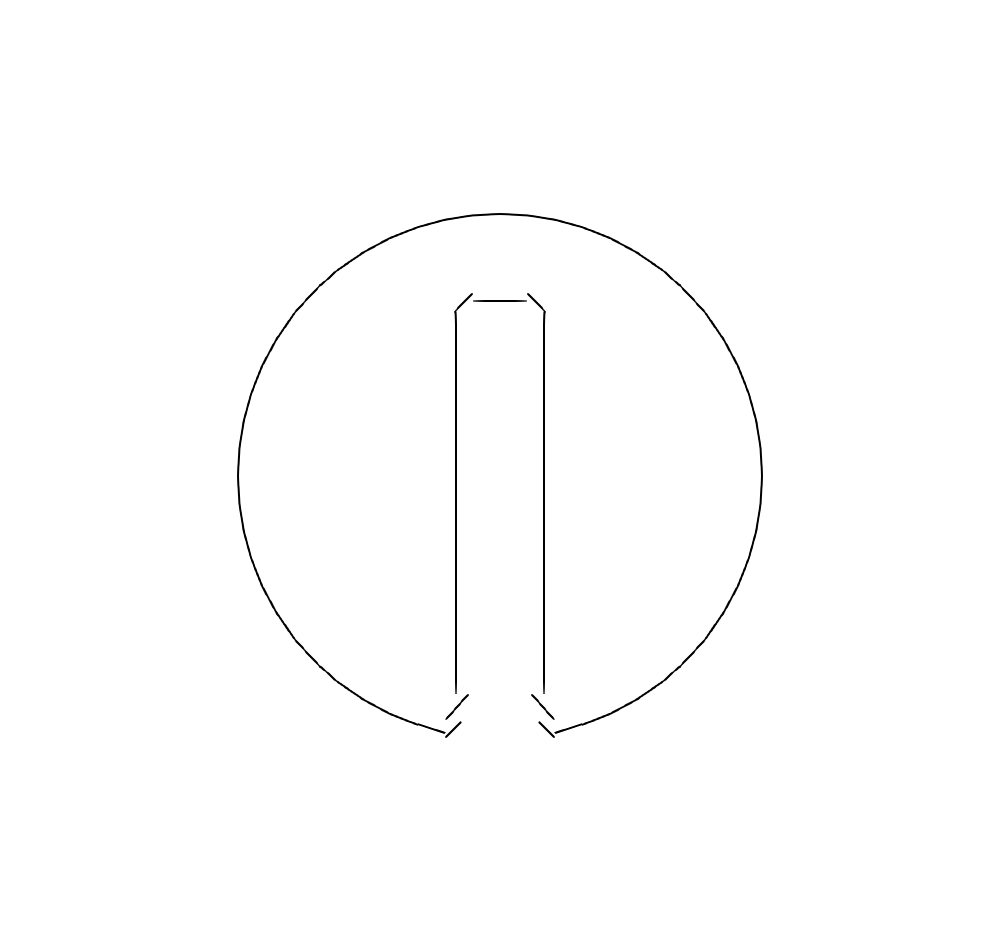}
        \caption*{$N=64,t=0$}
    \end{subfigure}\hfill
    \begin{subfigure}[b]{.33\textwidth}
        \centering\includegraphics[width=\textwidth]{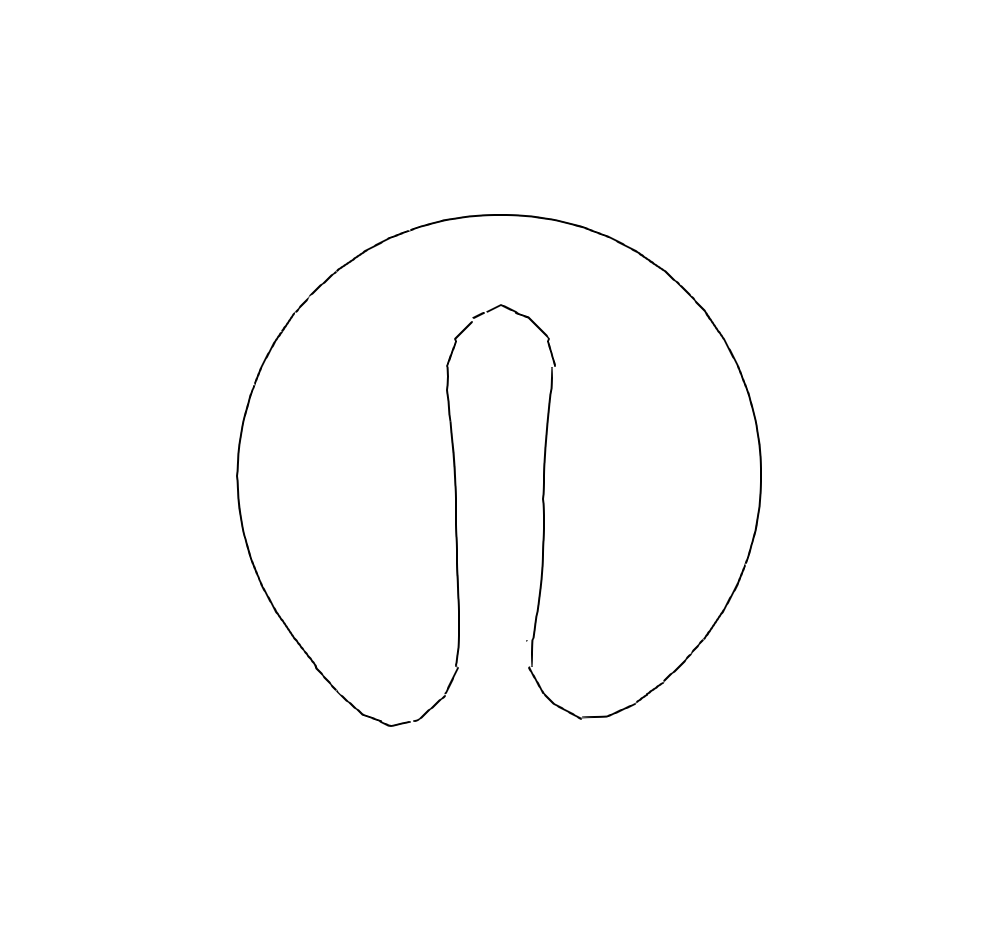}
        \caption*{$N=64,t=1,\text{ELVIRA}$}
    \end{subfigure}\hfill
    \begin{subfigure}[b]{.33\textwidth}
        \centering\includegraphics[width=\textwidth]{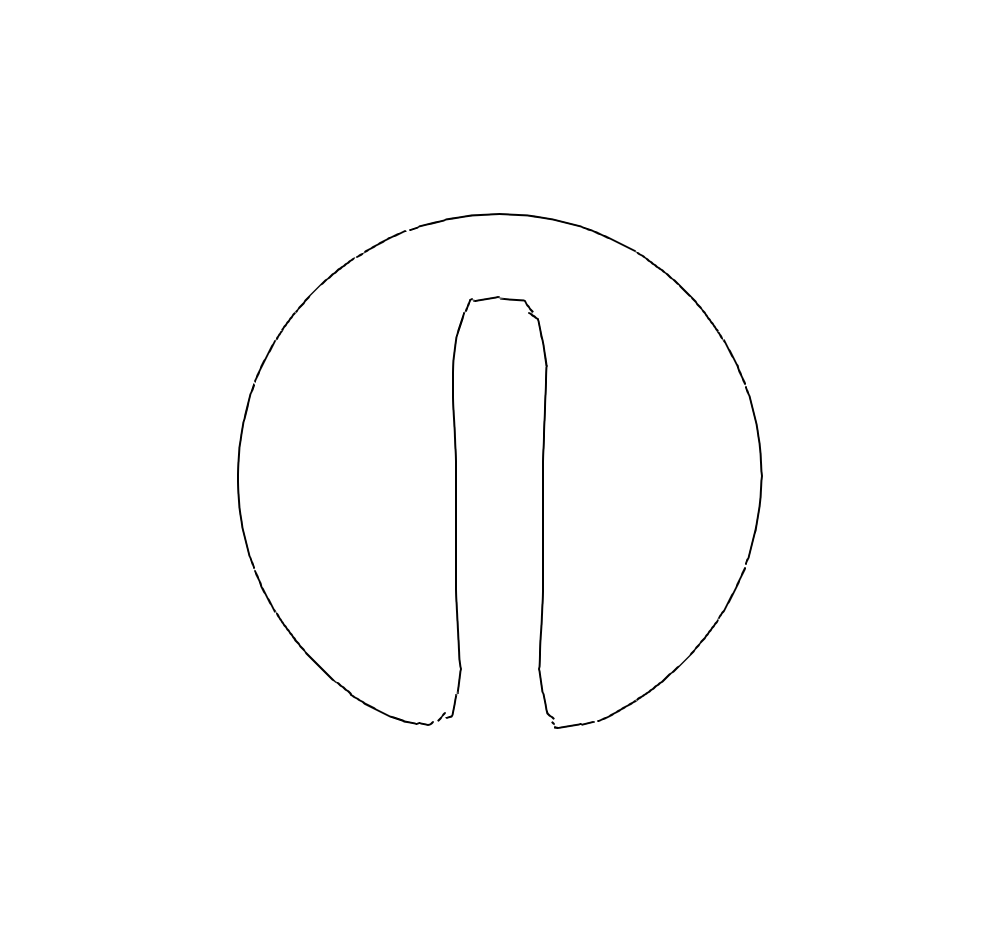}
        \caption*{$N=64,t=1,\text{R2P}$}
    \end{subfigure}\hfill
    \begin{subfigure}[b]{.33\textwidth}
        \centering\includegraphics[width=\textwidth]{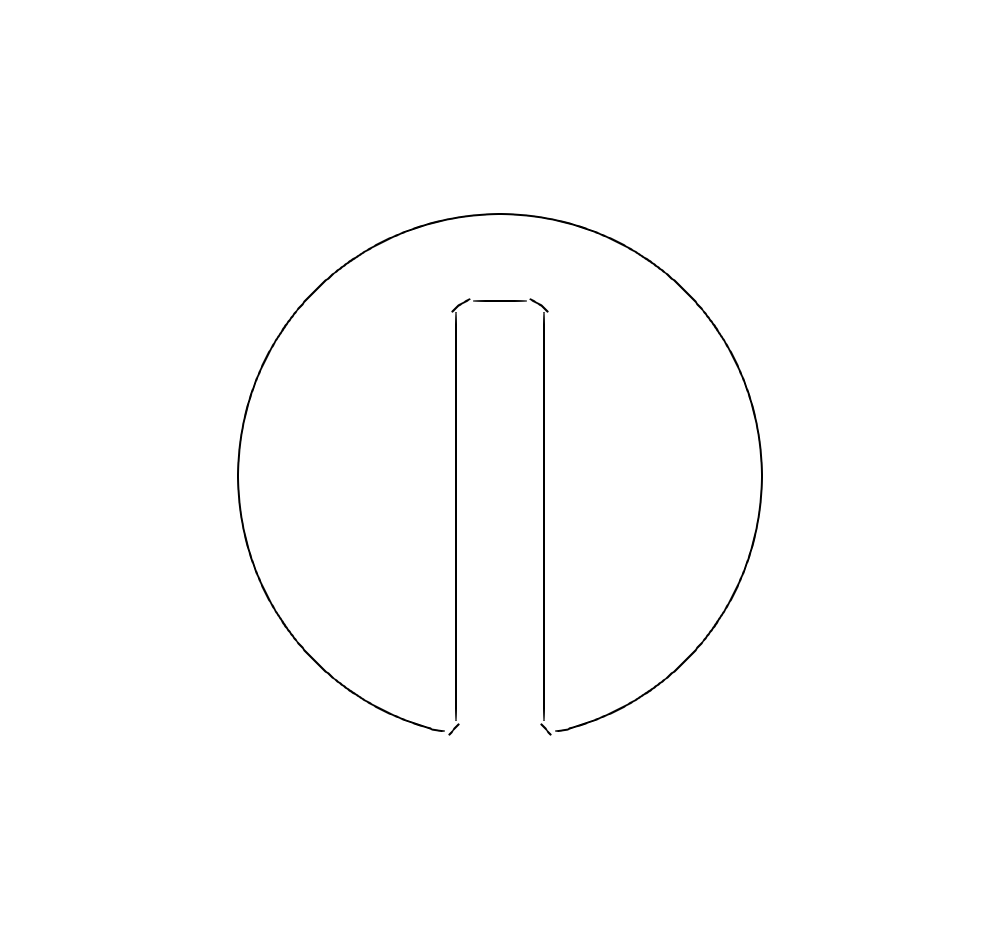}
        \caption*{$N=128,t=0$}
    \end{subfigure}\hfill
    \begin{subfigure}[b]{.33\textwidth}
        \centering\includegraphics[width=\textwidth]{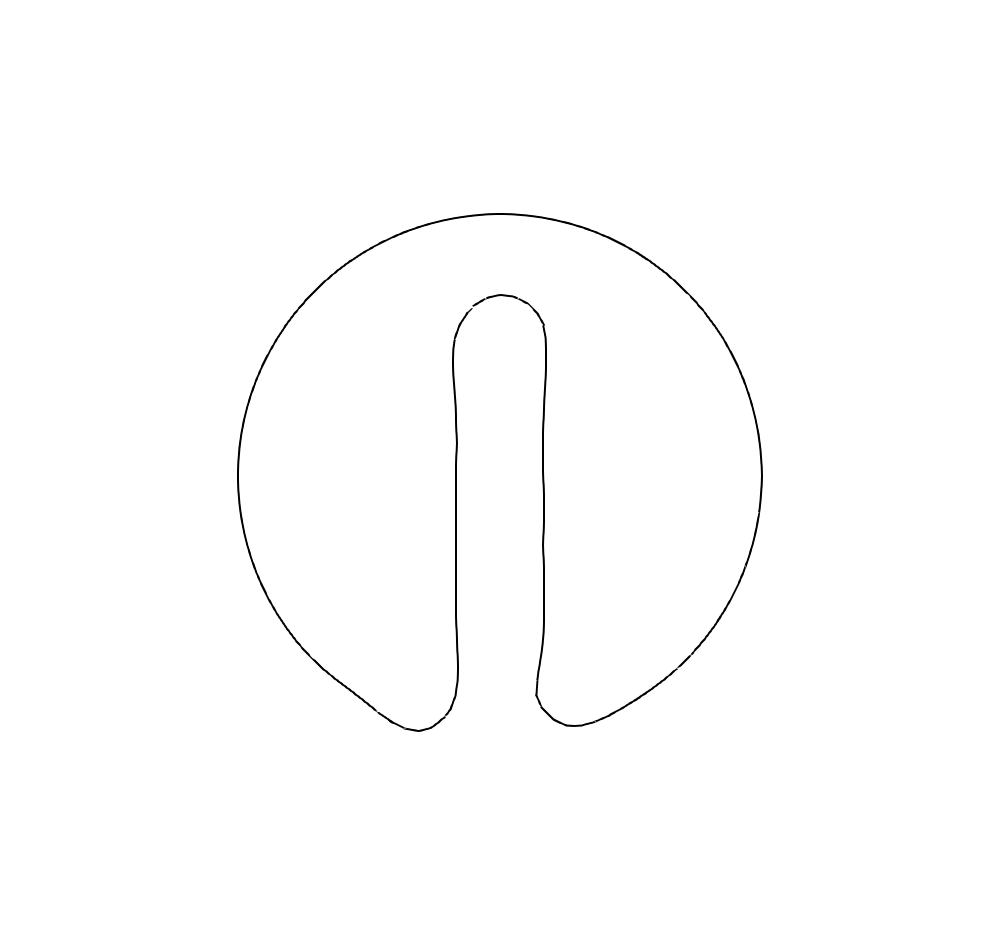}
        \caption*{$N=128,t=1,\text{ELVIRA}$}
    \end{subfigure}\hfill
    \begin{subfigure}[b]{.33\textwidth}
        \centering\includegraphics[width=\textwidth]{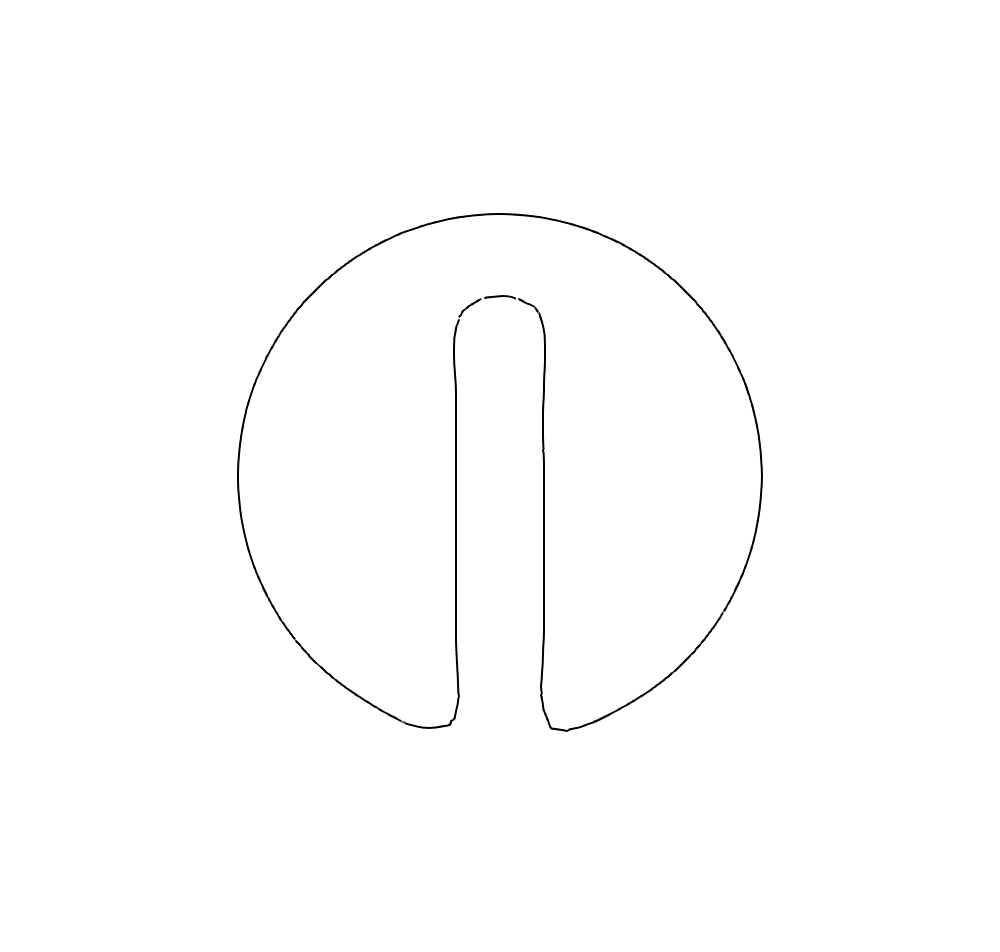}
        \caption*{$N=128,t=1,\text{R2P}$}
    \end{subfigure}\hfill  
    \caption{Interface shape for the Zalesak test case. The left column shows the initial condition, formed with an ELVIRA reconstruciton, which should be matched ideally after one revolution. The middle column shows the resulting shape after one revolution using ELVIRA interface reconstruction, while the right column shows the resulting shape after one revolution using R2P interface reconstruction.}
    \label{fig:zalesak-interface}
\end{figure}
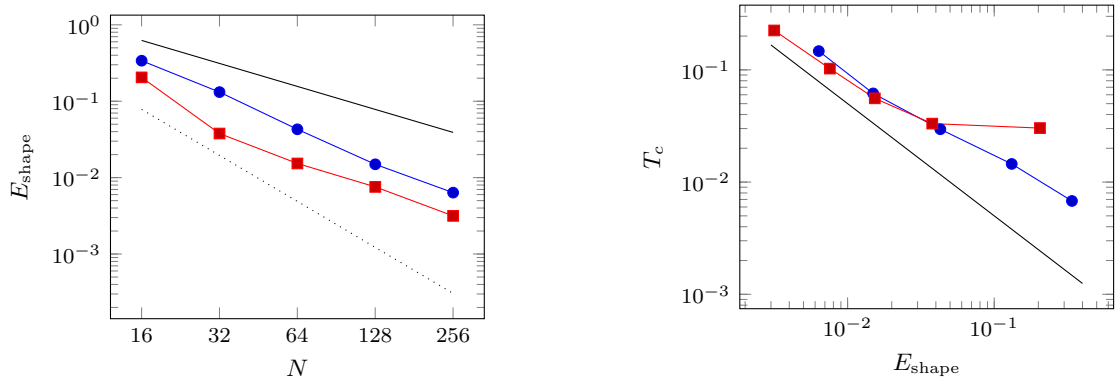
\begin{figure}
    \begin{subfigure}[b]{.5\textwidth}
        \centering
        \begin{tikzpicture}
            \begin{loglogaxis}[small,xlabel=$N$,ylabel=$E_{\text{shape}}$,log basis x=2,xticklabels={8,16,32,64,128,256}]
                \addplot table[x=nx,y=szalesak_elvira] {\timeshapeconstable};
                \label{plot:zalesak_elvira}
                \addplot table[x=nx,y=szalesak_r2p   ] {\timeshapeconstable};
                \label{plot:zalesak_r2p}
                \addplot[domain=16:256]{10/x};
                \label{plot:zalesak-1st_order}
                \addplot[dotted, domain=16:256]{20/x^2};
                \label{plot:zalesak-2nd_order}
            \end{loglogaxis}
        \end{tikzpicture}        
    \end{subfigure}
    \begin{subfigure}[b]{.5\textwidth}
        \centering
        \begin{tikzpicture}
            \begin{loglogaxis}[small,xlabel=$E_{\text{shape}}$,ylabel=$T_c$]
                \addplot table[x=szalesak_elvira,y=tzalesak_elvira] {\timeshapeconstable};
                \label{plot:zalesak_shapetime_elvira}
                \addplot table[x=szalesak_r2p,y=tzalesak_r2p   ] {\timeshapeconstable};
                \label{plot:zalesak_shapetime_r2p}
                \addplot[domain=3e-3:0.4]{.0005/x};
            \end{loglogaxis}
        \end{tikzpicture}        
    \end{subfigure}\hfill
    \caption{Zalesak case shape error (left) and computational time per time step  (right). ELVIRA (\ref{plot:zalesak_elvira}); R2P (\ref{plot:zalesak_r2p}); 1st order(\ref{plot:zalesak-1st_order}); 2nd order(\ref{plot:zalesak-2nd_order}).}
    \label{fig:zalesak-scatter}
\end{figure}
\begin{table}
    \centering
    \caption{Zalesak case conservation error.}
    \pgfplotstabletypeset[
        columns={nx,czalesak_r2p},
        columns/nx/.style={
            column name=$N$,
            int detect
        },
        columns/czalesak_r2p/.style={
            column name=$E_c$,
            dec sep align,
            precision=2
        },
    ]{\timeshapeconstable}
    \label{tab:zalesak-econs}
\end{table}

\subsection{Deformation of a Disk}
The 2D deformation case of \cite{Leveque1996} uses a cyclic velocity field to deform and recover a circular interface. The disk is deformed by the velocity field
\begin{subequations} 
    \begin{align}
        u&=+\sin^2(\pi x)\sin(2\pi y)\cos(\pi t/T),\\
        v&=-\sin^2(\pi y)\sin(2\pi x)\cos(\pi t/T),        
    \end{align}
\end{subequations}
where the disk returns to its original position and shape at $t=T$, and a value of $T=8$ is specified. The disk has a diameter $D=0.3$ and is centered at $(x,y)=(0.5,0.75)$ in a $[0,1]^2$ square domain. As in the Zalesak case, five mesh resolutions of $N=[16,32,64,128,256]$ are used, and the two-dimensional simulations are performed on a three-dimensional Cartesian mesh that is one cell in depth with cell thickness $1/N$. To produce a reference volume fraction field for the disk deformation at $t=T/2$, a high-resolution simulation is performed with $N=4096$ using ELVIRA. For each mesh resolution, the time step size is $\Delta t=64/(100N)$.
For all mesh sizes, the simulations are performed on nine cores within a single node of the Beowulf cluster.

\begin{figure}
    \begin{subfigure}[b]{.33\textwidth}
        \centering\includegraphics[width=\textwidth]{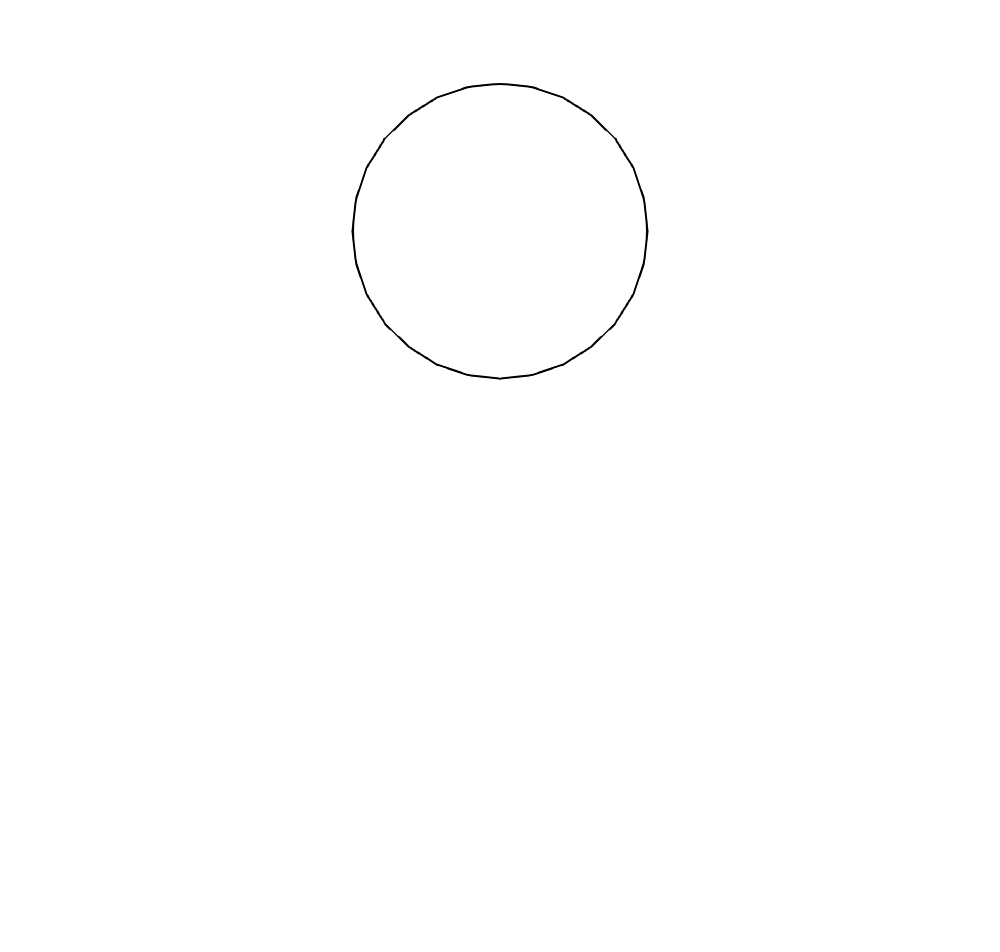}
        \caption*{$N=32,t=0$}
    \end{subfigure}\hfill
    \begin{subfigure}[b]{.33\textwidth}
        \centering\includegraphics[width=\textwidth]{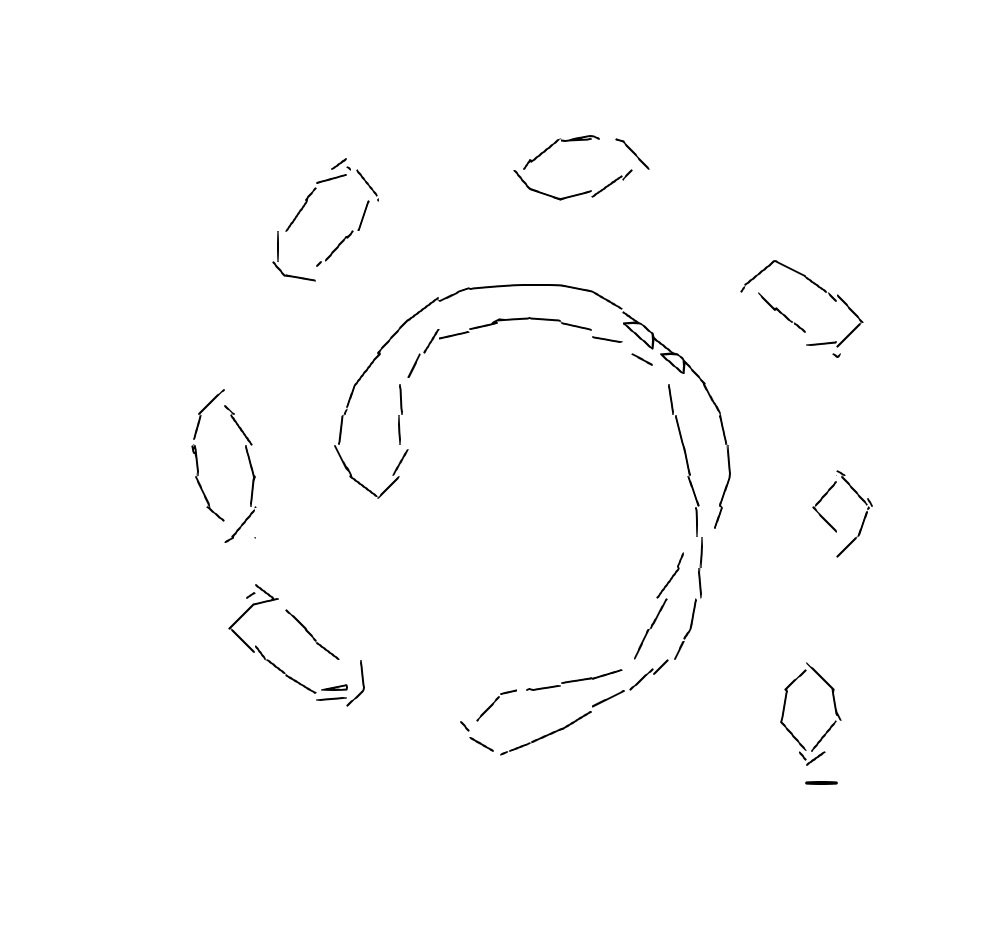}
        \caption*{$N=32,t=T/2$}
    \end{subfigure}\hfill
    \begin{subfigure}[b]{.33\textwidth}
        \centering\includegraphics[width=\textwidth]{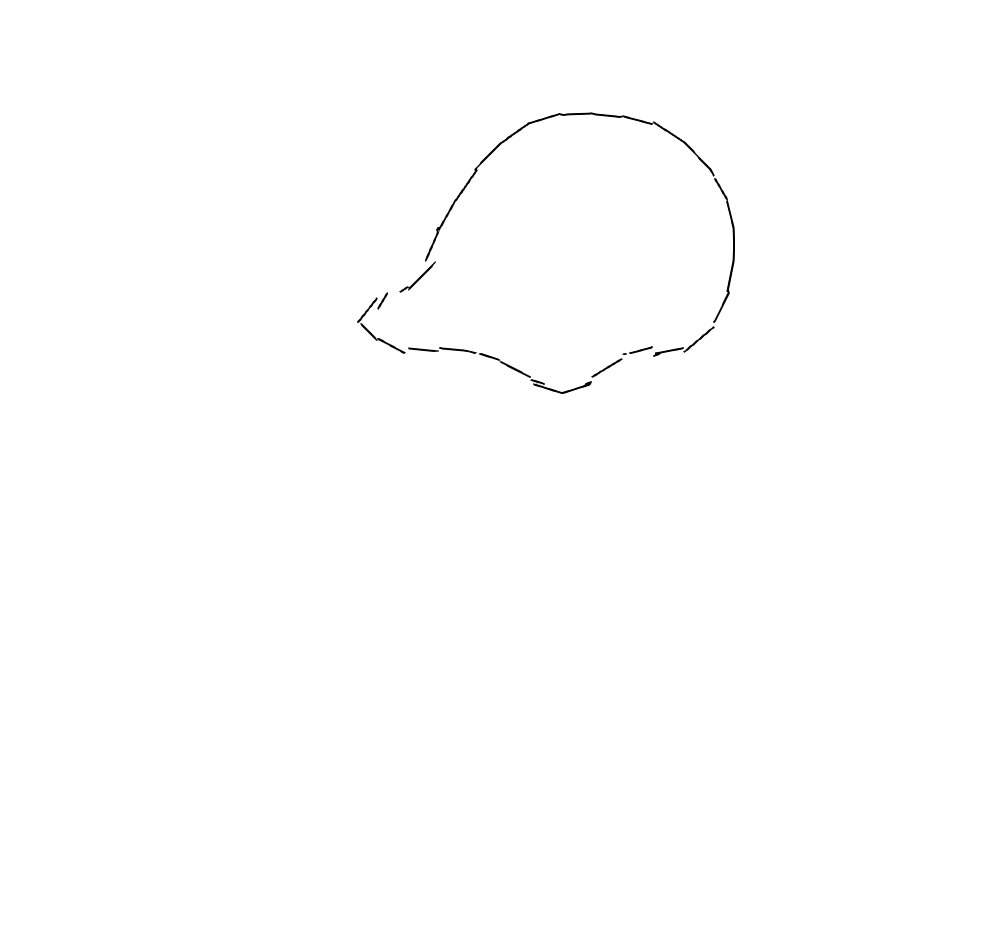}
        \caption*{$N=32,t=T$}
    \end{subfigure}\hfill
    \begin{subfigure}[b]{.33\textwidth}
        \centering\includegraphics[width=\textwidth]{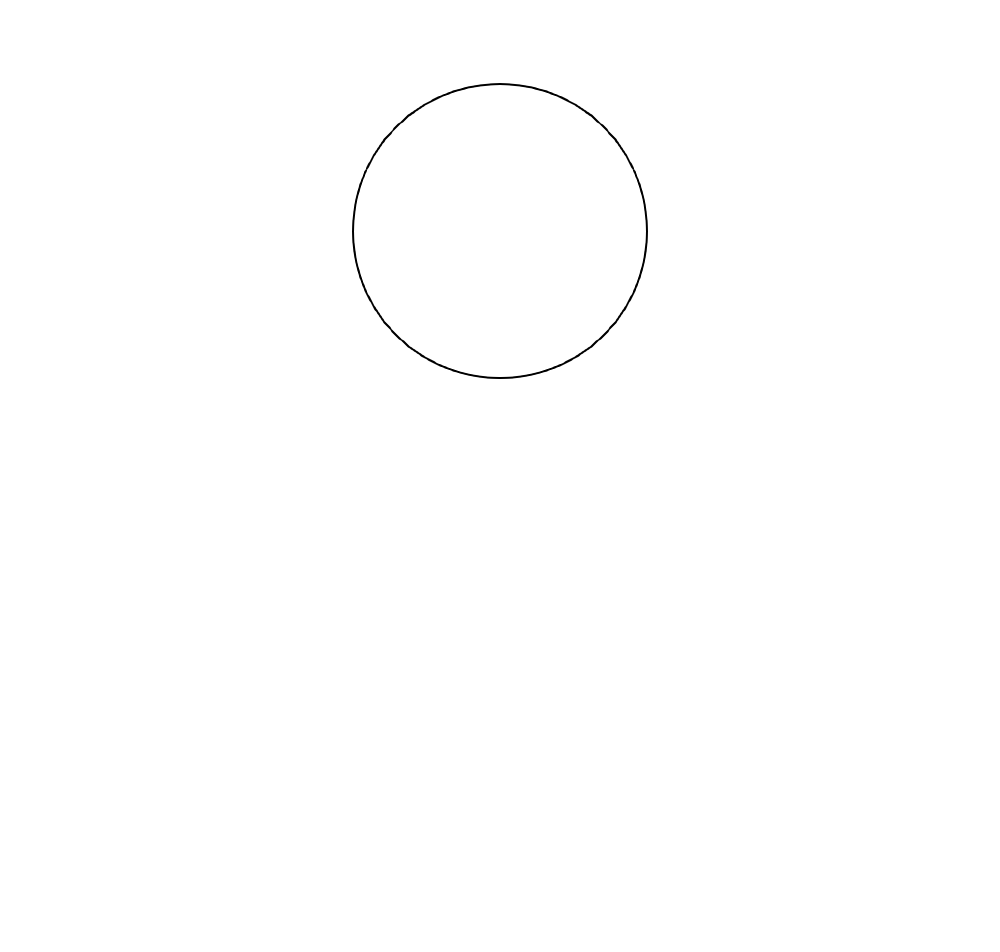}
        \caption*{$N=64,t=0$}
    \end{subfigure}\hfill
    \begin{subfigure}[b]{.33\textwidth}
        \centering\includegraphics[width=\textwidth]{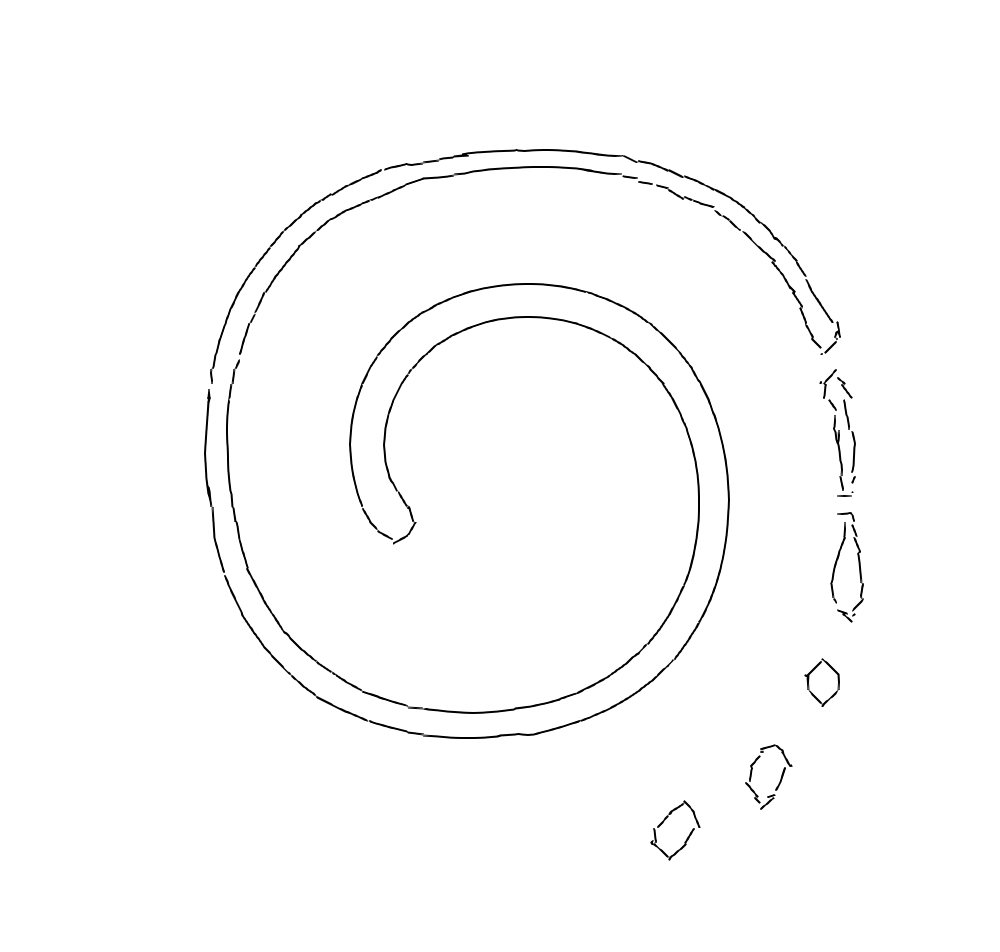}
        \caption*{$N=64,t=T/2$}
    \end{subfigure}\hfill
    \begin{subfigure}[b]{.33\textwidth}
        \centering\includegraphics[width=\textwidth]{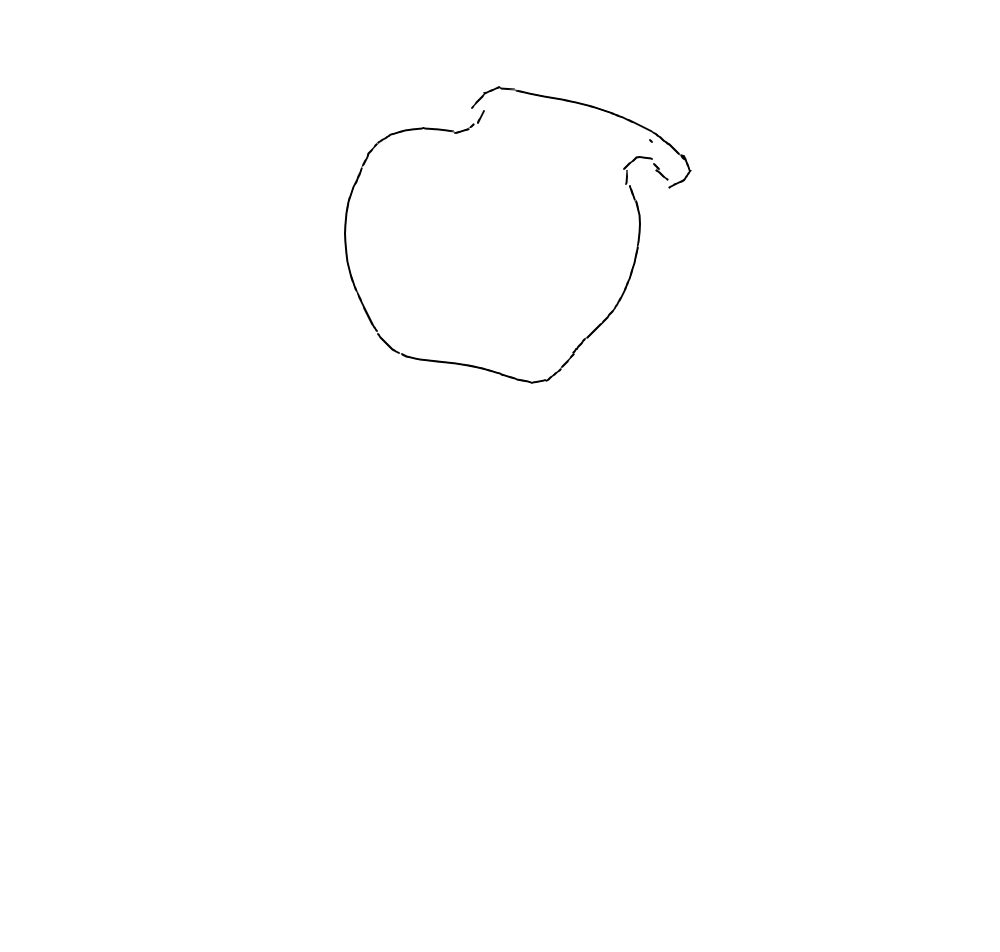}
        \caption*{$N=64,t=T$}
    \end{subfigure}\hfill
    \begin{subfigure}[b]{.33\textwidth}
        \centering\includegraphics[width=\textwidth]{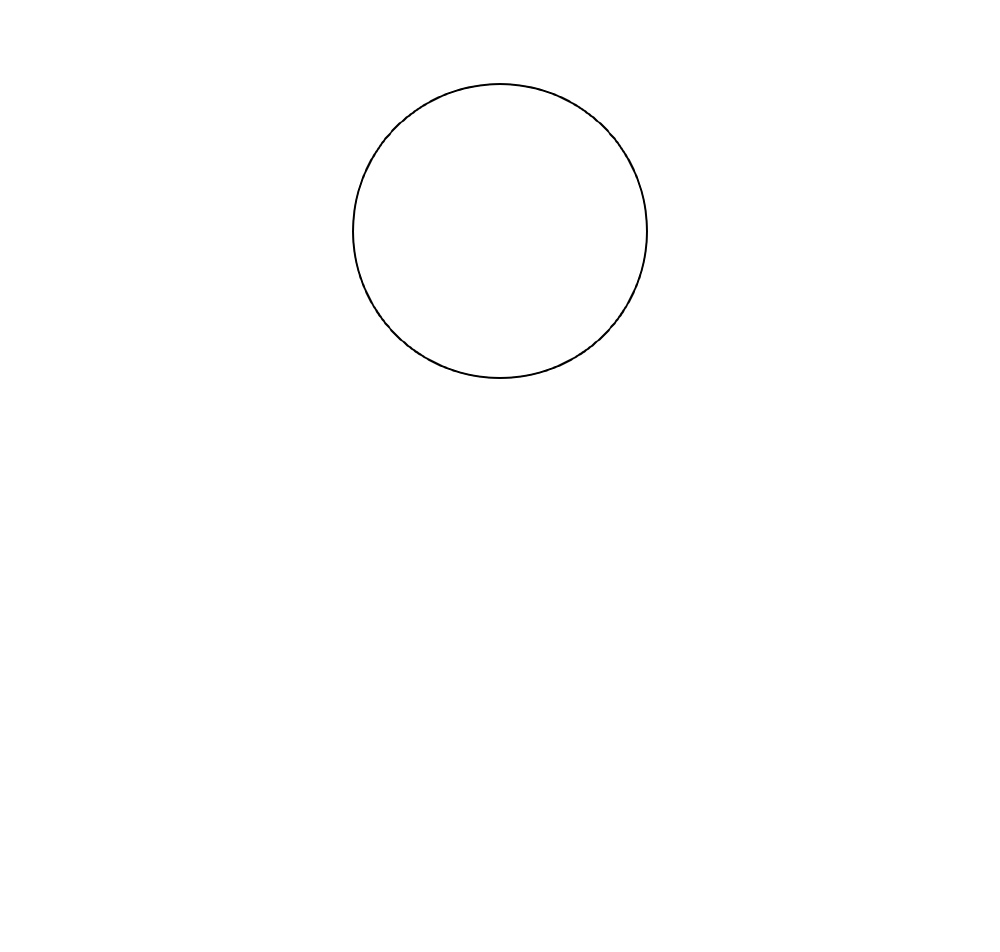}
        \caption*{$N=128,t=0$}
    \end{subfigure}\hfill
    \begin{subfigure}[b]{.33\textwidth}
        \centering\includegraphics[width=\textwidth]{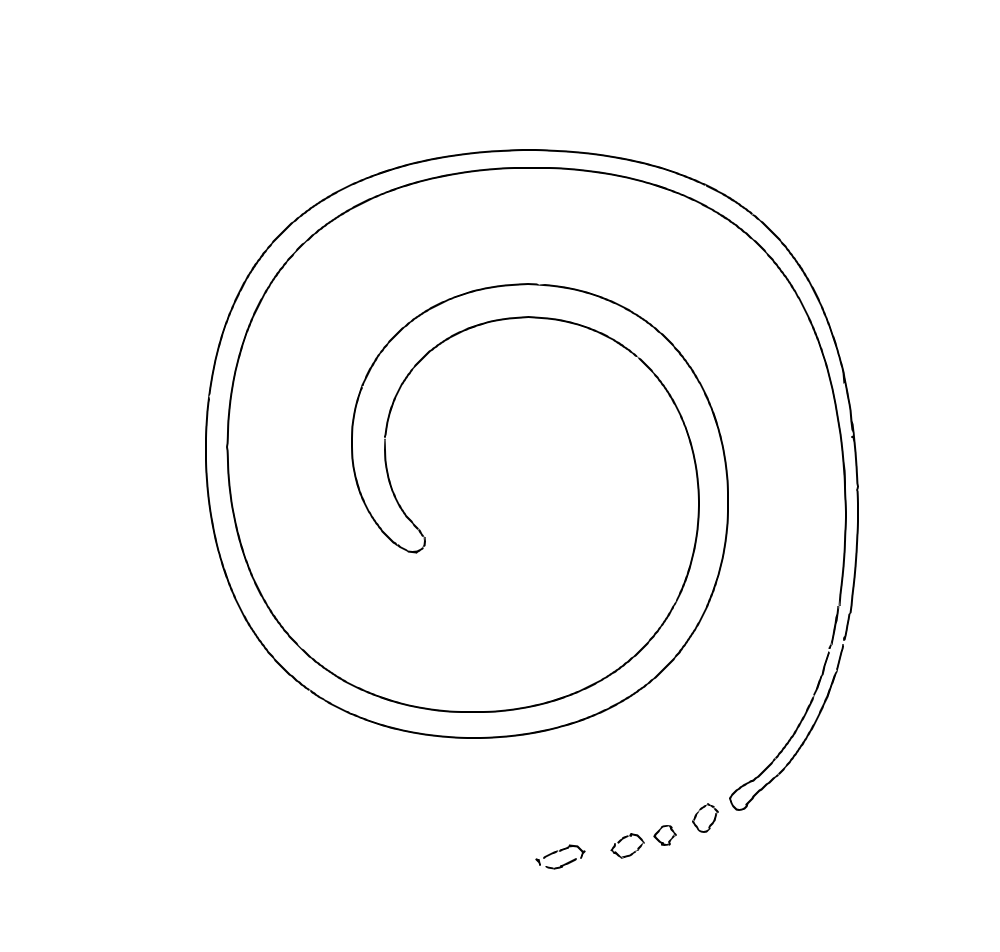}
        \caption*{$N=128,t=T/2$}
    \end{subfigure}\hfill
    \begin{subfigure}[b]{.33\textwidth}
        \centering\includegraphics[width=\textwidth]{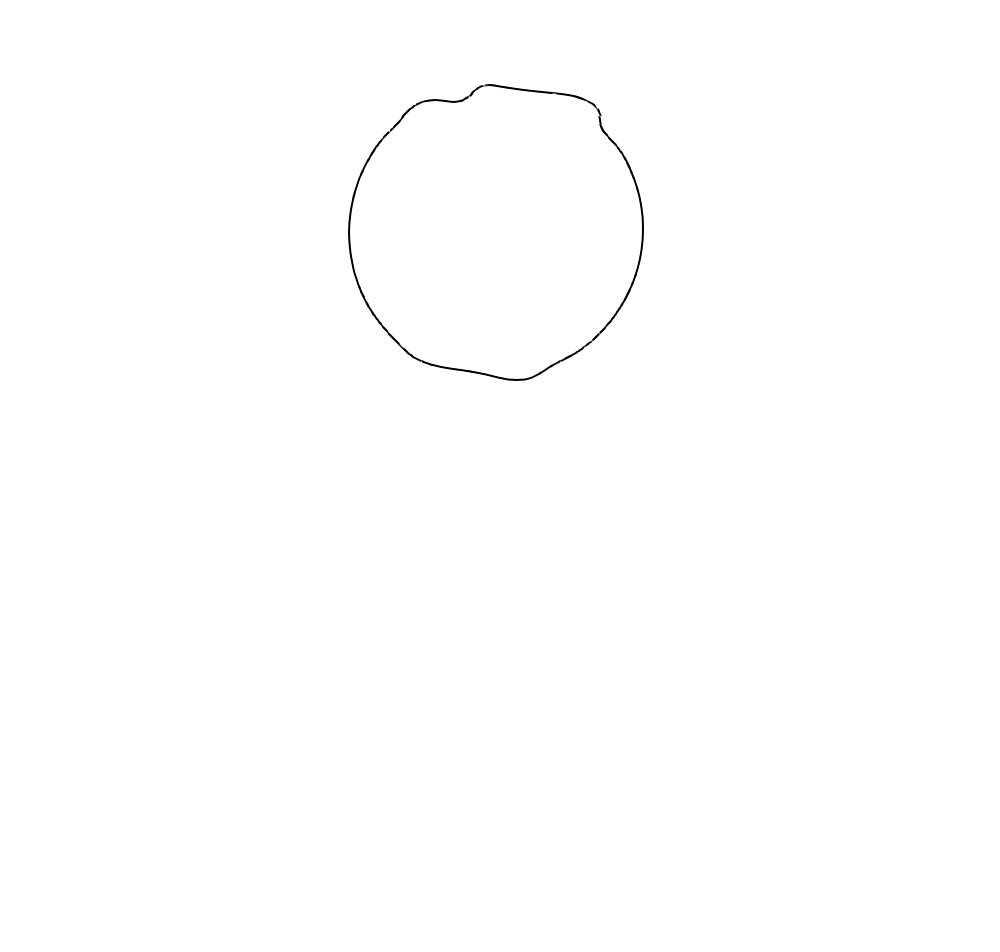}
        \caption*{$N=128,t=T$}
    \end{subfigure}\hfill
    \caption{Interface shape for the 2D deformation test case using ELVIRA interface reconstruction.}
    \label{fig:def2d-elvira}
\end{figure}

Figures \ref{fig:def2d-elvira} and \ref{fig:def2d-r2p} compare the interface shapes for the mesh resolutions $N=[32,64,128]$ using ELVIRA and R2P, respectively. When using the single-plane reconstruction, the thin film that forms by $t=T/2$ breaks up erroneously due to the lack of local resolution. While this breakup reduces with higher resolutions, the thinnest portion of the deformed disk still breaks up with $N=128$ and $256$. The erroneous interface breakup does not occur using R2P with any $N$, as the subgrid portions of the deformed disk are captured with a two-plane reconstruction.
Figure \ref{fig:def2d-scatter} shows the mesh dependence of the shape error from and computational cost of the compared reconstruction methods, where the shape error is measured at $t=T/2$ relative to the $N=4096$ reference volume fraction field. Both the ELVIRA and R2P methods exhibit first-order convergence in shape error, but the R2P method produces lower errors than those from the ELVIRA method. The computational cost $T_c$ of R2P relative to ELVIRA is comparable for a given shape error, where the cost difference between the two methods is greatest in the lowest resolution simulations where two-plane reconstruction is utilized more often. While the shape errors are similar for a given computational cost, Figures \ref{fig:def2d-elvira} and \ref{fig:def2d-r2p} clearly show that only R2P is able to preserve the interface topology for the tested mesh resolutions. Finally, the conservation error is shown in Table \ref{tab:def2d-econs}, where for all mesh resolutions, R2P preserves liquid volume to machine precision.

\begin{figure}
    \begin{subfigure}[b]{.33\textwidth}
        \centering\includegraphics[width=\textwidth]{figures/def2d/n32t0.png}
        \caption*{$N=32,t=0$}
    \end{subfigure}\hfill
    \begin{subfigure}[b]{.33\textwidth}
        \centering\includegraphics[width=\textwidth]{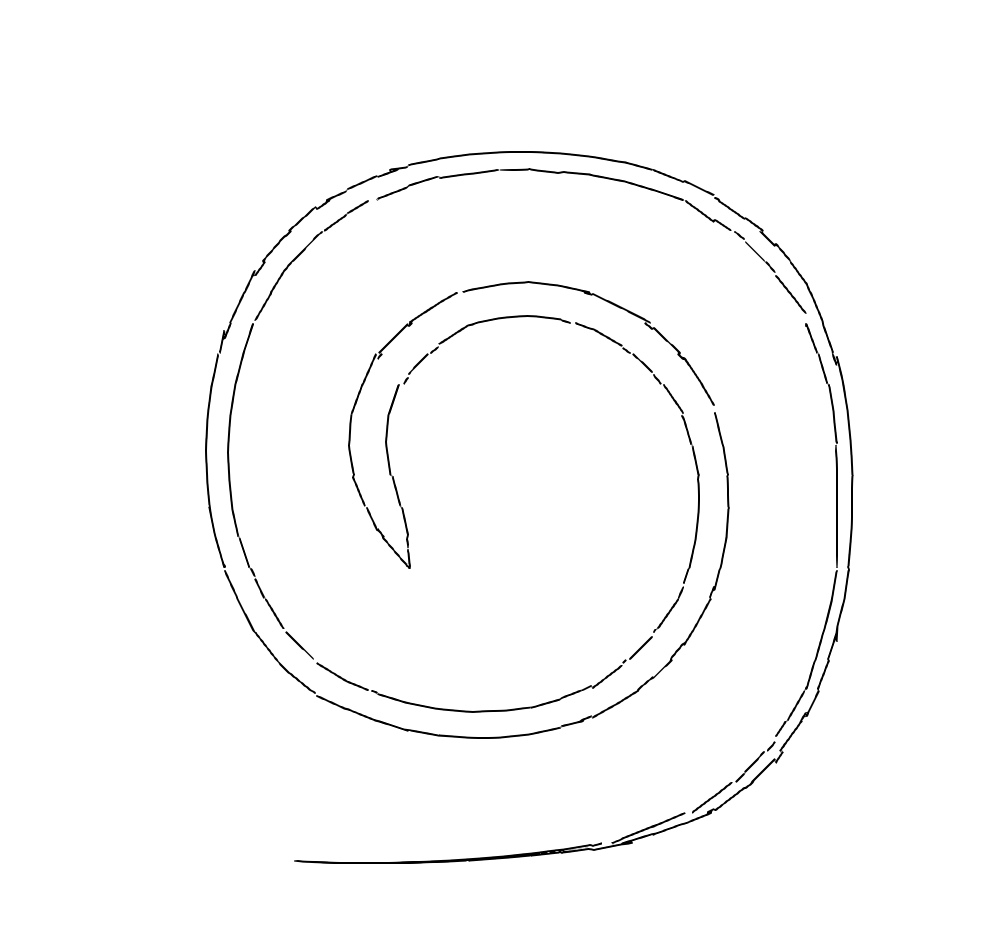}
        \caption*{$N=32,t=T/2$}
    \end{subfigure}\hfill
    \begin{subfigure}[b]{.33\textwidth}
        \centering\includegraphics[width=\textwidth]{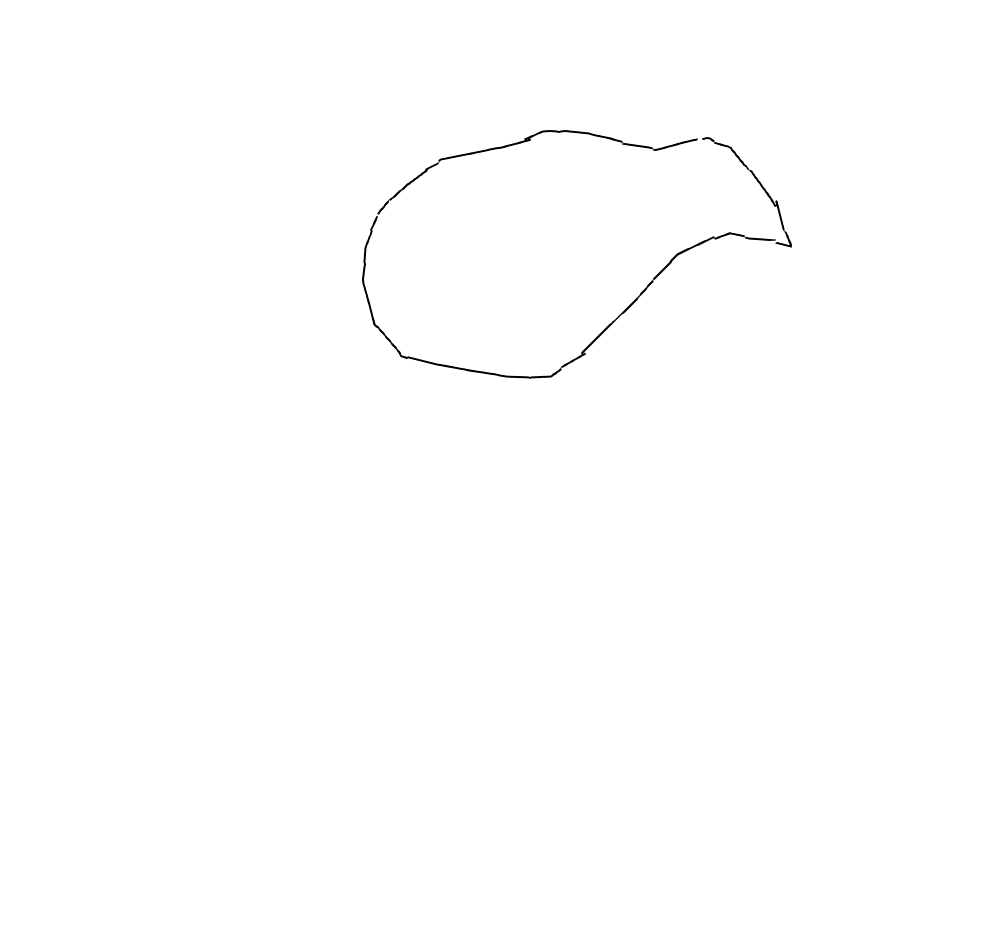}
        \caption*{$N=32,t=T$}
    \end{subfigure}\hfill
    \begin{subfigure}[b]{.33\textwidth}
        \centering\includegraphics[width=\textwidth]{figures/def2d/n64t0.png}
        \caption*{$N=64,t=0$}
    \end{subfigure}\hfill
    \begin{subfigure}[b]{.33\textwidth}
        \centering\includegraphics[width=\textwidth]{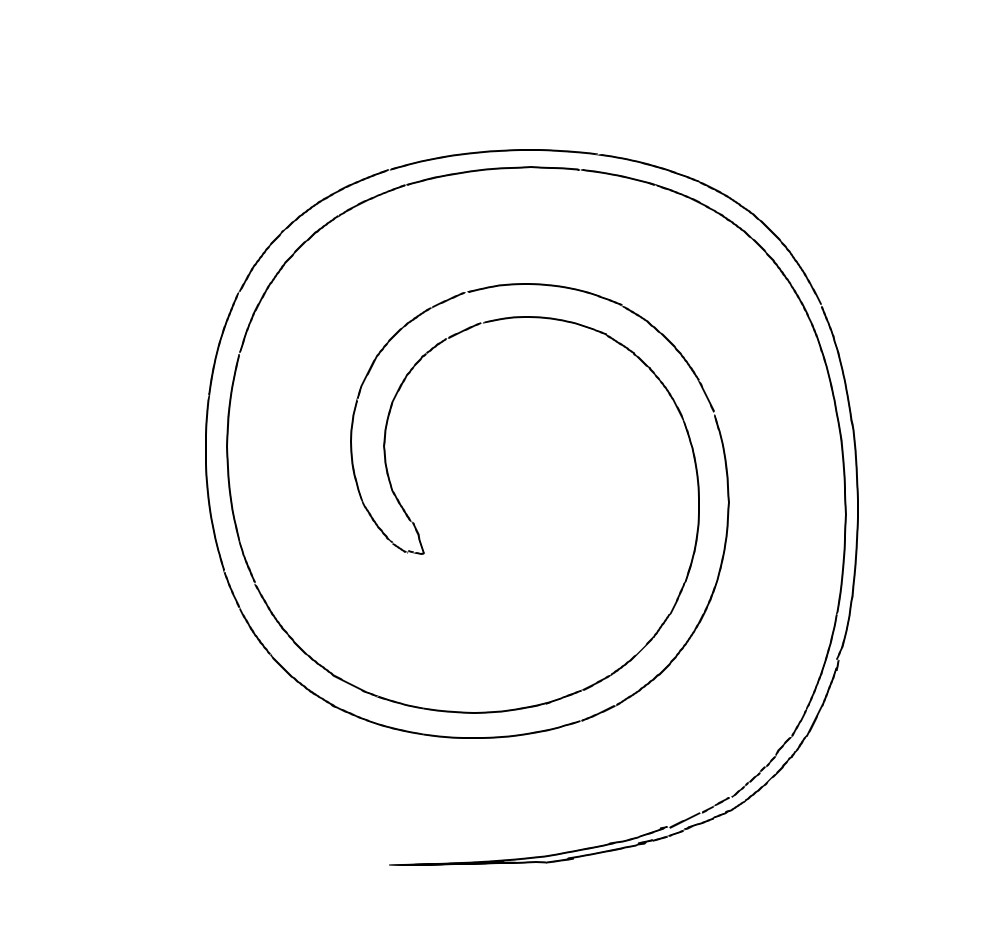}
        \caption*{$N=64,t=T/2$}
    \end{subfigure}\hfill
    \begin{subfigure}[b]{.33\textwidth}
        \centering\includegraphics[width=\textwidth]{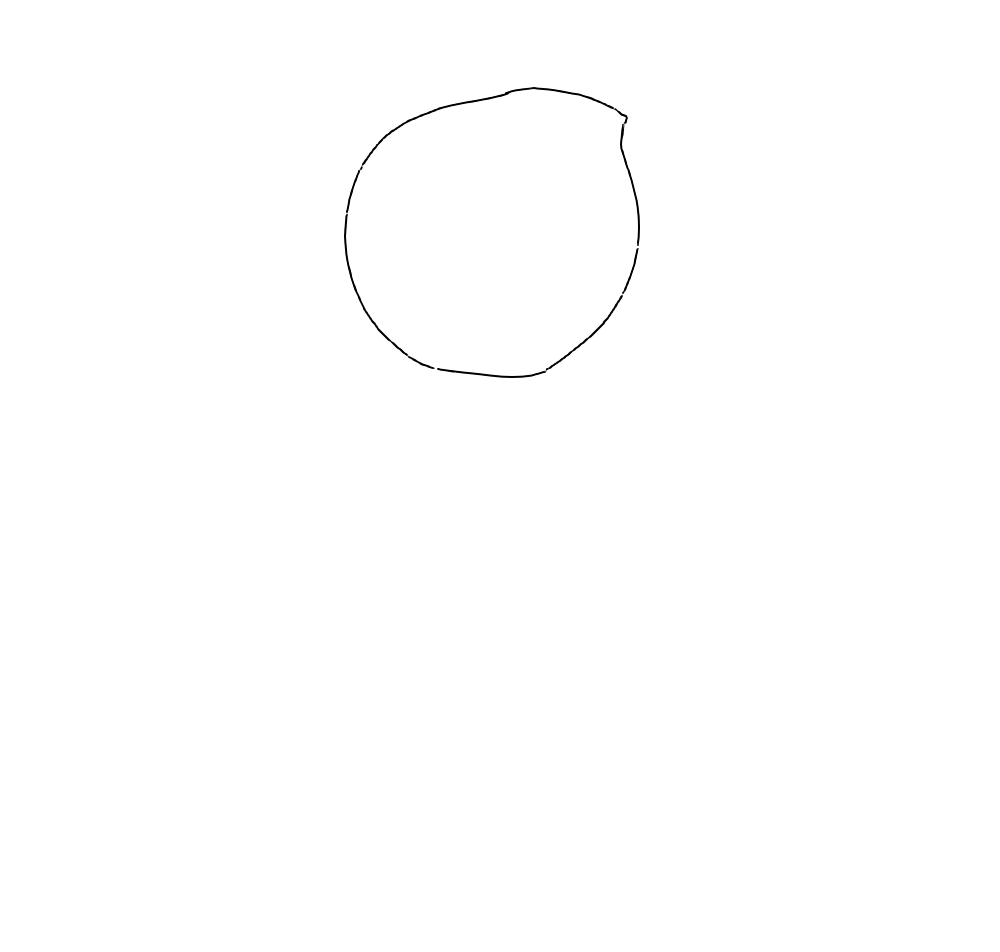}
        \caption*{$N=64,t=T$}
    \end{subfigure}\hfill
    \begin{subfigure}[b]{.33\textwidth}
        \centering\includegraphics[width=\textwidth]{figures/def2d/n128t0.png}
        \caption*{$N=128,t=0$}
    \end{subfigure}\hfill
    \begin{subfigure}[b]{.33\textwidth}
        \centering\includegraphics[width=\textwidth]{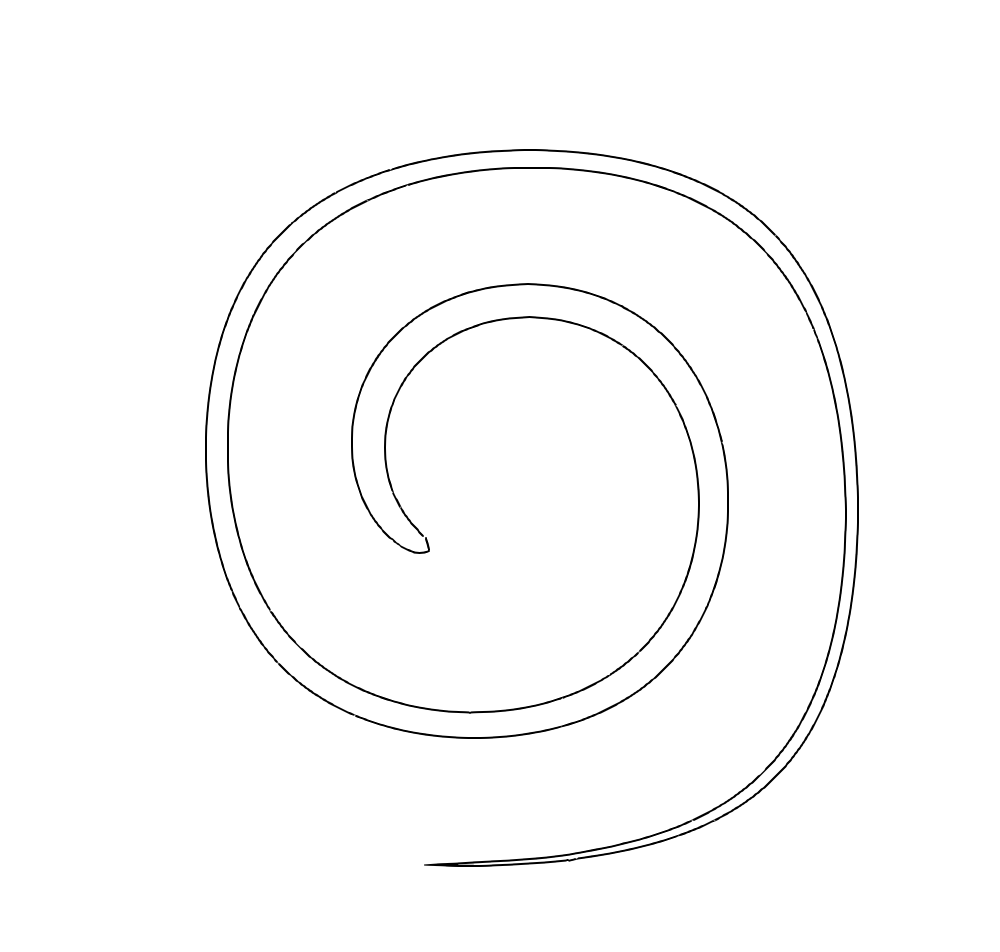}
        \caption*{$N=128,t=T/2$}
    \end{subfigure}\hfill
    \begin{subfigure}[b]{.33\textwidth}
        \centering\includegraphics[width=\textwidth]{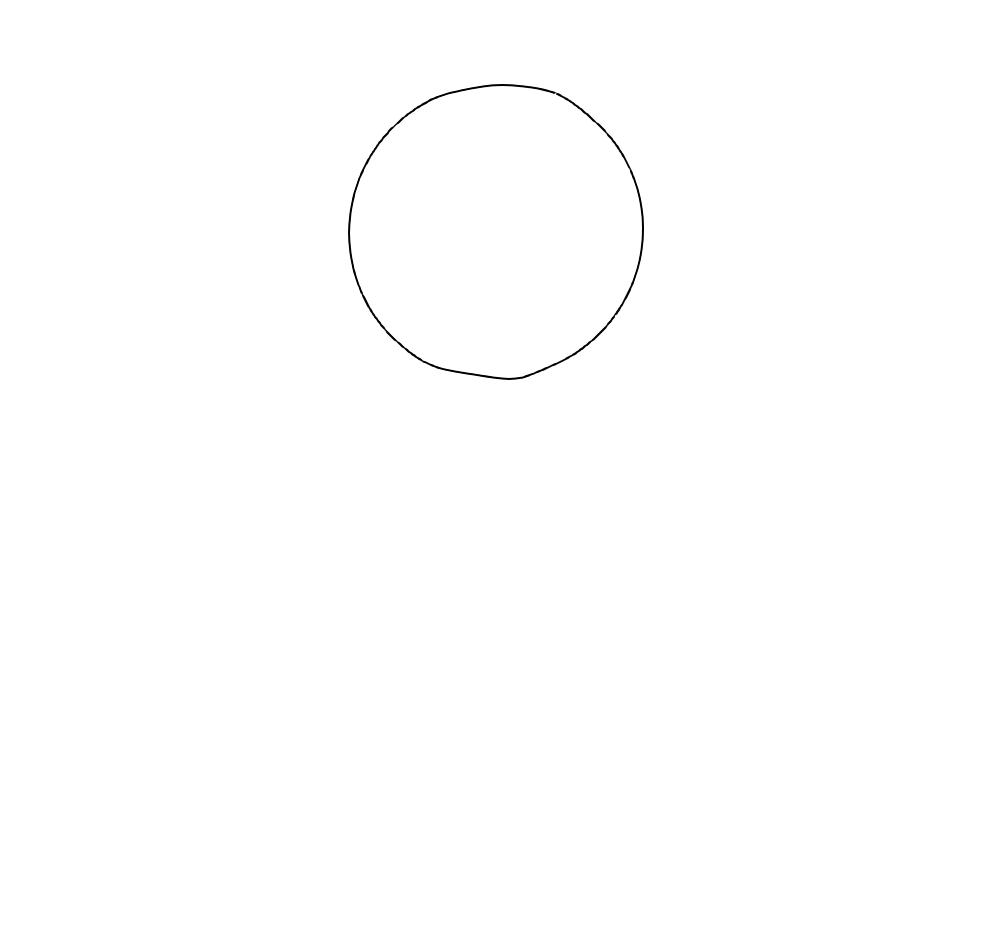}
        \caption*{$N=128,t=T$}
    \end{subfigure}\hfill  
    \caption{Interface shape for the 2D deformation test case using R2P interface reconstruction.}
    \label{fig:def2d-r2p}
\end{figure}
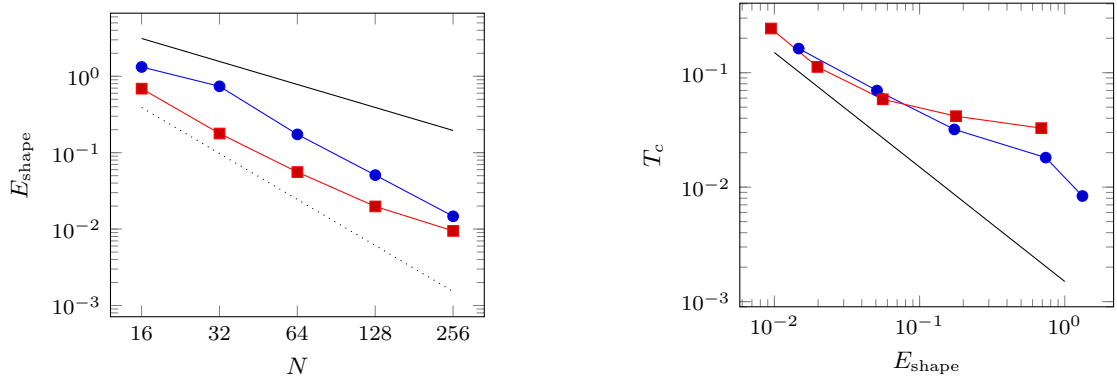
\begin{figure}
    \begin{subfigure}[b]{.5\textwidth}
        \centering
        \begin{tikzpicture}
            \begin{loglogaxis}[small,xlabel=$N$,ylabel=$E_{\text{shape}}$,log basis x=2,xticklabels={8,16,32,64,128,256}]
                \addplot table[x=nx,y=sdef2d_elvira] {\timeshapeconstable};
                \label{plot:def2d_elvira}
                \addplot table[x=nx,y=sdef2d_r2p   ] {\timeshapeconstable};
                \label{plot:def2d_r2p}
                \addplot[domain=16:256]{50/x};
                \label{plot:def2d-1st_order}
                \addplot[dotted, domain=16:256]{100/x^2};
                \label{plot:def2d-2nd_order}
            \end{loglogaxis}
        \end{tikzpicture}        
    \end{subfigure}
    \begin{subfigure}[b]{.5\textwidth}
        \centering
        \begin{tikzpicture}
            \begin{loglogaxis}[small,xlabel=$E_{\text{shape}}$,ylabel=$T_c$]
                \addplot table[x=sdef2d_elvira,y=tdef2d_elvira] {\timeshapeconstable};
                \label{plot:def2d_shapetime_elvira}
                \addplot table[x=sdef2d_r2p,y=tdef2d_r2p   ] {\timeshapeconstable};
                \label{plot:def2d_shapetime_r2p}
                \addplot[domain=1e-2:1]{.0015/x};
            \end{loglogaxis}
        \end{tikzpicture}        
    \end{subfigure}\hfill
    \caption{2D deformation case shape error (left) and computational time per time step  (right) at $t=T/2$. ELVIRA (\ref{plot:def2d_elvira}); R2P (\ref{plot:def2d_r2p}); 1st order(\ref{plot:def2d-1st_order}); 2nd order(\ref{plot:def2d-2nd_order}).}
    \label{fig:def2d-scatter}
\end{figure}
\begin{table}
    \centering
    \caption{2D deformation case conservation error.}
    \pgfplotstabletypeset[
        columns={nx,cdef2d_r2p},
        columns/nx/.style={
            column name=$N$,
            int detect
        },
        columns/cdef2d_r2p/.style={
            column name=$E_c$,
            dec sep align,
            precision=2
        },
    ]{\timeshapeconstable}
    \label{tab:def2d-econs}
\end{table}   
\subsection{Deformation of a Sphere}
In the 3D deformation case, also proposed by \cite{Leveque1996}, a sphere is deformed by the time-dependent velocity field
\begin{subequations}
    \begin{align}
        u&=2\sin^2(\pi x)\sin(2\pi y)\sin(2\pi z)\cos(\pi t/T),\\
        v&=-\sin(2\pi x)\sin^2(\pi y)\sin(2\pi z)\cos(\pi t/T),\\
        w&=-\sin(2\pi x)\sin(2\pi y)\sin^2(\pi z)\cos(\pi t/T), 
    \end{align}        
\end{subequations}
where the sphere returns to its original position and shape at $t=T$. In this work, a final time of $T=3$ is used. The sphere has a diameter $D=0.3$ and is centered at $(x,y,z)=(0.35,0.35,0.35)$ in a $[0,1]^3$ cubic domain. In each spatial dimension, $N=[16,32,64,128,256]$ cells are used, and the time step size is $\Delta t=32/(100N)$. To produce a reference volume fraction field for the sphere deformation at $t=T/2$, a high-resolution simulation is performed with $N=768$ using R2P. For all mesh sizes, the simulations are performed on eight nodes (using eight cores each) of the Beowulf cluster.

\begin{figure}
    \begin{subfigure}[b]{.33\textwidth}
        \centering\includegraphics[width=\textwidth]{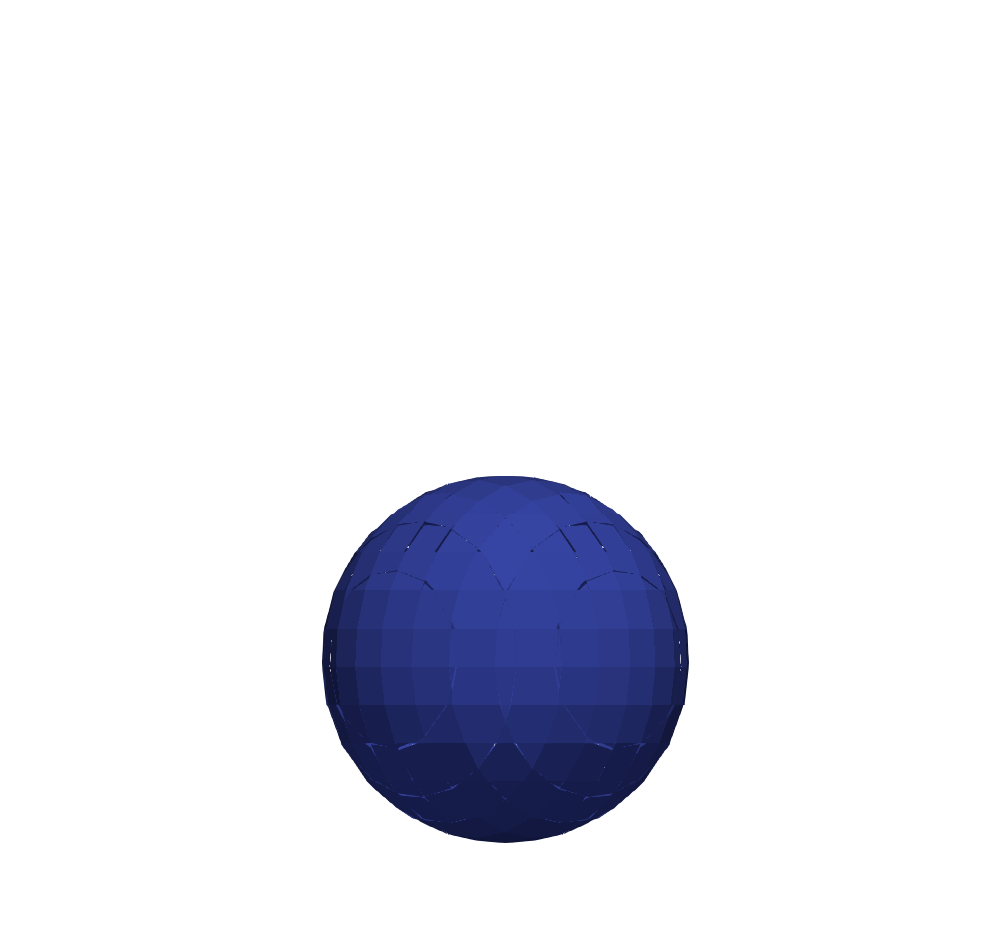}
        \caption*{$N=32,t=0$}
    \end{subfigure}\hfill
    \begin{subfigure}[b]{.33\textwidth}
        \centering\includegraphics[width=\textwidth]{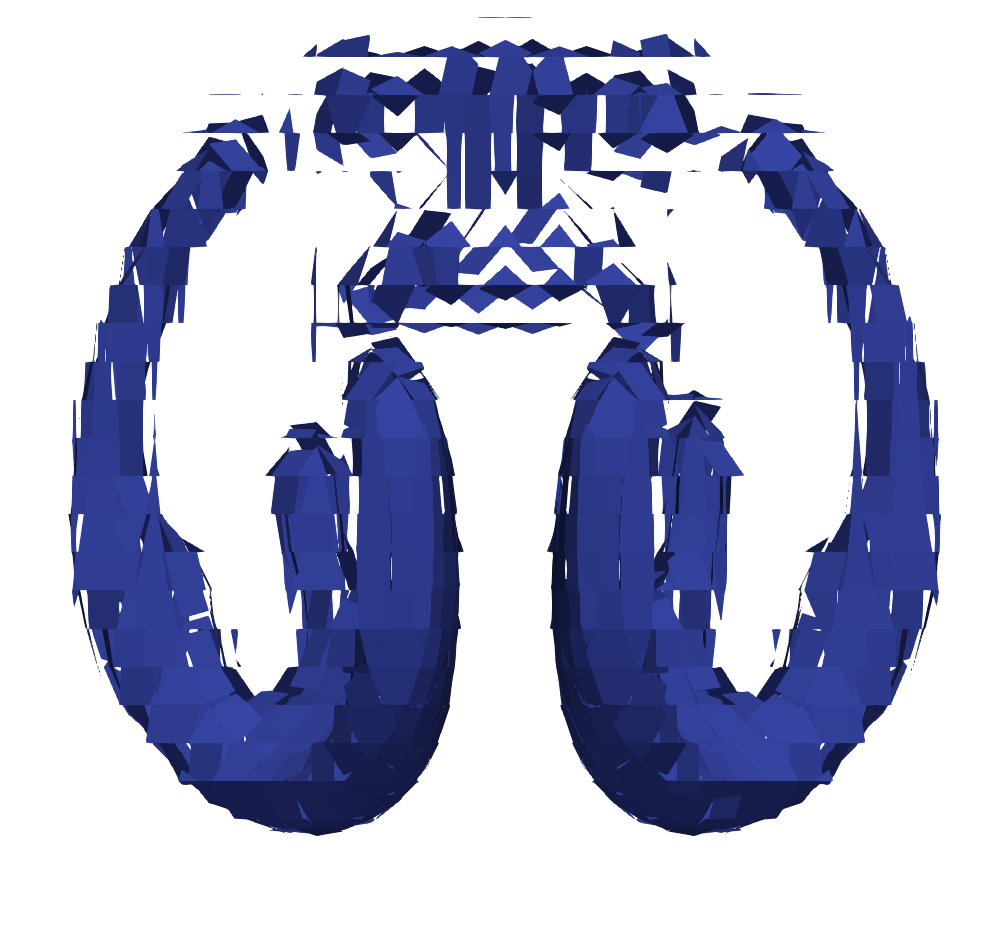}
        \caption*{$N=32,t=T/2$}
    \end{subfigure}\hfill
    \begin{subfigure}[b]{.33\textwidth}
        \centering\includegraphics[width=\textwidth]{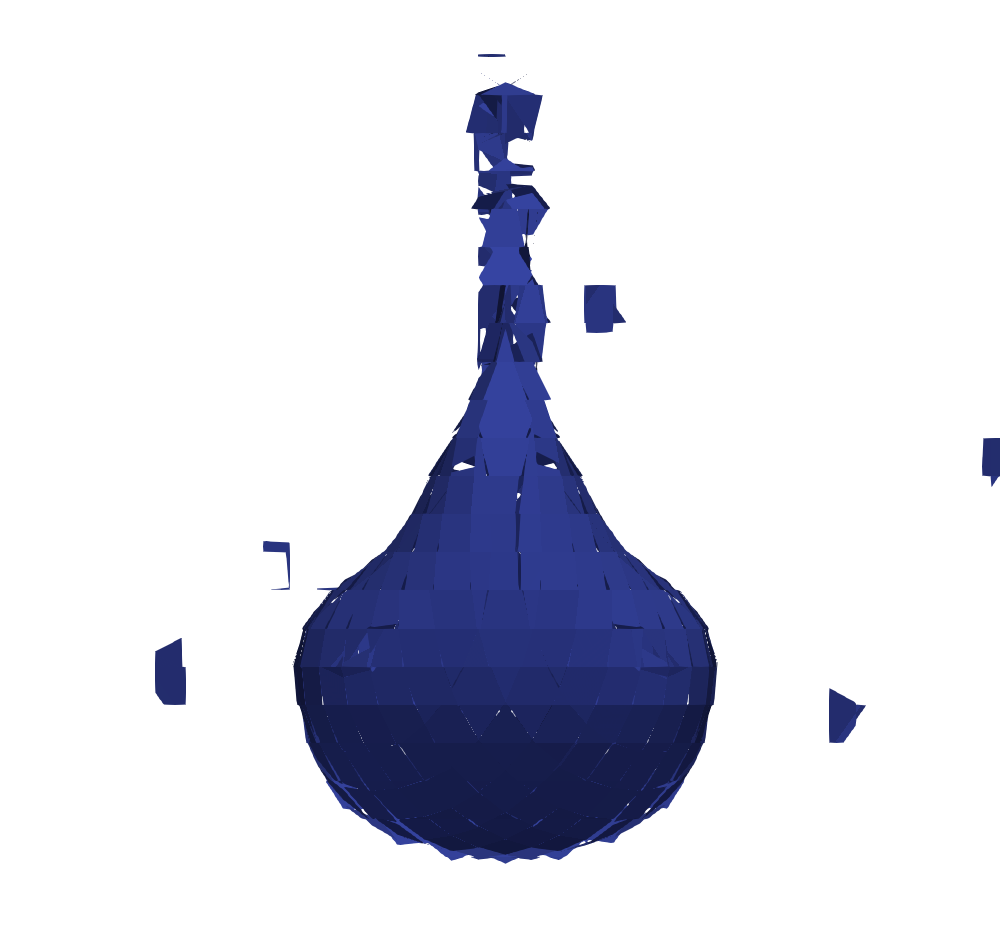}
        \caption*{$N=32,t=T$}
    \end{subfigure}\hfill
    \par\medskip
    \begin{subfigure}[b]{.33\textwidth}
        \centering\includegraphics[width=\textwidth]{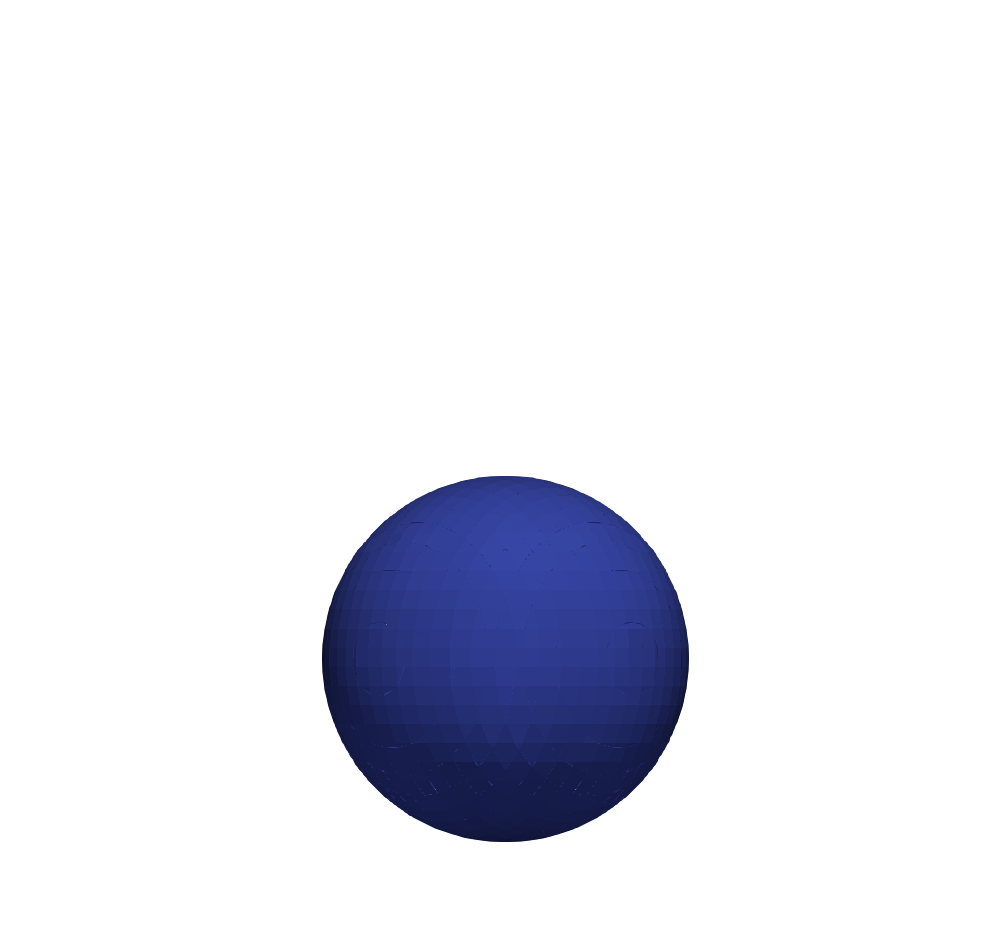}
        \caption*{$N=64,t=0$}
    \end{subfigure}\hfill
    \begin{subfigure}[b]{.33\textwidth}
        \centering\includegraphics[width=\textwidth]{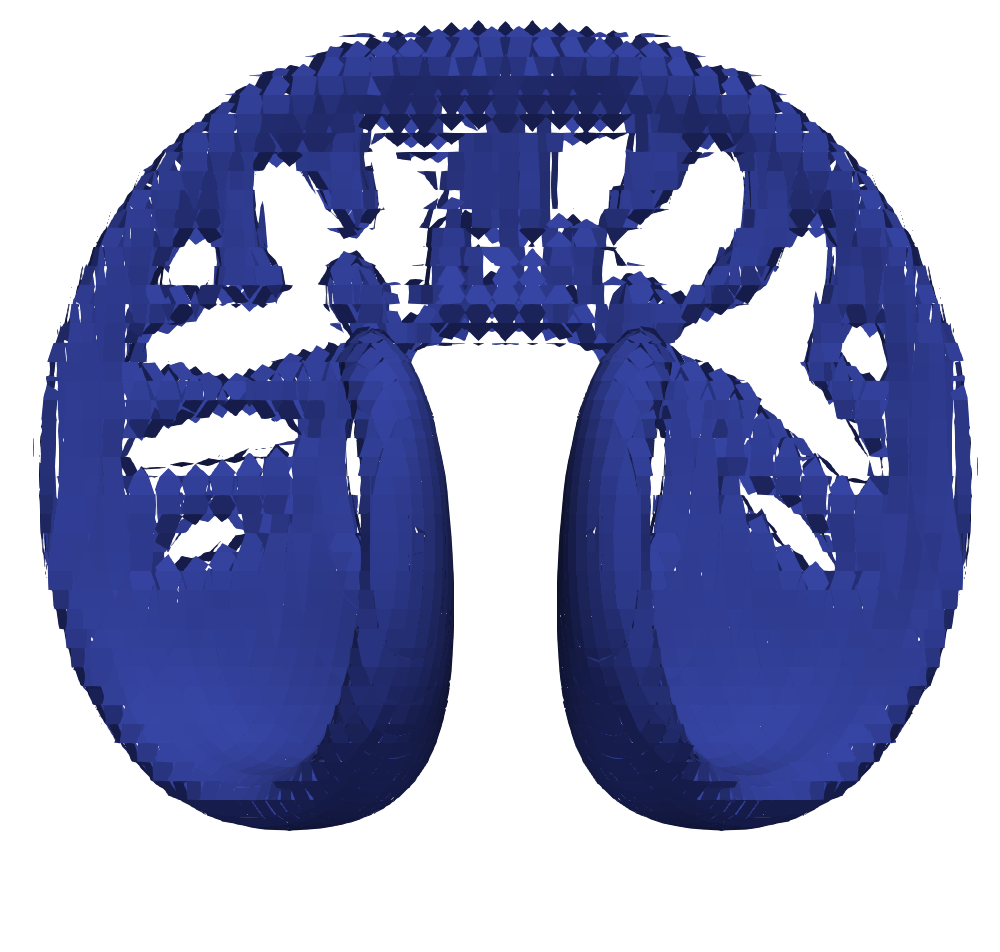}
        \caption*{$N=64,t=T/2$}
    \end{subfigure}\hfill
    \begin{subfigure}[b]{.33\textwidth}
        \centering\includegraphics[width=\textwidth]{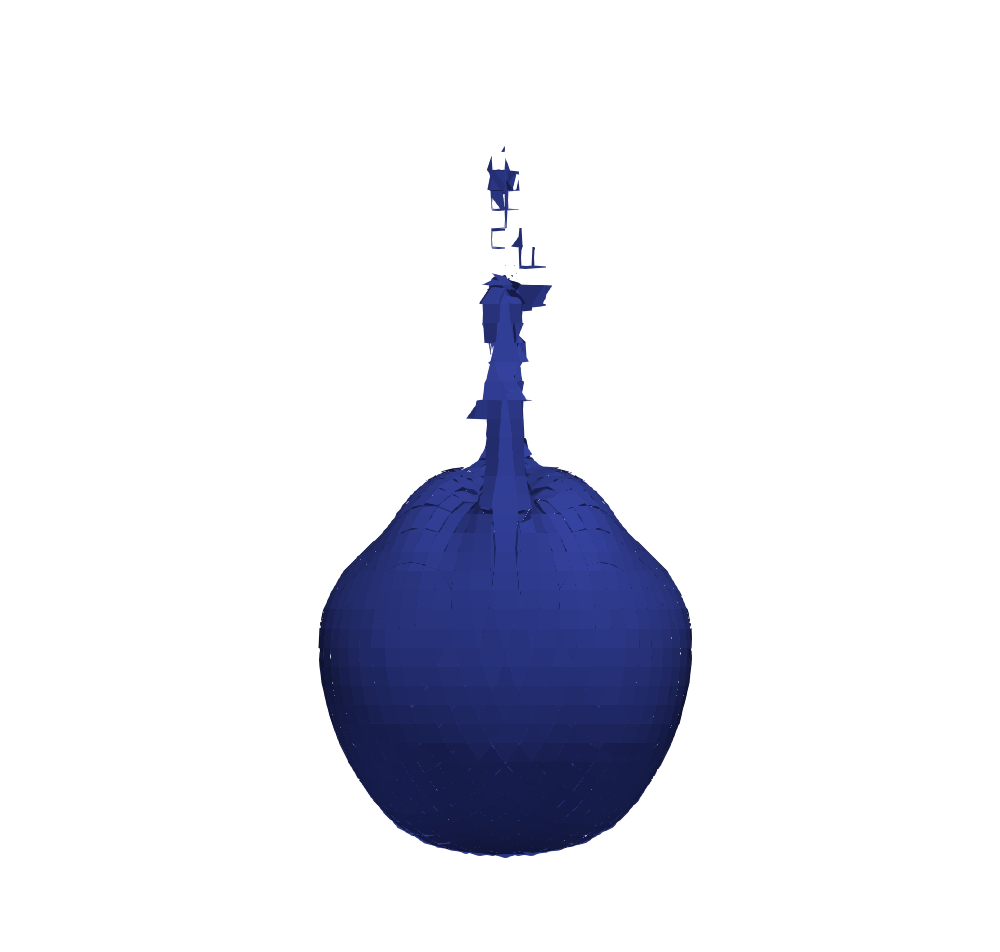}
        \caption*{$N=64,t=T$}
    \end{subfigure}\hfill
    \par\medskip
    \begin{subfigure}[b]{.33\textwidth}
        \centering\includegraphics[width=\textwidth]{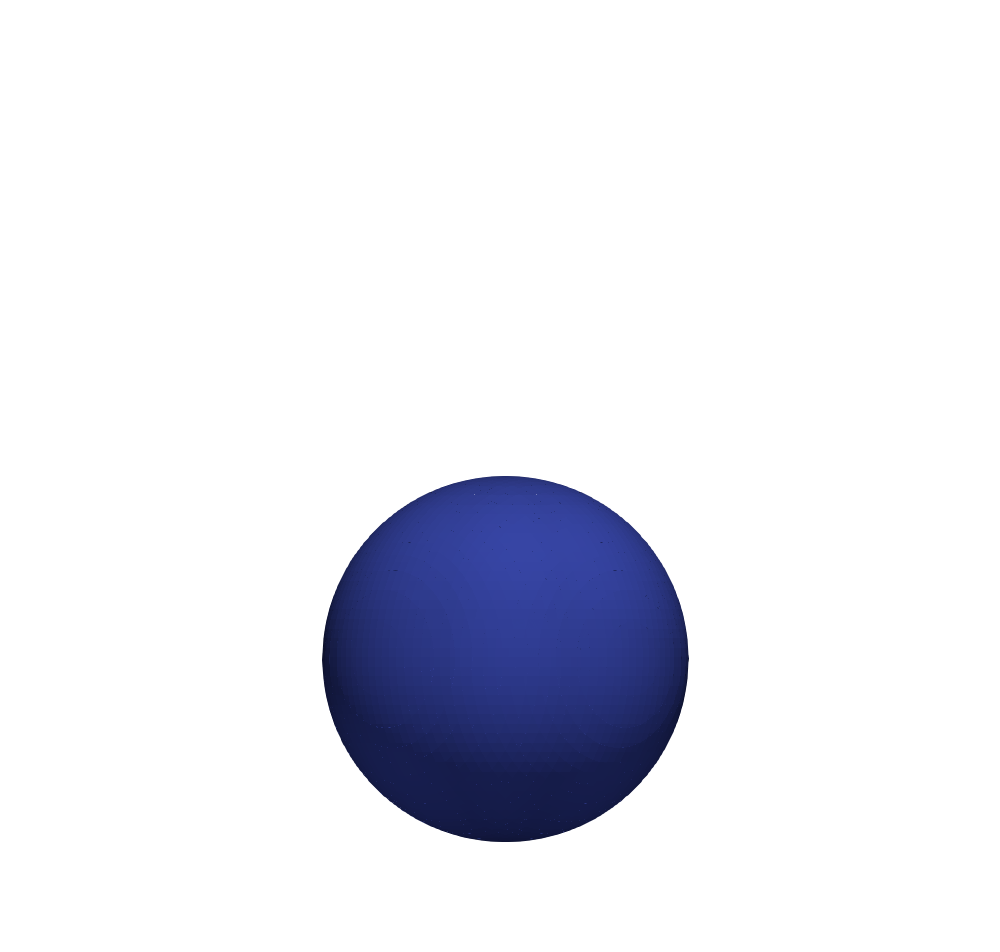}
        \caption*{$N=128,t=0$}
    \end{subfigure}\hfill
    \begin{subfigure}[b]{.33\textwidth}
        \centering\includegraphics[width=\textwidth]{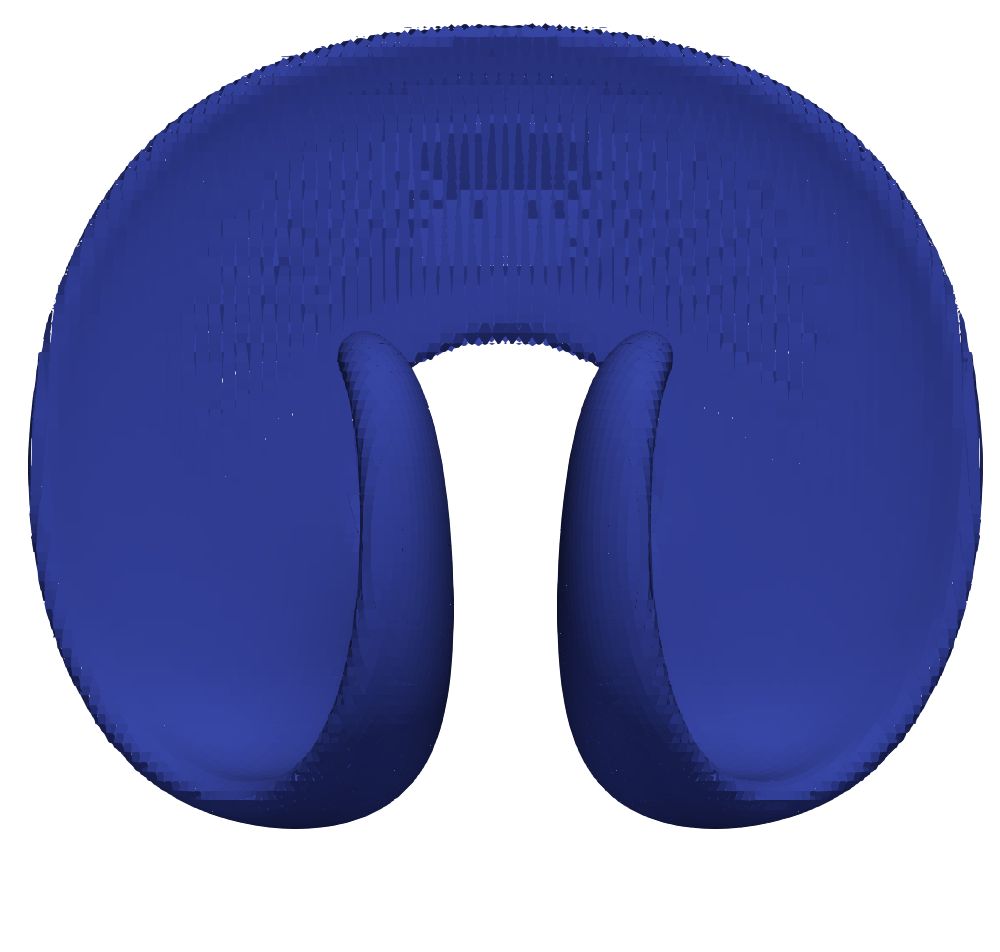}
        \caption*{$N=128,t=T/2$}
    \end{subfigure}\hfill
    \begin{subfigure}[b]{.33\textwidth}
        \centering\includegraphics[width=\textwidth]{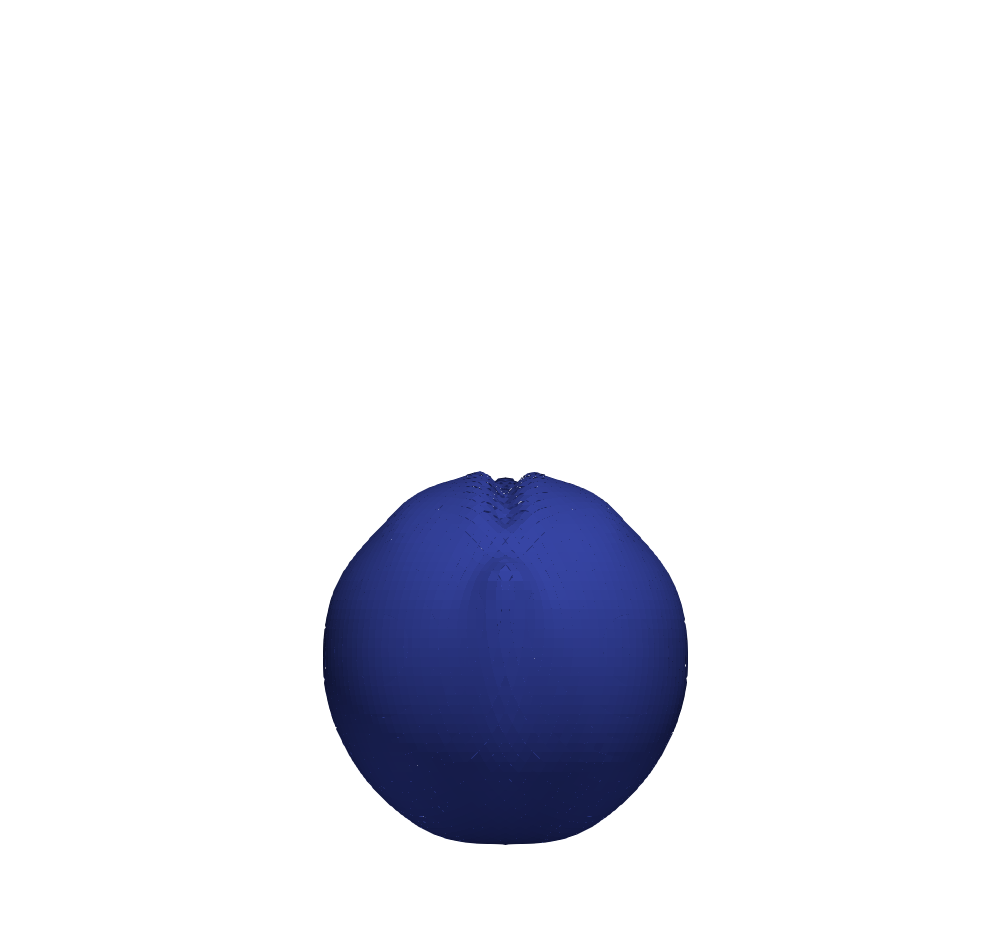}
        \caption*{$N=128,t=T$}
    \end{subfigure}\hfill
    \caption{Interface shape for the 3D deformation test case using ELVIRA interface reconstruction.}
    \label{fig:def3d-elvira}
\end{figure}

\begin{figure}
    \begin{subfigure}[b]{.33\textwidth}
        \centering\includegraphics[width=\textwidth]{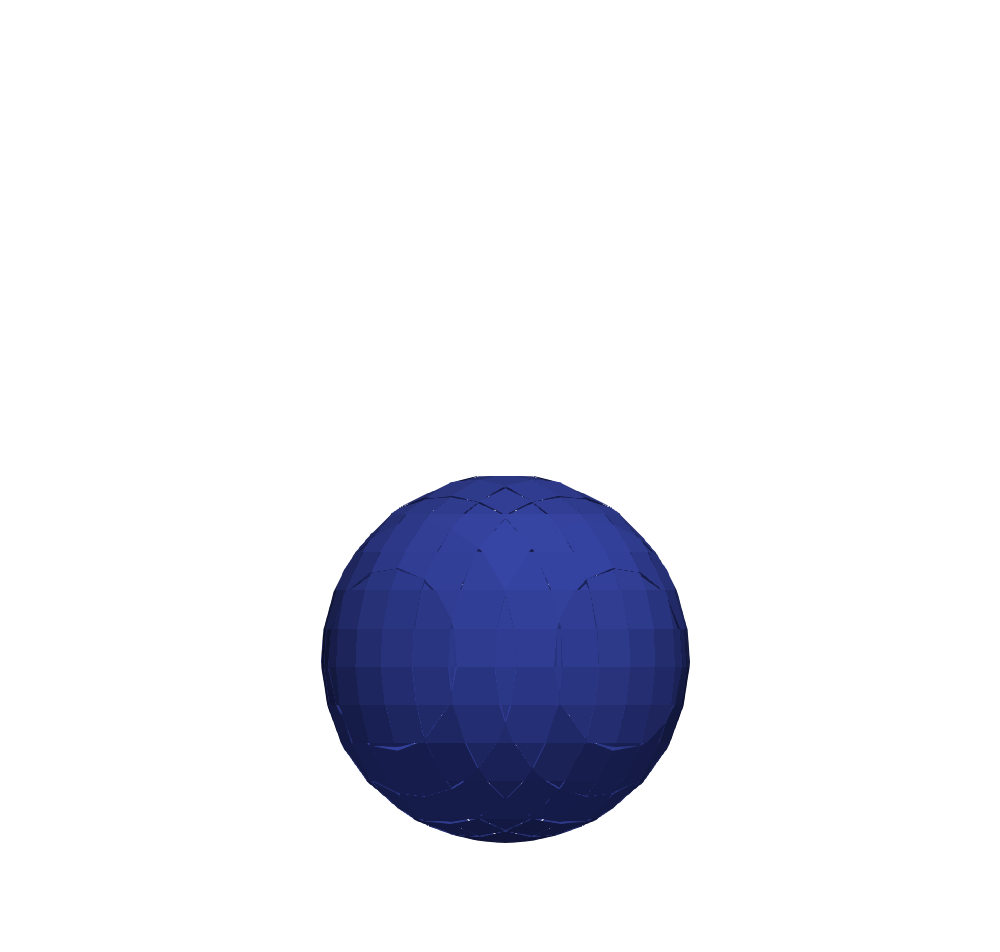}
        \caption*{$N=32,t=0$}
    \end{subfigure}\hfill
    \begin{subfigure}[b]{.33\textwidth}
        \centering\includegraphics[width=\textwidth]{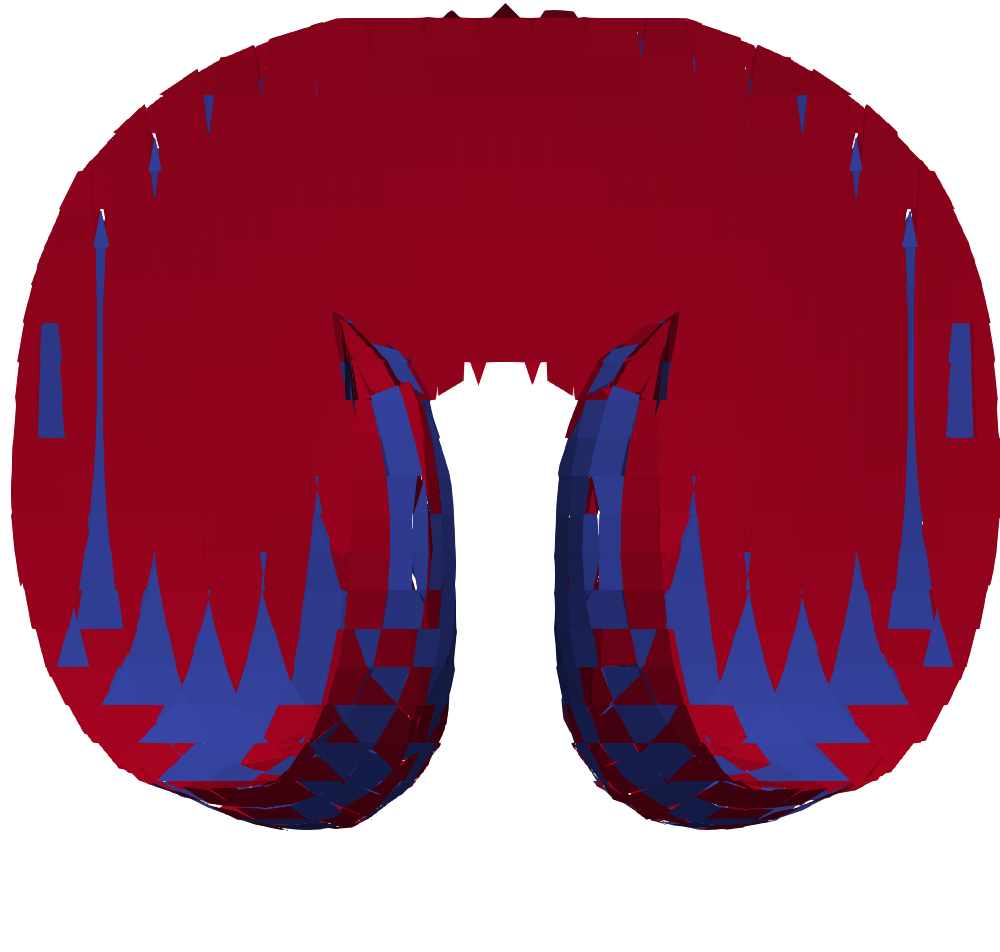}
        \caption*{$N=32,t=T/2$}
    \end{subfigure}\hfill
    \begin{subfigure}[b]{.33\textwidth}
        \centering\includegraphics[width=\textwidth]{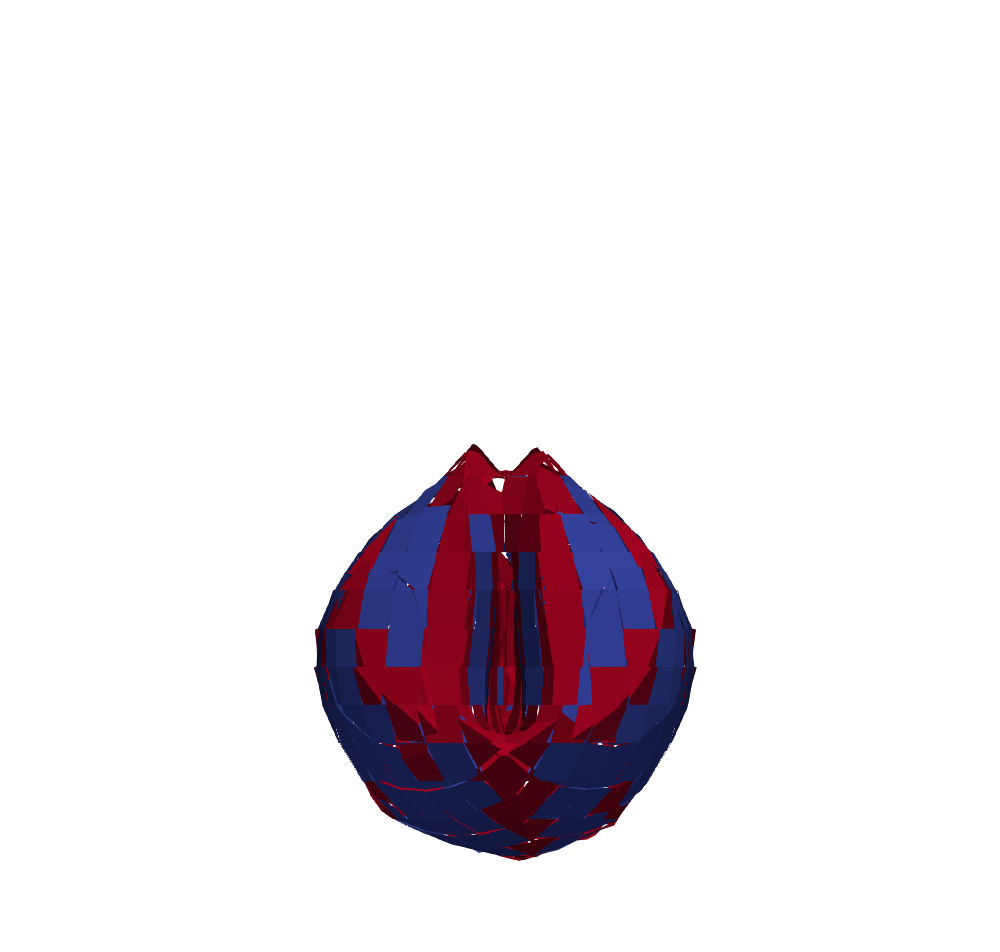}
        \caption*{$N=32,t=T$}
    \end{subfigure}\hfill
    \par\medskip
    \begin{subfigure}[b]{.33\textwidth}
        \centering\includegraphics[width=\textwidth]{figures/def3d/n64t0.png}
        \caption*{$N=64,t=0$}
    \end{subfigure}\hfill
    \begin{subfigure}[b]{.33\textwidth}
        \centering\includegraphics[width=\textwidth]{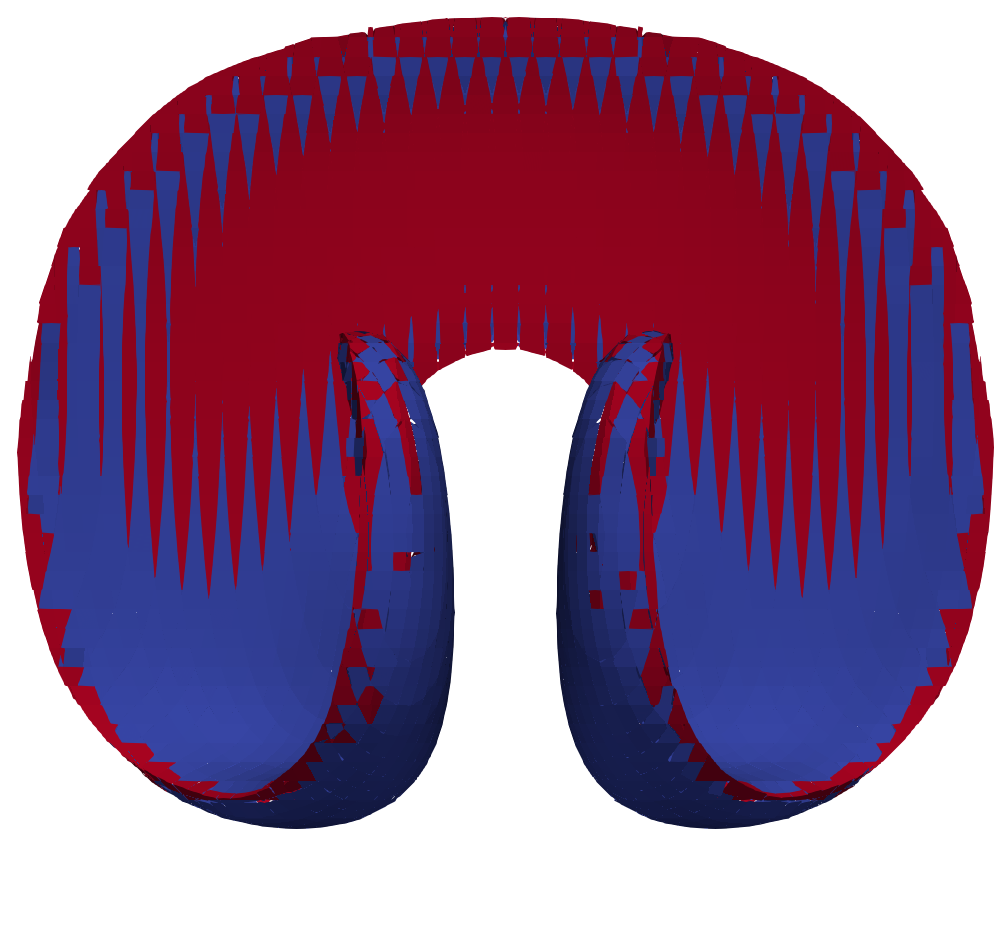}
        \caption*{$N=64,t=T/2$}
    \end{subfigure}\hfill
    \begin{subfigure}[b]{.33\textwidth}
        \centering\includegraphics[width=\textwidth]{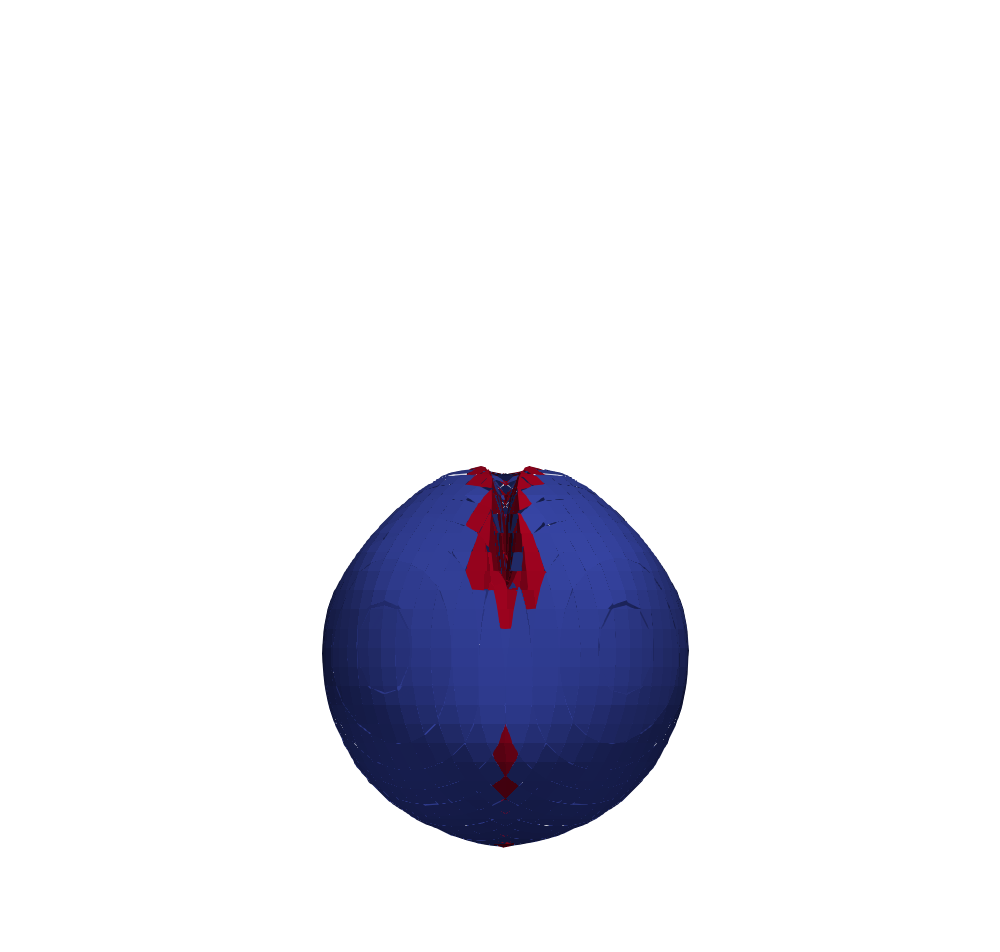}
        \caption*{$N=64,t=T$}
    \end{subfigure}\hfill
    \par\medskip
    \begin{subfigure}[b]{.33\textwidth}
        \centering\includegraphics[width=\textwidth]{figures/def3d/n128t0.png}
        \caption*{$N=128,t=0$}
    \end{subfigure}\hfill
    \begin{subfigure}[b]{.33\textwidth}
        \centering\includegraphics[width=\textwidth]{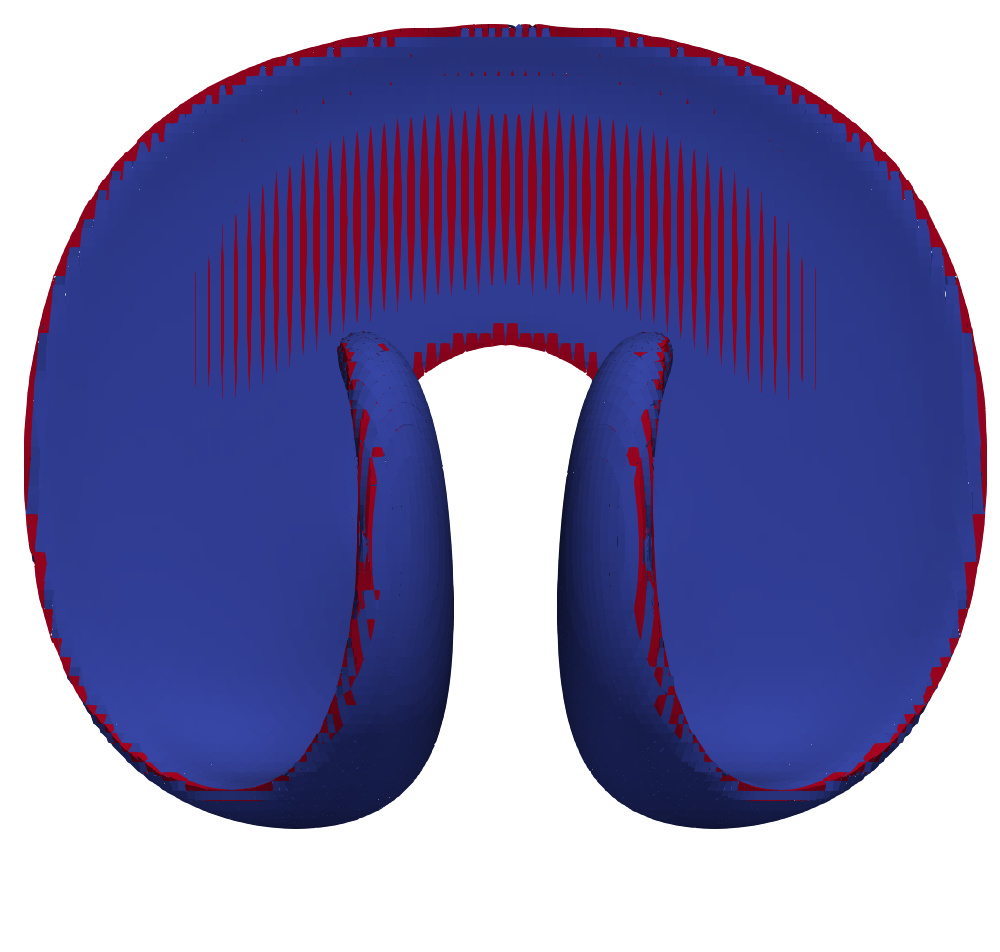}
        \caption*{$N=128,t=T/2$}
    \end{subfigure}\hfill
    \begin{subfigure}[b]{.33\textwidth}
        \centering\includegraphics[width=\textwidth]{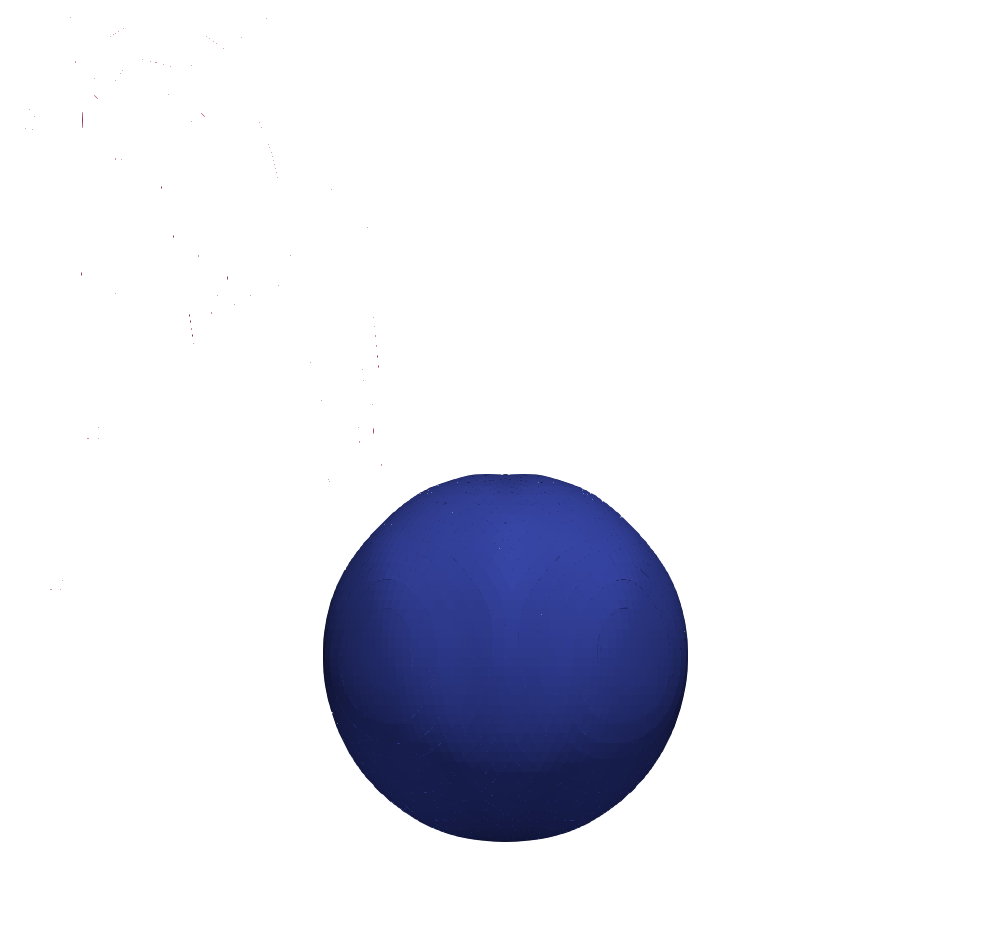}
        \caption*{$N=128,t=T$}
    \end{subfigure}\hfill
    \caption{Interface shape for the 3D deformation test case using R2P interface reconstruction. Single-plane reconstructions are shown in blue, while two-plane reconstructions are shown in red.}
    \label{fig:def3d-r2p}
\end{figure}

Figures \ref{fig:def3d-elvira} and \ref{fig:def3d-r2p} compare the interface shapes for the mesh resolutions $N=[32,64,128]$ using ELVIRA and R2P, respectively. Just as in the 2D deformation case, the thin film that forms by $t=T/2$ breaks up erroneously when using a single-plane reconstruction due to the lack of local resolution. The fragments produced by the mid-time breakup remain at $t=T$, and the final sphere interface departs from the initial interface. The erroneous interface breakup does not occur with R2P, as the subgrid portions of the deformed sphere are captured with a two-plane reconstruction. Furthermore, the final interface shape, when using R2P, is much closer to the initial spherical interface.
Figure \ref{fig:def3d-scatter} shows the mesh dependence of the shape error from and computational cost of the compared reconstruction methods, where the shape error is computed at $t=T/2$ relative to the $N=768$ reference volume fraction field. Both the ELVIRA and R2P methods exhibit first-order convergence in shape error, but the R2P method produces lower errors than those from the ELVIRA method. The ELVIRA and R2P methods incur similar computational costs relative to the shape error, although R2P requires a lower mesh resolution to achieve any given shape error.
Finally, the conservation error is shown in Table \ref{tab:def3d-econs}, where for all mesh resolutions, R2P preserves liquid volume to machine precision.

\begin{figure}
    \begin{subfigure}[T]{.5\textwidth}
        \centering
        \begin{tikzpicture}
            \begin{loglogaxis}[small,xlabel=$N$,ylabel=$E_{\text{shape}}$,log basis x=2,xticklabels={8,16,32,64,128,256}]
                \addplot table[x=nx,y=sdef3d_elvira] {\timeshapeconstable};
                \label{plot:def3d_elvira}
                \addplot table[x=nx,y=sdef3d_r2p   ] {\timeshapeconstable};
                \label{plot:def3d_r2p}
                \addplot[domain=16:256]{20/x};
                \label{plot:def3d-1st_order}
                \addplot[dotted, domain=16:256]{50/x^2};
                \label{plot:def3d-2nd_order}
            \end{loglogaxis}
        \end{tikzpicture}        
    \end{subfigure}\hfill
    \begin{subfigure}[T]{.5\textwidth}
        \centering
        \begin{tikzpicture}
            \begin{loglogaxis}[small,xlabel=$E_{\text{shape}}$,ylabel=$T_c$]
                \addplot table[x=sdef3d_elvira,y=tdef3d_elvira] {\timeshapeconstable};
                \addplot table[x=sdef3d_r2p,y=tdef3d_r2p   ] {\timeshapeconstable};
                \addplot[domain=1e-2:1]{.07/x};
            \end{loglogaxis}
        \end{tikzpicture}        
    \end{subfigure}\hfill
    \caption{3D deformation case shape error (left) and computational time per time step  (right) at $t=T/2$. ELVIRA (\ref{plot:def3d_elvira}); R2P (\ref{plot:def3d_r2p}); 1st order(\ref{plot:def3d-1st_order}); 2nd order(\ref{plot:def3d-2nd_order}).}
    \label{fig:def3d-scatter}
\end{figure}
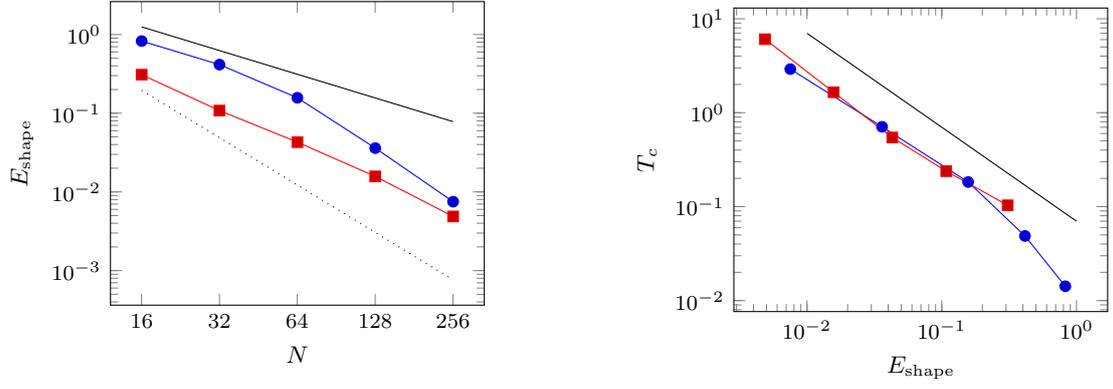   
\begin{table}
    \centering
    \caption{3D deformation case conservation error.}
    \pgfplotstabletypeset[
        columns={nx,cdef3d_r2p},
        columns/nx/.style={
            column name=$N$,
            int detect
        },
        columns/cdef3d_r2p/.style={
            column name=$E_c$,
            dec sep align,
            precision=2
        },
    ]{\timeshapeconstable}
    \label{tab:def3d-econs}
\end{table}

\subsection{Transport of a 2D Thin-Film Bubble}\label{rotatedbubble}
\begin{figure}[tbp]
    \begin{subfigure}[b]{.3\textwidth}
        \centering\includegraphics[width=\textwidth]{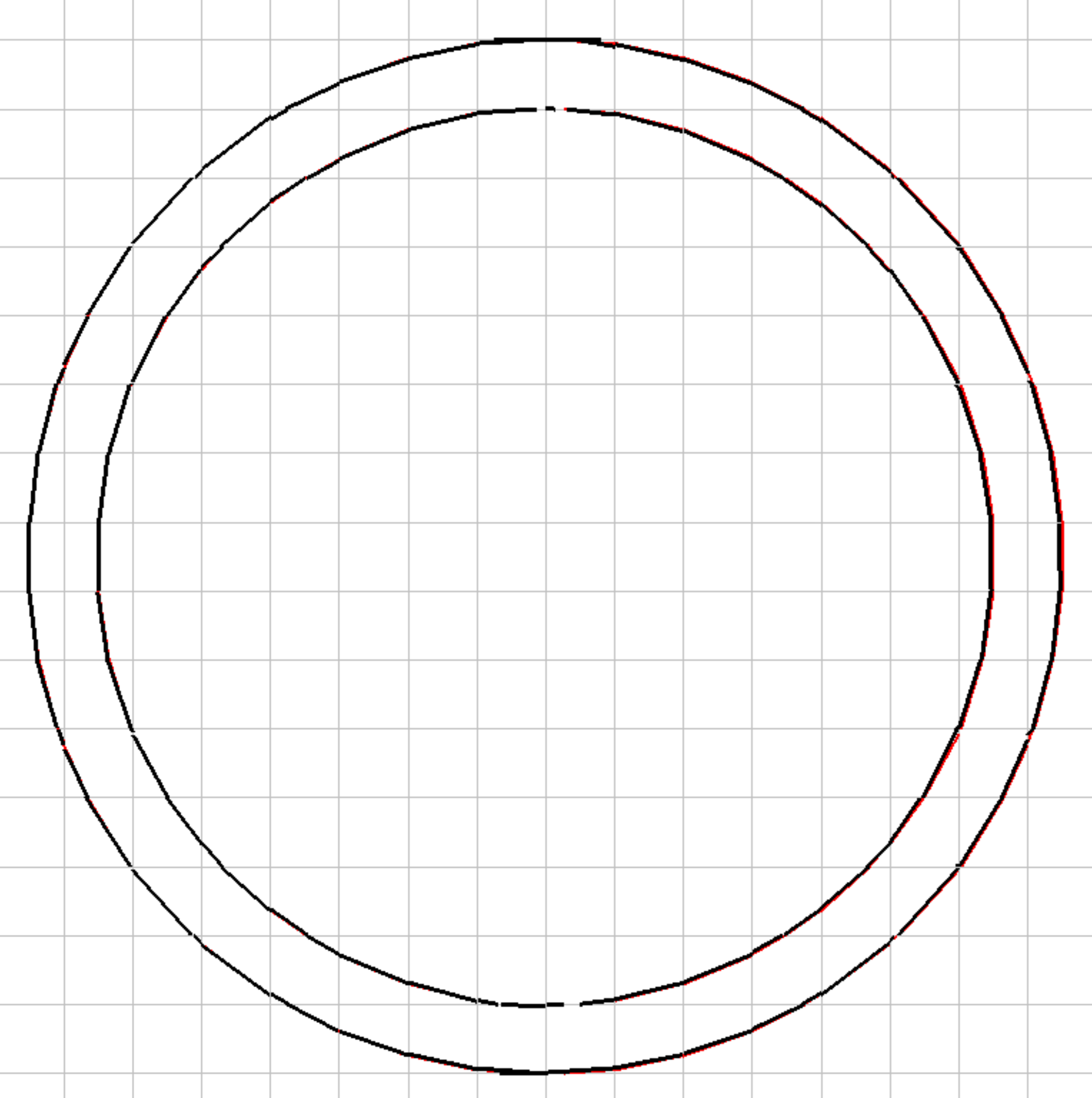}
        \caption*{$h/\Delta x=1$}
    \end{subfigure}\hfill
    \begin{subfigure}[b]{.3\textwidth}
        \centering\includegraphics[width=\textwidth]{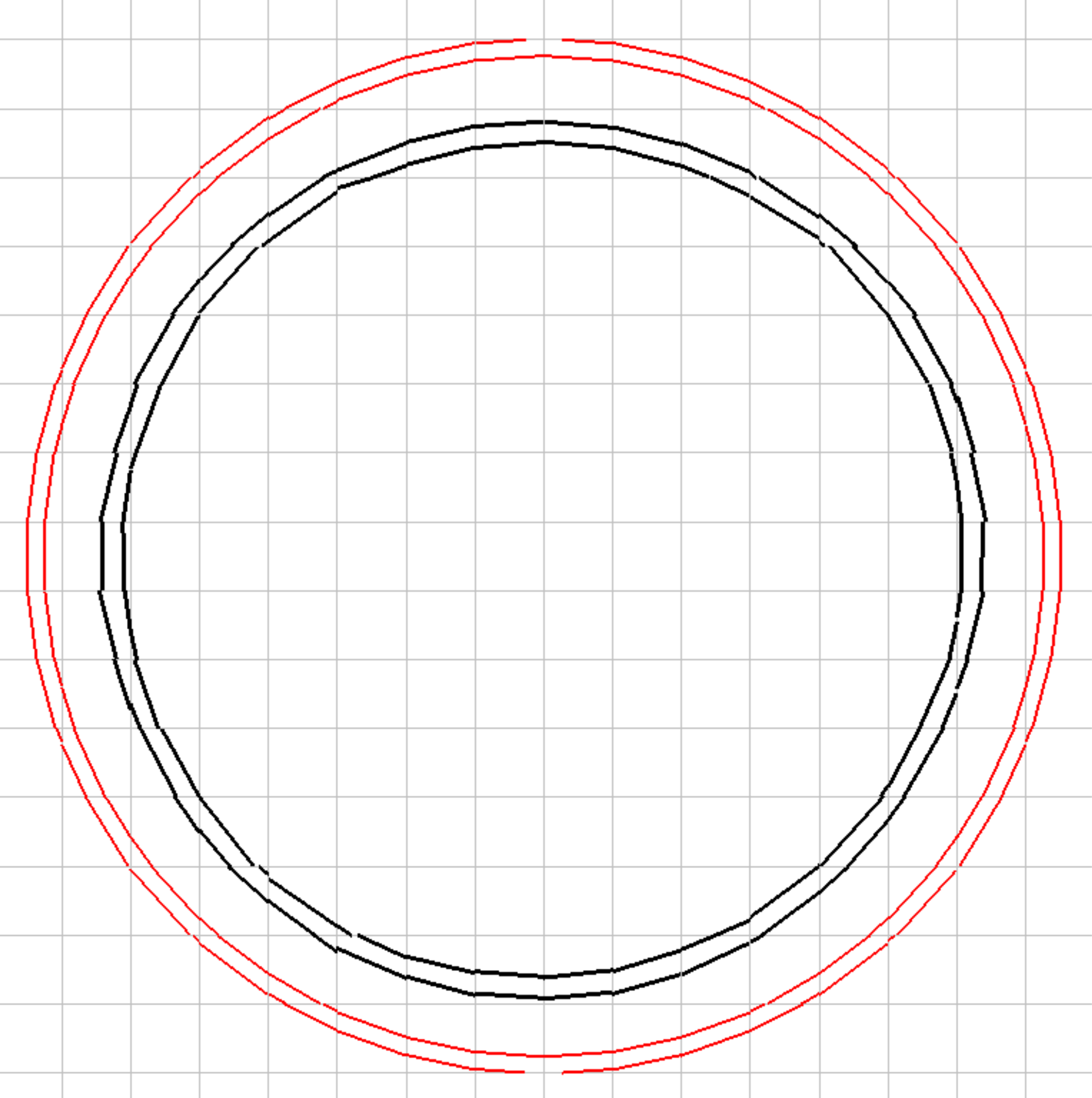}
        \caption*{$h/\Delta x=1/4$}
    \end{subfigure}\hfill
    \begin{subfigure}[b]{.3\textwidth}
        \centering\includegraphics[width=\textwidth]{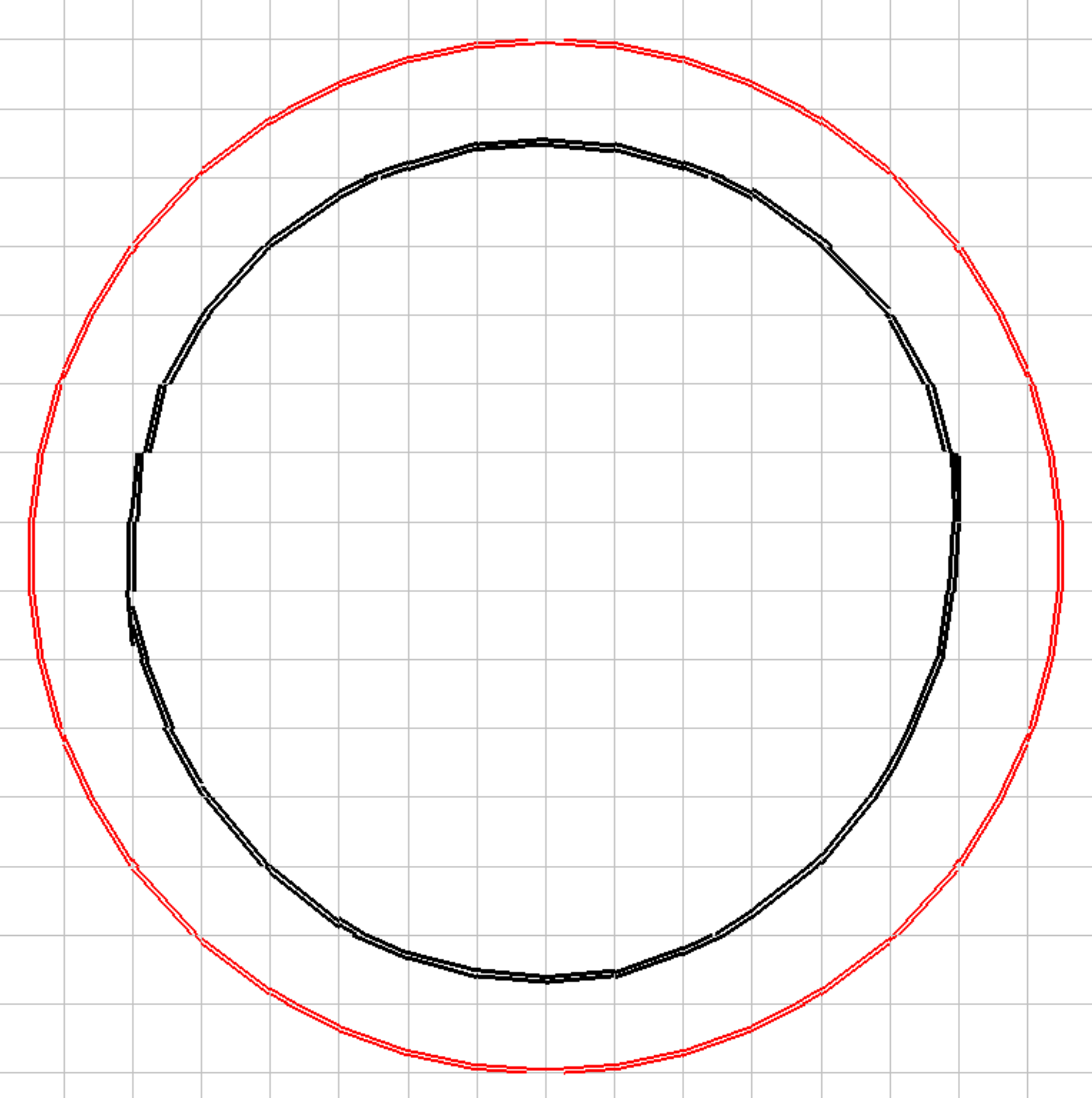}
        \caption*{$h/\Delta x=1/16$}
    \end{subfigure}\hfill        
    \caption{Initial interface (\fullred) and interface after one rotation (\fullblack) of a thin-film bubble.}
    \label{fig:rotatedbubble}
\end{figure}

R2P greatly improves the capturing of the sheet formation in the canonical deformation test cases. To directly evaluate the ability of R2P to transport thin sheets, a thin-film bubble is initialized and transported in the rigid body velocity field
\begin{equation}
    (u,v)=(-2\pi y,2\pi x)
\end{equation}
at a CFL of 0.96 for one full rotation. This bubble is initially centered at $(x,y)=(0.0,0.25)$ in a $L=[-0.5,0.5]^2$ domain and parameterized by an outer radius $R=0.15$ and film thickness $h$. To study the effect of film resolution on the results, the mesh spacing is held constant at $\Delta x=1/50$ and $h$ is varied from $\Delta x$ to $\Delta x/16$. As Figure \ref{fig:rotatedbubble} shows, the thin-film bubble is maintained at all mesh resolutions, forming no gaps during the rotation. When $h/\Delta x<1$ such that the bubble is formed from mostly two-plane reconstructions, the bubble diameter decreases after one rotation due to erroneous transfer of gas from inside to outside the bubble. The drift of the interface occurs because while the transported liquid and gas centroid locations are included in the cost function in Eq.\ \eqref{eq:cost-fcn}, they are not used as constraints on the interface position, unlike the liquid volume fraction. Therefore, the centroid locations can move with each instantiation of the interface reconstruction, even in a zero velocity field, as the centroids are recomputed to be consistent with the new reconstruction. Mass is conserved to machine precision ($E_c=\order{10^{-14}}$), however, as can be seen in the thickening of the film after one rotation alongside the diameter decrease.

\section{Surface Tension Modeling}\label{surface_tension}
For subgrid-thickness films, the accurate estimation of surface tension forces requires special treatment compared to fluid features that can be captured with one-plane interfaces. The continuum surface force (CSF) method \citep{Brackbill1992}, commonly used in VOF simulations, approximates the volumetric surface tension force at interfacial cells with 
\begin{equation}
    \sigma\kappa\delta\bm{n}\approx\sigma\kappa\nabla\alpha/[\chi],
\end{equation}
where $\kappa$ is the interfacial curvature, $\delta$ is the interface Dirac delta function, and $[\chi]=\chi_2-\chi_1=1$ is the jump in indicator function across the interface. The volume fraction gradient $\nabla\alpha$ is an approximation of $\delta\bm{n}$, and as such, the CSF method assumes that $\alpha$ transitions between $0$ and $1$ within a distance $\Delta x$ about the interface. This assumption is violated in the case of a subgrid-thickness film, where $\alpha$ can be much less than 1, leading to a significant underprediction of $\nabla \alpha$ and thereby the surface tension-induced pressure jump across the film. 

To mitigate this issue, the film is ``thickened" with the introduction of a ficticious volume fraction field
\begin{equation}
    \gamma=
    \begin{cases}
         1 & \text{for } \Omega \in \mathcal{L} \\
         0 & \text{for } \Omega \in \mathcal{G} \\
    \alpha & \text{for } \Omega \notin \mathcal{L} \cup \mathcal{G},
    \end{cases}
\end{equation}
where $\mathcal{L}$ is the set of liquid film cells, and $\mathcal{G}$ is the set of gas film cells as defined in \ref{classification}. The volume fraction field enforces a volume fraction jump of 1 when computing $\nabla \gamma$ between an interfacial cell and a non-interfacial cell. The volumetric surface tension force is then approximated as
\begin{equation}
    \sigma\kappa\delta\bm{n}\approx\sigma\kappa\nabla\gamma.
\end{equation}
A staggered discretization of the force at the face between two cells $A$ and $B$ is
\begin{equation}
    \sigma\kappa\nabla\gamma\approx\left(\sigma\kappa\right)_{AB}\frac{\gamma_B-\gamma_A}{\Delta x},
\end{equation}
where $\square_{AB}$ signifies a face-centered value. By construction, this scheme is well-balanced provided that the discrete gradient operators for $\gamma$ and the pressure are equivalent \citep{Popinet2018}.
\subsection{Face-centered Curvature}\label{curvature}
To compute $\kappa_{AB}$, the curvature is interpolated at the cell faces from the cell-centered curvatures of the bounding cells. Each interfacial polygon in a cell contains a representative local interfacial curvature value, and these curvatures are denoted $\kappa_1$ and $\kappa_2$ in the case of a two-plane cell. The discrete curvatures can be estimated through a number of methods that utilize polygonal or pointwise interfacial data \citep{Popinet2009,Jibben2019,hanComparisonMethodsCurvature2024}.
The face-centered curvature $\kappa^f$ is computed as the average of the cell-centered curvatures, weighted by the surface area $w$ of the corresponding polygon. For the face between two cells $A$ and $B$,
\begin{equation}
    \kappa^f_{AB}=\frac{\sum_{i=1}^{N_A}w_{A,i}\kappa_{A,i}+\sum_{i=1}^{N_B}w_{B,i}\kappa_{B,i}}{\sum_{i=1}^{N_A}w_{A,i}+\sum_{i=1}^{N_B}w_{B,i}},
\end{equation} 
where $N_A$ and $N_B$ are the number of planes in cells $A$ and $B$, respectively. In the special case where $\max(N_A,N_B)=2$ and $\min(N_A,N_B)=0$, however, this surface averaging may be inappropriate because it does not capture the individual pressure jumps across the film interfaces. In the case that $\kappa_1\approx-\kappa_2$, as in a thin-film bubble, the pressure jump would be severely underestimated, as $\kappa^f_{AB}\approx 0$. Instead, the face-centered curvature is chosen to be equal to the curvature corresponding to the interfacial polygon whose normal is most aligned with the face normal. The direction of the face normal used to determine the curvature is chosen such that it points from the cell with higher $\gamma$ to the cell with lower $\gamma$, since the interface normal points from the liquid to the gas phase. 
In the case where $N_A=2$ and $N_B=0$, the face-centered curvature is
\begin{equation}
    \begin{aligned}
        \kappa^f_{AB}&=\kappa_{A,i}, \\
                    i&=\argmax_i \left(\bm{n}^p_{A,i}\cdot\left(\sign(0.5-\gamma_B)\bm{n}^f_{AB}\right)\right),
    \end{aligned}
\end{equation}
where $\bm{n}^p_{A,i}$ is the normal vector of the $i$-th plane in cell $A$, $\gamma_B$ is the volume fraction of the non-interfacial cell, and $\bm{n}^f_{AB}$ is the normal vector of the cell face pointing from cell $A$ to cell $B$. 

\subsection{Surface Tension at Film Edges}\label{subsec:edge_surfacet}
The film curvature model discussed in Section \ref{curvature}, however, does not apply to the edges of these films. In the absence of a curvature model for film edges, it is necessary to resolve the edge on the mesh in order to maintain the local surface tension force. At low mesh resolutions, this can be achieved by utilizing single-plane PLIC reconstructions near the film edge that introduce numerical surface tension to maintain a cylindrical rim shape. The LVIRA method \citep{Pilliod2004} is used in this work as the PLIC reconstruction. To determine whether to switch to an LVIRA representation for a given cell $\Omega_i$, the magnitude of a surface-averaged normal vector, 
\begin{equation}
    \mathscr{M}=\left\Vert\frac{\sum_{\Omega_j\in\mathscr{N}_i}\sum^{Q_j}_{k=1}w_k\bm{N}_k}{\sum_{\Omega_j\in\mathscr{N}_i}\sum^{Q_j}_{i=1}w_k}\right\Vert,
\end{equation}
is computed from the normals of the advected polygons within the neighborhood $\mathscr{N}_i$ of $\Omega_i$. Then, a smoothing operation is performed on $\mathscr{M}$ by weighting it by $A$, the surface area of the interface reconstruction $\Gamma$, and averaging over the cell neighborhood again to produce
\begin{equation}
    \Bar{\mathscr{M}}=\frac{\sum_{\Omega_j\in\mathscr{N}_i}A_j \mathscr{M}_j}{\sum_{\Omega_i\in\mathscr{N}_i}A_j}.
\end{equation}
The smoothed, surface-averaged normal magnitude $\Bar{\mathscr{M}}$ is computed as a measure of the unimodality of the surface normal distribution across the cell neighborhood. As $0 \le \Bar{\mathscr{M}} \le 1 $, a magnitude $\Bar{\mathscr{M}}$ close to 1 indicates that the distribution is strongly unimodal and that LVIRA should be used to compute the interface normal, while lower values of $\Bar{\mathscr{M}}$ indicate that the interface should be represented by two planes with R2P.

\subsection{Thin-Film Bubble in Equilibrium}
The proposed surface tension model, in conjunction with R2P, is used to simulate a 2D thin-film bubble in static equilibrium. This test demonstrates the ability of the method to capture the correct pressure jump across the thin film and establishes that the modified CSF method does not induce spurious velocities around the interface \citep{Francois2006,Popinet2009} when the bubble is in theoretical equilibrium, where the interfacial curvature is fixed to the exact value.
The inner diameter is $D=0.3$, and the outer diameter is $D+h$, where the film thickness $h=10^{-3}$. The bubble is placed in a two-dimensional square domain of length $L=1$ with periodic boundary conditions and a uniform Cartesian grid of cell length $\Delta x=0.02$. The center of the bubble is placed at the center of the domain. The viscosity ratio is unity with a viscosity of $\mu=10^{-3}$, the density is a constant $\rho=1$, and the surface tension coefficient is $\sigma=1$. The Laplace number $La=\rho\sigma D/\mu^2=3.0\times 10^5$. The time step size is $\Delta t=8\times 10^{-4}$, and the simulation is performed up to the time $T=25$.

\begin{figure}[tbp]
    \centering
    \begin{subfigure}[t]{.33\textwidth}
        \centering\includegraphics[width=\textwidth]{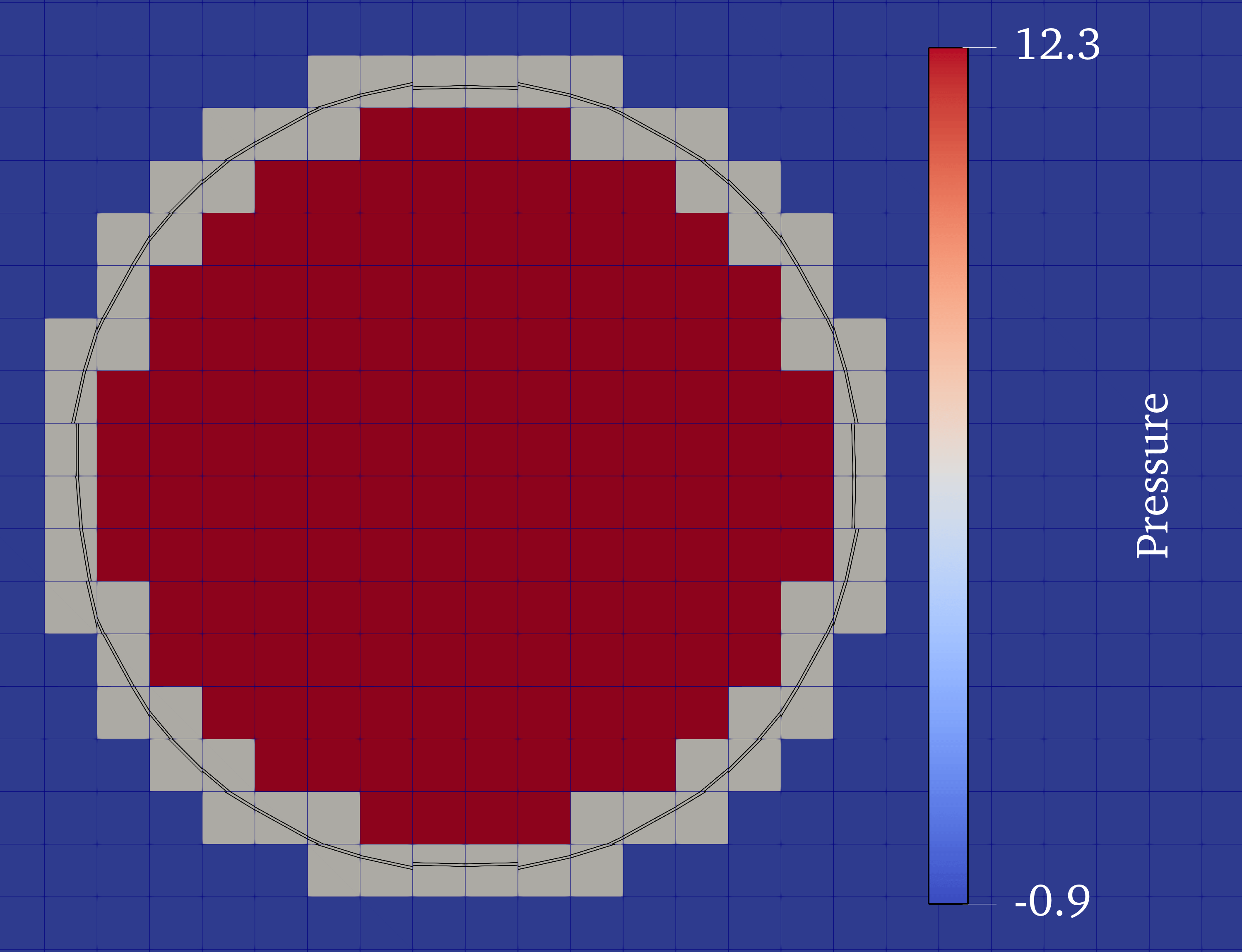}
        \caption{Color map of thin-film bubble pressure field relative to the domain average. The two-plane interface representation is shown in black.}
    \end{subfigure}\hfill
    \begin{subfigure}[t]{.33\textwidth}
        \centering\includegraphics[width=\textwidth]{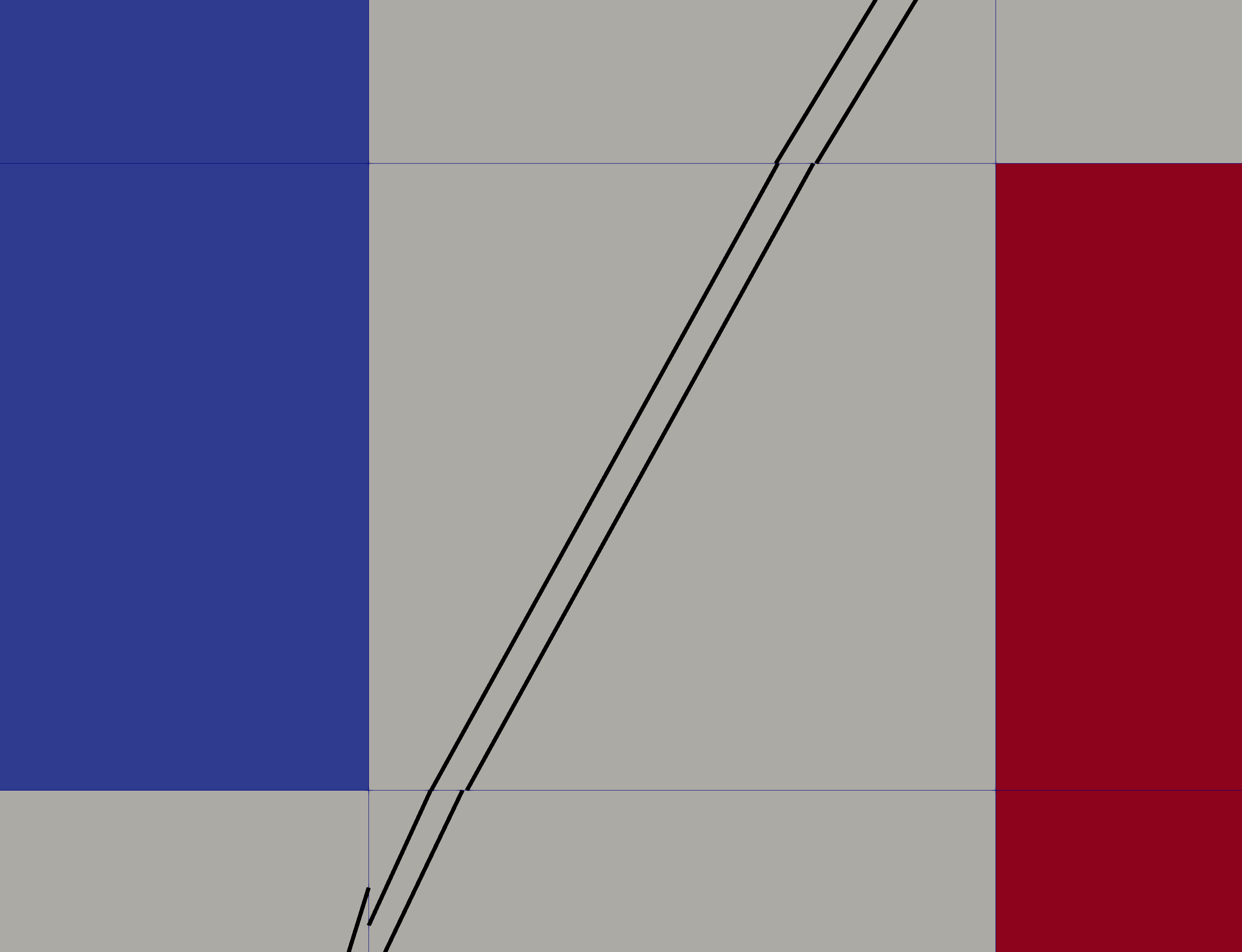}
        \caption{Closeup of thin-film bubble.}
    \end{subfigure}\hfill
    \begin{subfigure}[t]{.33\textwidth}
        \centering\includegraphics[width=\textwidth]{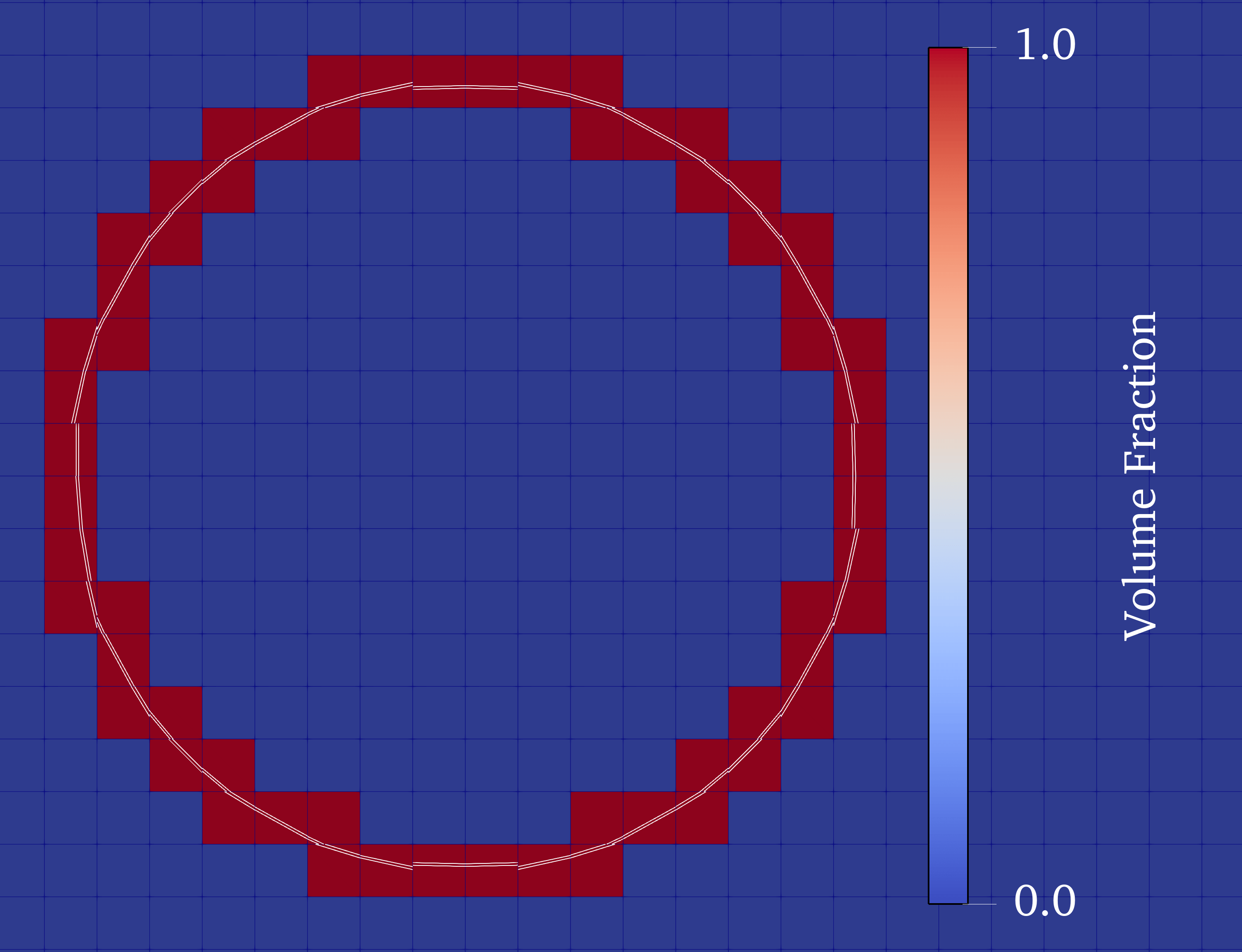}
        \caption{Color map of thickned volume fraction field $\gamma$ with interface in white.}
    \end{subfigure}
    \caption{Thin-film bubble with fixed curvature in static equilibrium.}
    \label{fig:equilibriumbubble}    
\end{figure}

The expected pressure jump across the thin-film can be computed from the Young-Laplace equation by summing the pressure jumps across the inner and outer interfaces as
\begin{equation}\label{eq:younglaplace}
    \Delta P=\sigma(\kappa_{\text{inner}}+\kappa_{\text{outer}})=\frac{1}{0.150}+\frac{1}{0.151}\approx 13.289183223.
\end{equation}
The pressure field is shown in Figure \ref{fig:equilibriumbubble}, along with the thickened volume fraction field $\gamma$ and the time evolution of the maximum velocity in the domain. Table \ref{table:pressure} compares the theoretical pressure jump across the thin film from Eq.\ \eqref{eq:younglaplace} and the numerical result shown Figure \ref{fig:equilibriumbubble}. Note that due to the artificial thickening of the film for the surface tension calculation, the pressure jump is stairstepped between the outer and inner pressure values such that the pressure takes an intermediate value inside the film cells. Figure \ref{fig:equilibriumbubble} and Table \ref{table:pressure} both show that R2P is capable of both representing the thin film and capturing the correct pressure jump. Furthermore, the proposed thin-film surface tension calculation recovers exact numerical balance with the pressure force.
\begin{table}
    \centering
    \begin{tabular}{c|c|c}
        Theoretical & Numerical & \% error \\ \hline
        13.289183223 & 13.289183080 & \qty{1.08e-6}{} \\
    \end{tabular}
    \caption{Pressure jump across the thin film.}
    \label{table:pressure}
\end{table}

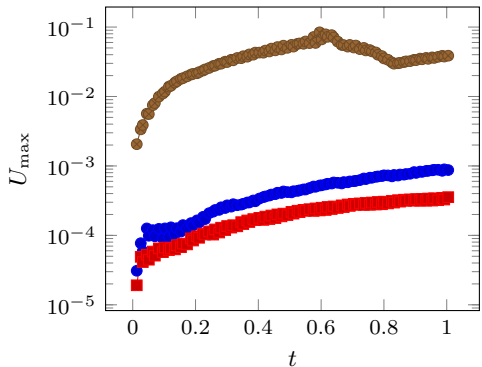
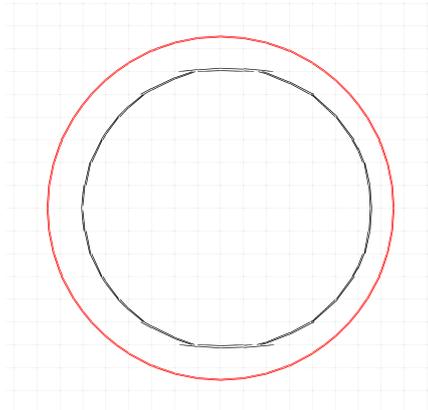
\begin{figure}[tbp]
    \begin{subfigure}[b]{.49\textwidth}
        \centering
        \begin{tikzpicture}
            \begin{semilogyaxis}[small,xlabel=$t$,ylabel=$U_{\max}$]
                \addplot table[x=Time,y=Umax] {\loadedumaxdrop};
                \label{plot:umaxdrop}
                \addplot table[x=Time,y=Umax] {\loadedumaxbubble};
                \label{plot:umaxbubble}
                \addplot table[x=Time,y=Umax] {\loadedumaxfilm};
                \label{plot:umaxfilm}
            \end{semilogyaxis}
        \end{tikzpicture}
        \caption{Maximum velocity.}
    \end{subfigure}\hfill
    \begin{subfigure}[b]{.49\textwidth}
        \centering\includegraphics[width=0.7\textwidth]{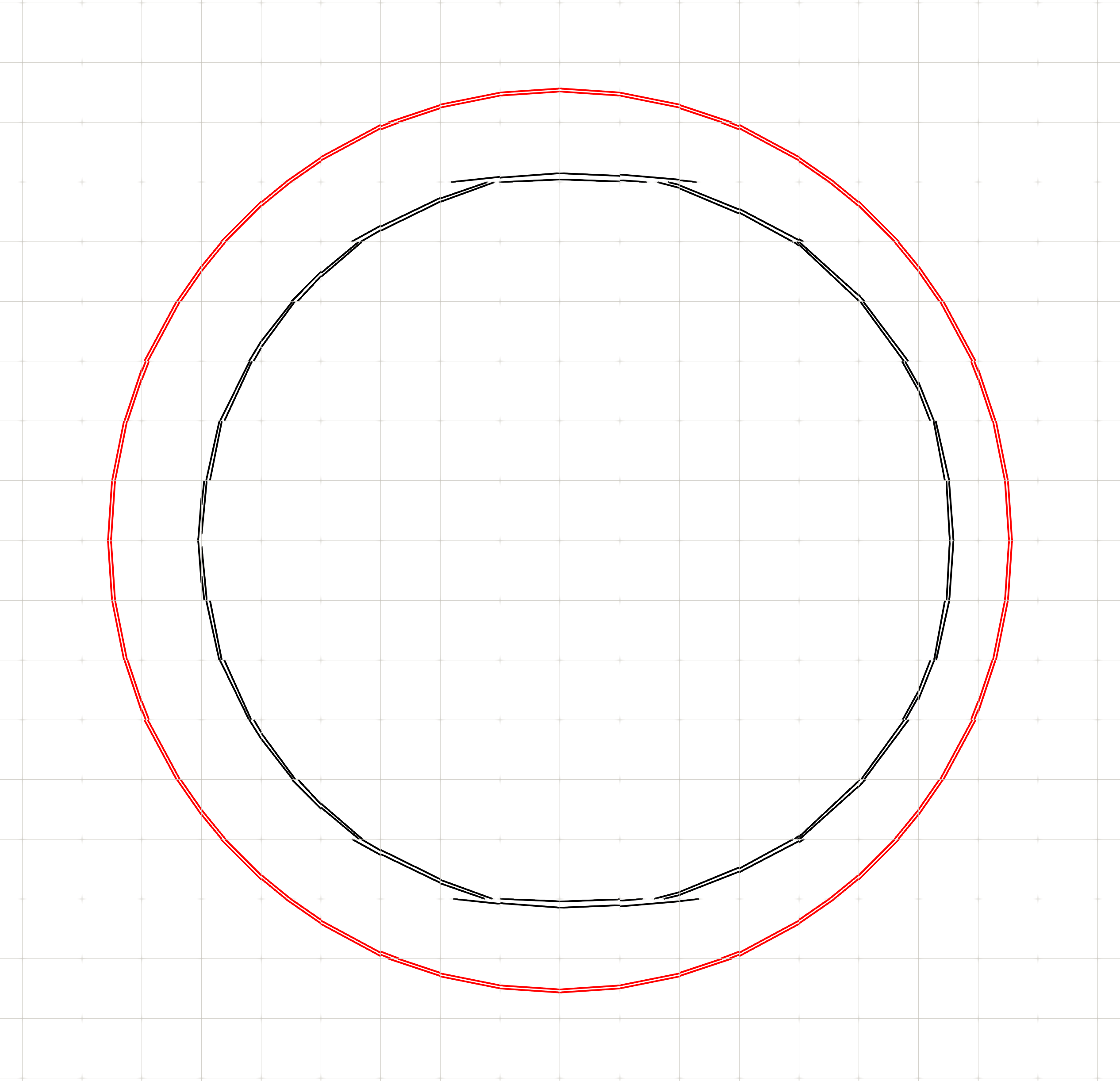}
        \caption{Thin-film bubble interface.}
        \label{subfig:bubble_interface}
    \end{subfigure}
    \caption{Time evolution of the maximum velocity around the phase interface for a liquid droplet (\ref{plot:umaxdrop}), gas bubble (\ref{plot:umaxbubble}), and thin-film bubble (\ref{plot:umaxfilm}). Phase interface of the thin-film bubble in (\subref{subfig:bubble_interface}) at $t=0$ (red) and $t=T$ (black). }
    \label{fig:translatedbubble}
\end{figure}

Next, the spurious velocities are measured in the case of a translating 2D thin-film bubble and compared to those from a translating liquid droplet and a translating gas bubble. Unlike in the static equilibrium case, the curvature used to compute the surface tension force is estimated by parabolic fitting onto the interface reconstruction \citep{Jibben2019,hanComparisonMethodsCurvature2024}, and therefore, exact numerical balance is not expected. The geometry is identical to that of the static equilibrium case with the exception of the droplet and bubble each having a single diameter of $D=0.3$. The viscosity ratio $\mu_a/\mu_b$ is 50, the density ratio $\rho_a/\rho_b$ is 1000, the gas Reynolds number $\text{Re}_b=\rho_b U_0 D/\mu_b$ is 100, the Weber number $\We=\rho_b U^2_0 D/\sigma$ is 20, the Laplace number is 60, and the translation velocity $U_0=1$ is in the $x$-direction. The cases are run with a time step size $\Delta t=6.4\times 10^{-3}$ to a final time of $T=1$ such that the interface traverses the periodic domain once and returns to its initial position. Figure \ref{fig:translatedbubble} shows the time evolution of the maximum velocity for the liquid droplet, gas bubble, and thin-film bubble cases, along with the initial and final interface shapes for the translated thin-film bubble. While the droplet and gas bubble exhibit lower spurious velocities than the thin-film bubble by over one order of magnitude, the spurious velocities remain at below $0.1U_0$ throughout the thin-film bubble translation. As discussed in Section \ref{rotatedbubble}, the interface exhibits drift towards the center of the bubble due to the movement of gas from the inside of the bubble.

\subsection{Bag Formation on an Impulsively Accelerated Droplet}
The following test of the proposed method simulates the aerodynamic deformation of a liquid droplet. At moderate Weber numbers of $\We\approx15$, the droplet evolves in the ``bag breakup'' regime, where the droplet flattens into a disk, deforms into a bag-shaped liquid sheet, and eventually breaks up into smaller droplets. The initial droplet diameter is $D=1$, the domain size is $[L_x,L_y,L_z]=[20,10,10]$, and the uniform computational mesh size is $\Delta x=D/12.8$. The simulation is performed with a time step size of $\Delta t= 5\times 10^{-2}$ to a time of $T=82$, at which point the liquid bag has a thickness of $h<2\times10^{-3}\Delta x$ at its thinnest point. The film thickness inside a cell $\Omega_i$ is defined as
\begin{equation}
        h_i=\frac{2\min\left(\sum_{\Omega_j \in \mathscr{N}_i}\alpha_j V_j,\sum_{\Omega_j \in \mathscr{N}_i}(1-\alpha_j) V_j\right)}{\sum_{\Omega_j \in \mathscr{N}_i}A_j},
        \label{eq:thickness-s4}
\end{equation}
where $\mathscr{N}_i$ is the set of all cells that share a vertex with cell $\Omega_i$, $V_i$ is the cell volume, and $A_i$ is the surface area of the liquid-gas interface $\Gamma_i$ in the cell.
As the modeling of the sheet breakup is outside the scope of this work, this test serves to demonstrate the ability of R2P to capture the formation of the subgrid-thickness liquid bag. The simulation is performed at $\text{Re}_b=100$ and $\We=20$, with $\rho_b=1$, $\rho_a/\rho_b=1000$, and $\mu_a/\mu_b=50$. The gas inflow is generated from a precursor simulation of homogeneous isotropic turbulence with $\text{R}_{\lambda}=45$, and it has a turbulence intensity of $5\%$ and a mean inflow velocity of $U=1$.
A threshold value of $\Bar{\mathscr{M}}=0.5$ determines the use of a PLIC reconstruction at the edge of the liquid sheet.

\begin{figure}[p]
    \begin{subfigure}[b]{\textwidth}
        \centering\includegraphics[width=0.8\textwidth]{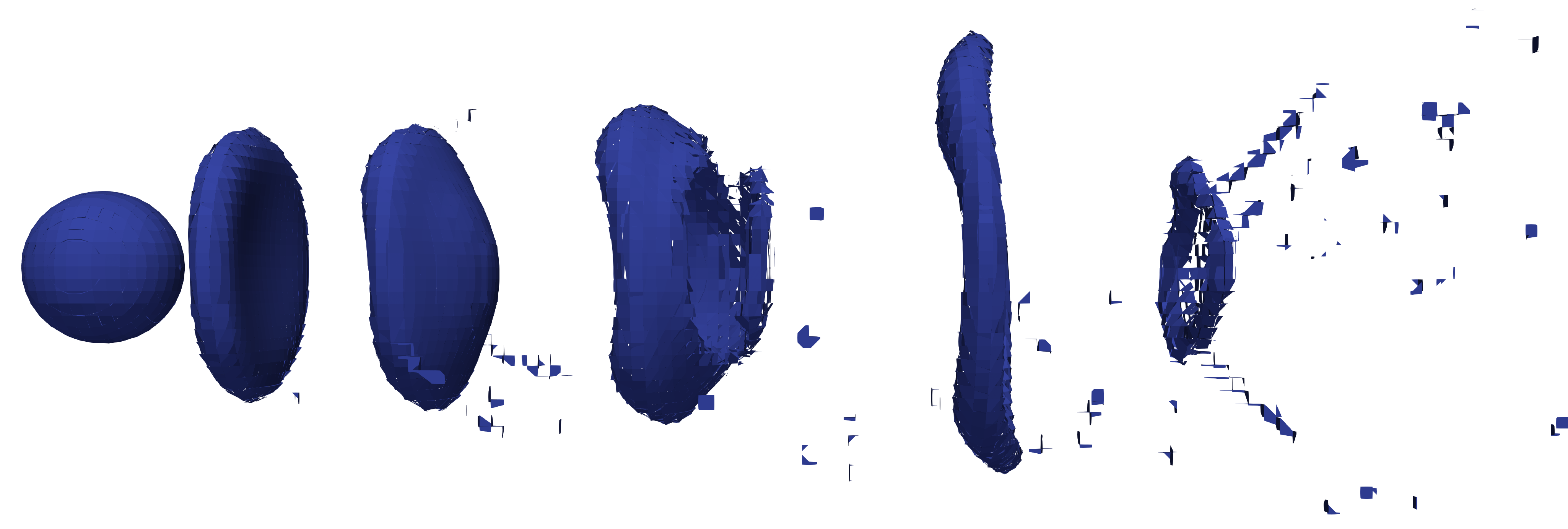}
        \caption{ELVIRA.}
    \end{subfigure}\hfill
    \par
    \begin{subfigure}[b]{\textwidth}
        \centering\includegraphics[width=0.8\textwidth]{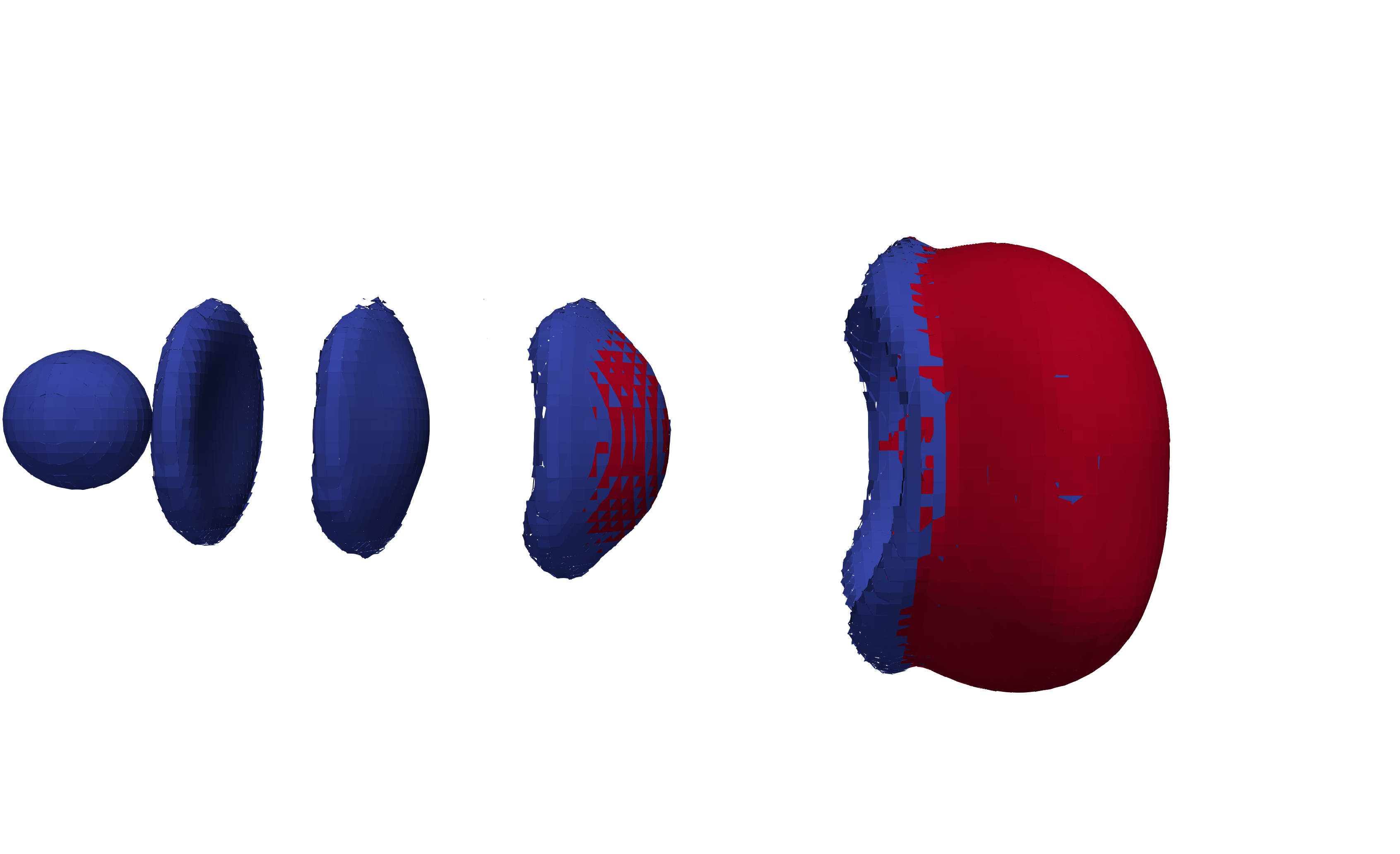}
        \caption{R2P with film surface tension model.}
    \end{subfigure}\hfill
    \par
    \begin{subfigure}[b]{\textwidth}
        \centering\includegraphics[width=0.8\textwidth]{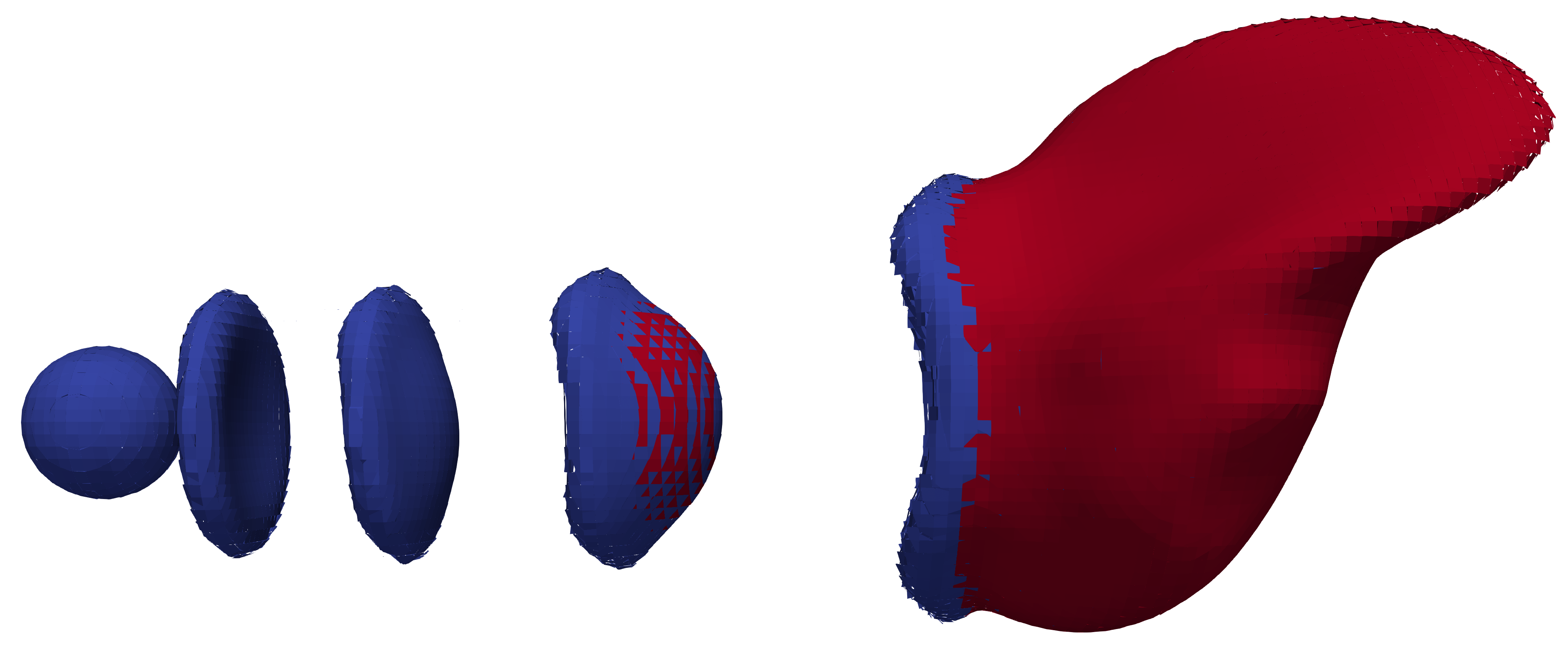}
        \caption{R2P with standard CSF surface tension.}
    \end{subfigure}\hfill
    \caption{Interface shape for the deformation of an impulsively accelerated droplet with $D/\Delta x=12.8$. Single-plane reconstructions are shown in blue, while two-plane reconstructions are shown in red.}
    \label{fig:bag}    
\end{figure}

\begin{figure}[ht]
        \centering\includegraphics[width=\textwidth]{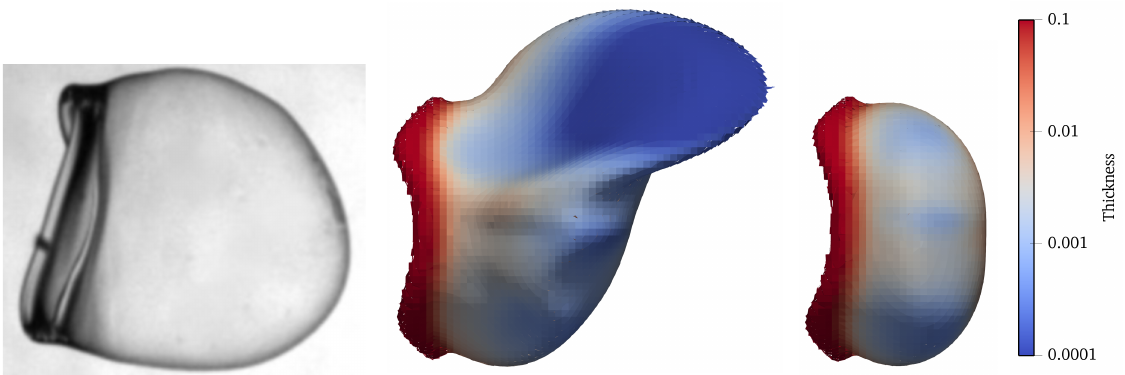}
    \caption{Interface shape for the deformation of an impulsively accelerated droplet with $D/\Delta x=12.8$ using CSF surface tension (middle) and film surface tension model (right). Interface elements are colored by film thickness. Experimental image from \cite{Opfer2014} shown on the left, reproduced with permission from the publisher.}
    \label{fig:bagthickness}    
\end{figure}

Figure \ref{fig:bag} compares the interface shapes of the deformed droplet at $t=T$ resulting from ELVIRA and R2P interface reconstruction.
The respective computational costs of the simulations utilizing ELVIRA and R2P reconstructions are 46.7 and 54.2 core-hours, reflecting a modest 16\% increase from ELVIRA to R2P.
The two-plane interface representation is able to capture the formation and evolution of the liquid film without erroneous tearing, while the ELVIRA representation causes the deforming droplet to puncture due to the lack of numerical resolution and produce additional liquid structures unphysically. 
To demonstrate the effect of the film surface tension model, the simulation with two-plane reconstruction is performed with and without the surface tension model, the latter being the standard CSF method. The film thickness profile and interface shape using both surface tension models are compared in Figure \ref{fig:bagthickness} alongside an image from the experiments of \cite{Opfer2014}.
When using R2P with the standard CSF surface tension model, the interface deforms unevenly due to the surrounding turbulent gas,
whereas when using the film surface tension model, the interface deforms relatively uniformly, as the surface tension stabilizes the interface by counteracting the perturbations from the turbulent velocity field. The spherical bag shape produced by the simulation using the film surface tension model is closer to the experimentally observed shape. Additionally, the selective use of a PLIC reconstruction where $\Bar{\mathscr{M}}=0.5$ maintains the toroidal rim at the edge of the sheet. Without the numerical surface tension introduced from the PLIC reconstruction, i.e., with $\Bar{\mathscr{M}}=1$, the entire drop has been observed to deform into a sheet. Since the two-plane reconstruction maintains the liquid film at arbitrary thicknesses, subgrid-scale breakup models could be applied at any point to produce child droplets from the liquid film volume.

\subsection{Binary Collision of Offset Droplets}
\begin{figure}[htbp]
    \begin{subfigure}[b]{0.28\textwidth}
        \centering\includegraphics[width=\textwidth]{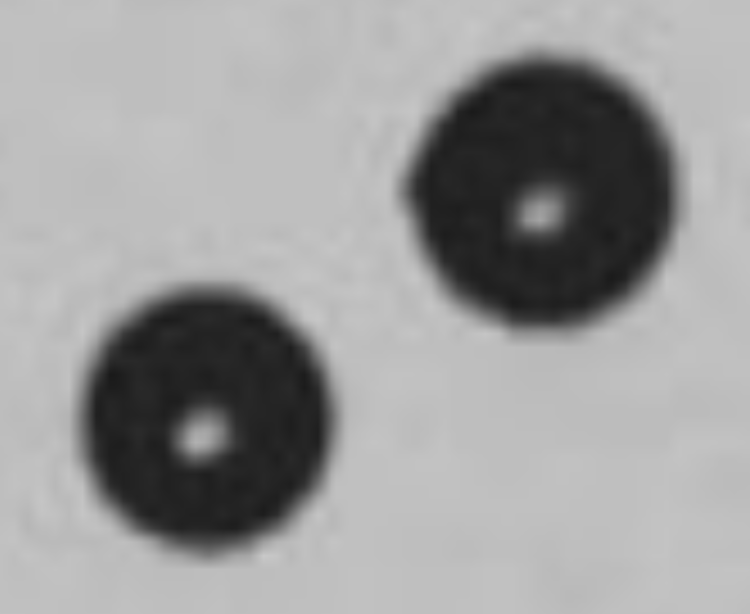}
        \caption{$t=\qty{0}{\ms}$}
    \end{subfigure}\hfill
    \begin{subfigure}[b]{0.28\textwidth}
        \centering\includegraphics[width=\textwidth]{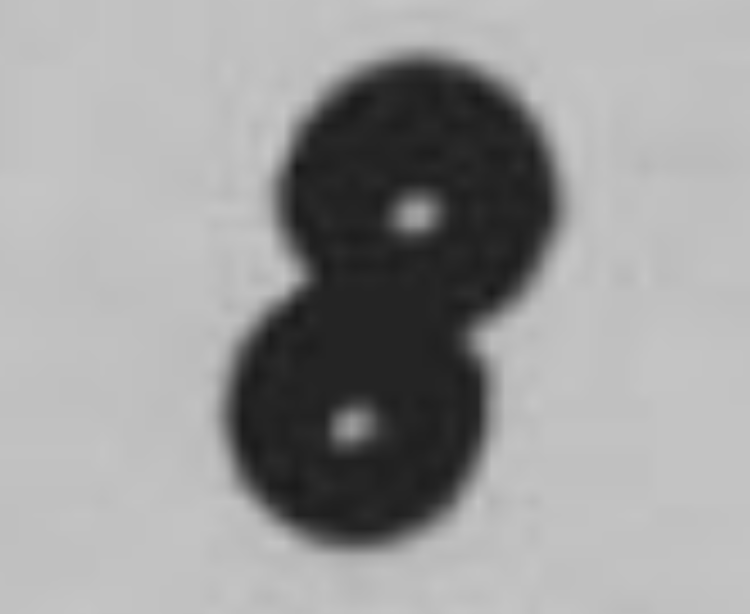}
        \caption{$t=\qty{18}{\ms}$}
    \end{subfigure}\hfill
    \begin{subfigure}[b]{0.28\textwidth}
        \centering\includegraphics[width=\textwidth]{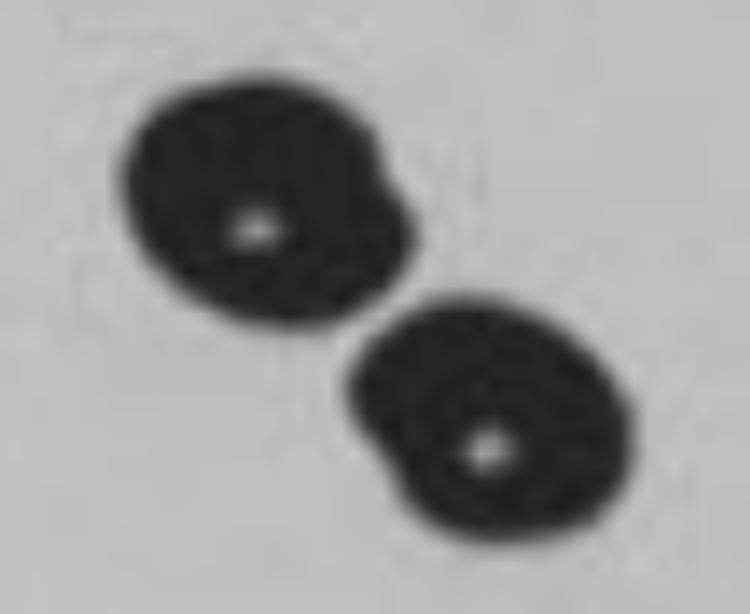}
        \caption{$t=\qty{38}{\ms}$}
    \end{subfigure}\hfill
    \par
    \begin{subfigure}[b]{0.25\textwidth}
        \centering\includegraphics[width=\textwidth]{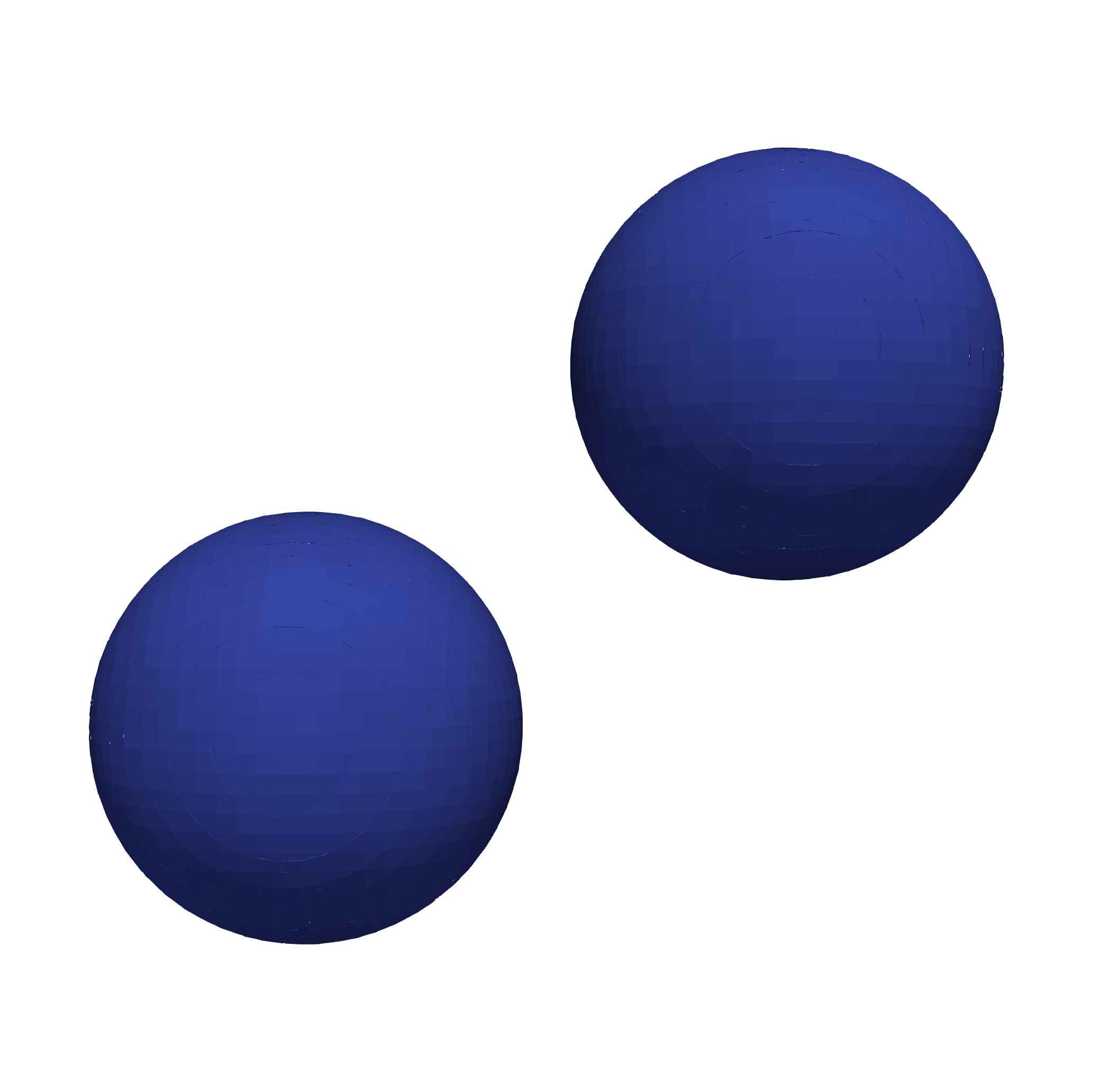}
        \caption{$t=\qty{0}{\ms}$}
    \end{subfigure}\hfill
    \begin{subfigure}[b]{0.25\textwidth}
        \centering\includegraphics[width=\textwidth]{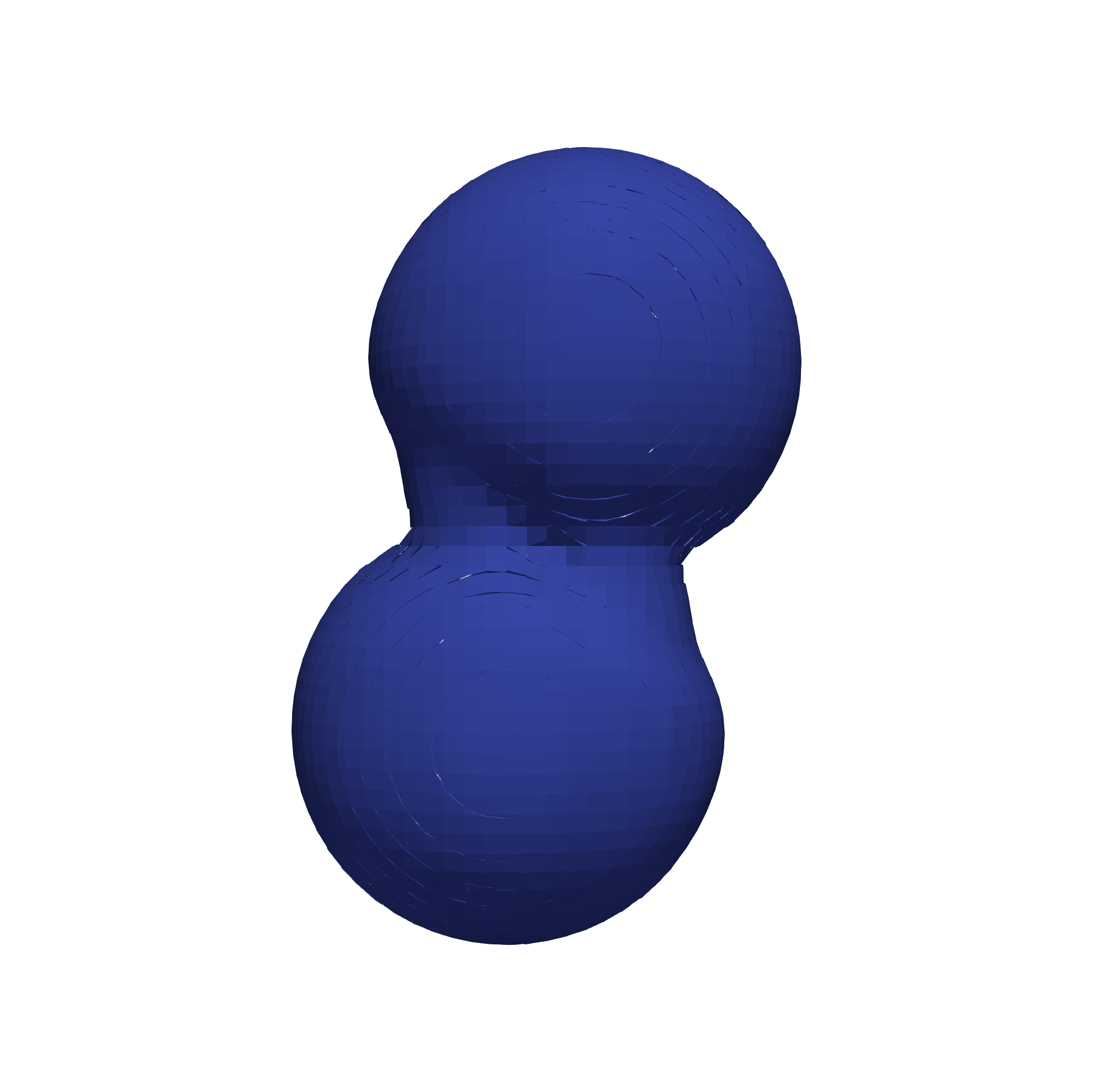}
        \caption{$t=\qty{29}{\ms}$}
    \end{subfigure}\hfill
    \begin{subfigure}[b]{0.25\textwidth}
        \centering\includegraphics[width=\textwidth]{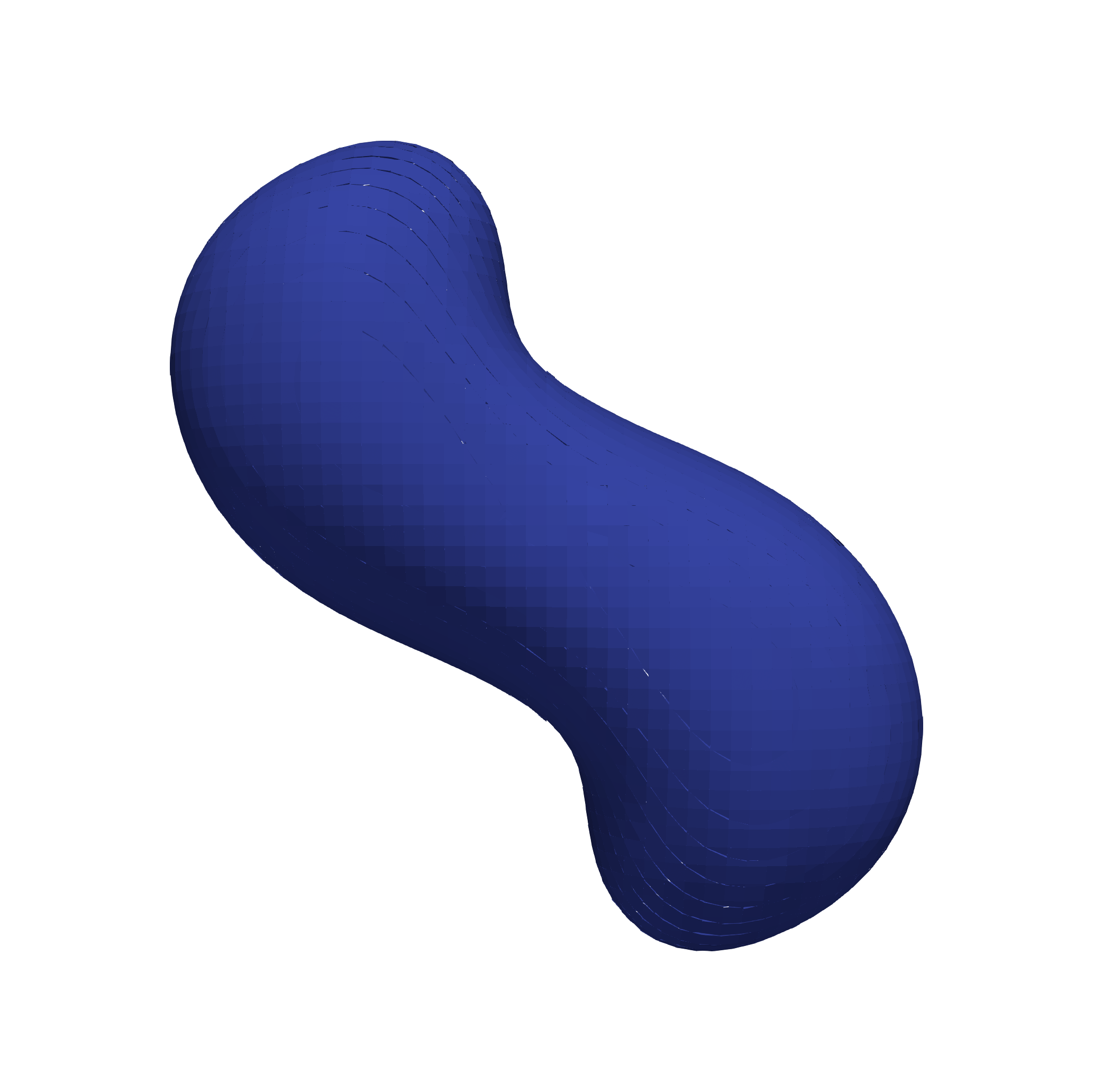}
        \caption{$t=\qty{61}{\ms}$}
    \end{subfigure}\hfill
    \par
    \begin{subfigure}[b]{0.25\textwidth}
        \centering\includegraphics[width=\textwidth]{figures/collision/t00.png}
        \caption{$t=\qty{0}{\ms}$}
    \end{subfigure}\hfill
    \begin{subfigure}[b]{0.25\textwidth}
        \centering\includegraphics[width=\textwidth]{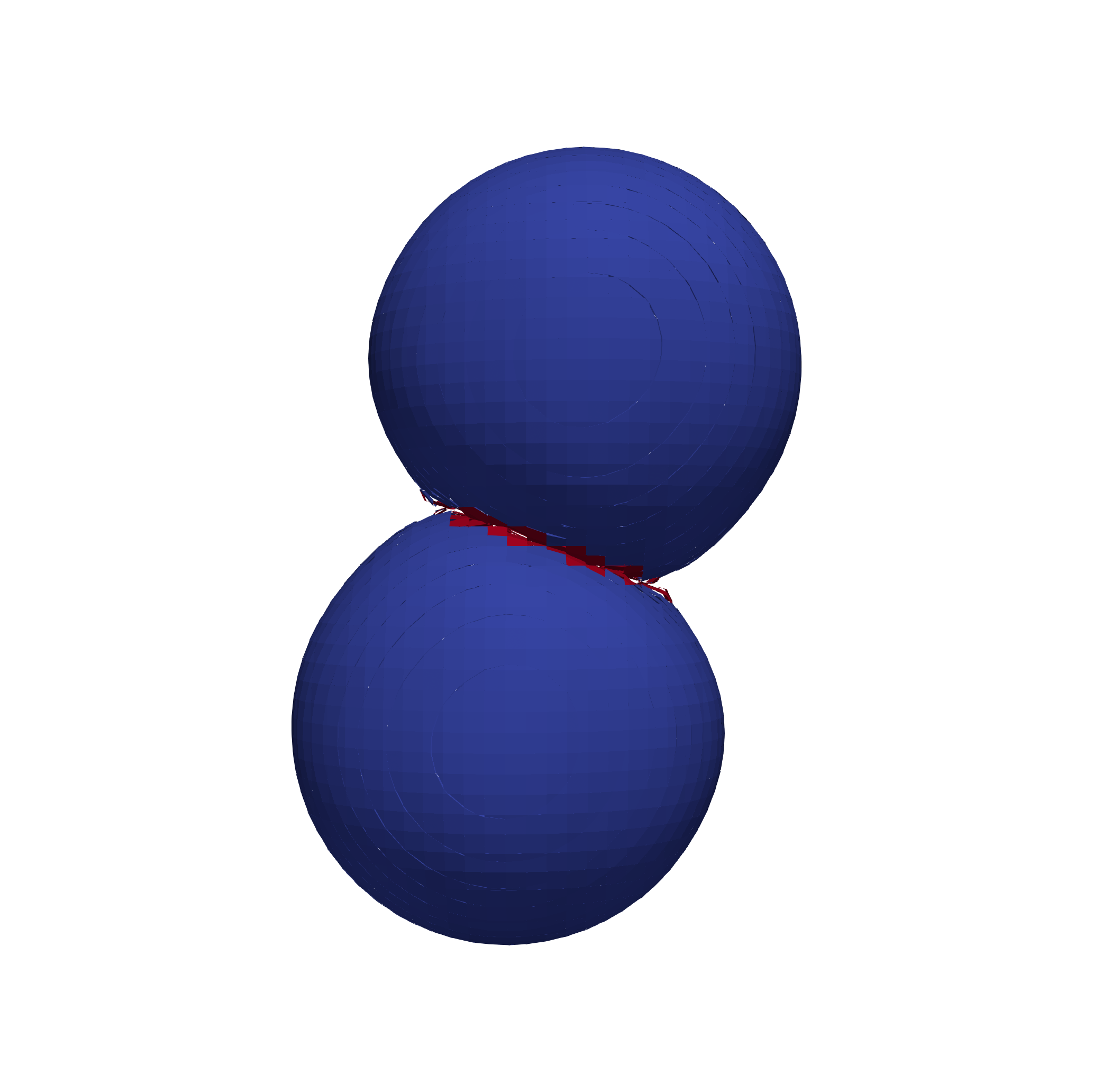}
        \caption{$t=\qty{29}{\ms}$}
        \label{subfig:gasfilm}
    \end{subfigure}\hfill
    \begin{subfigure}[b]{0.25\textwidth}
        \centering\includegraphics[width=\textwidth]{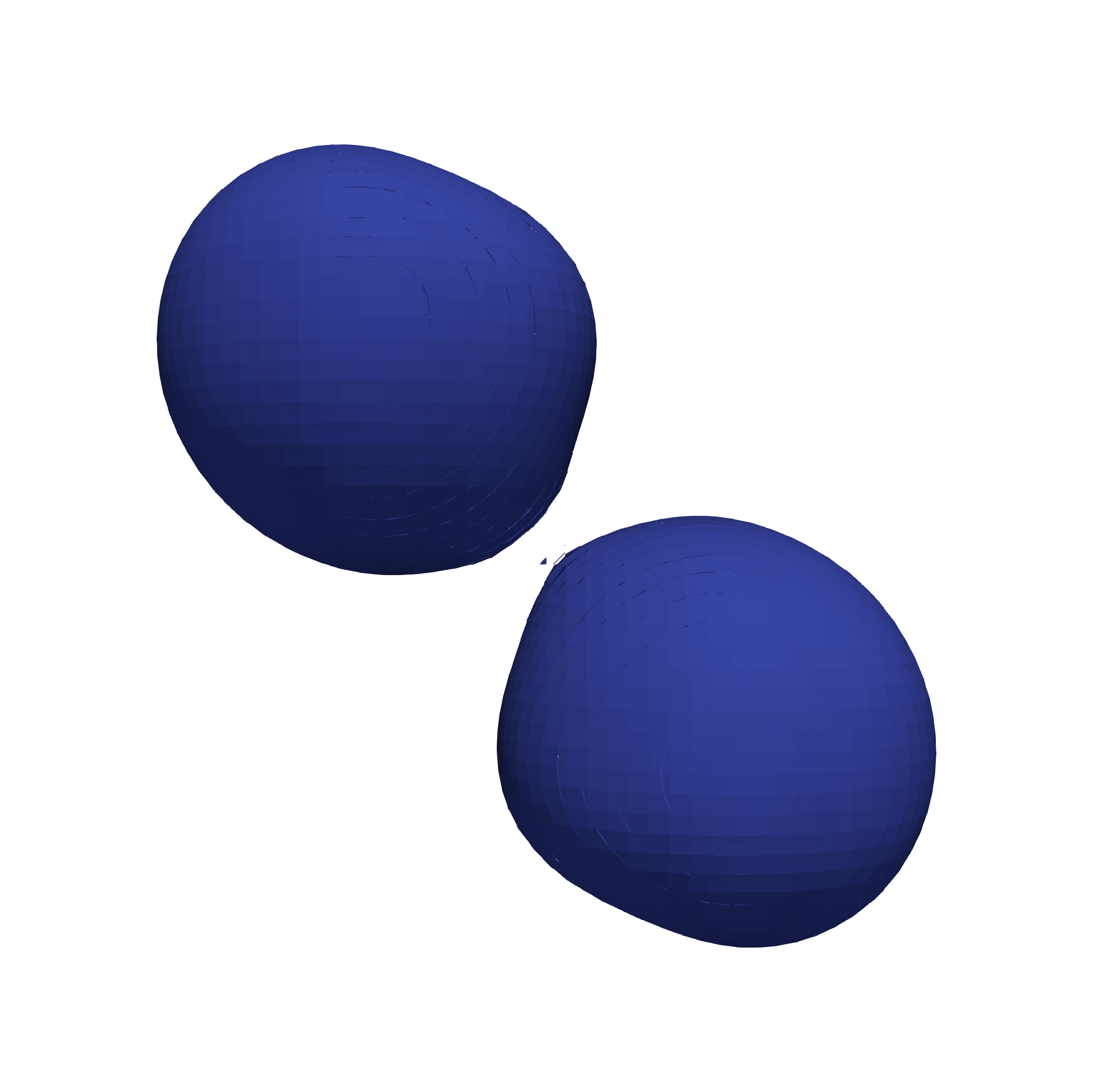}
        \caption{$t=\qty{61}{\ms}$}
    \end{subfigure}\hfill
    \caption{Interface shape for the binary collision of offset droplets using ELVIRA (middle) and R2P (bottom) with $D/\Delta x=23.04$. Single-plane reconstructions are shown in blue, while two-plane reconstructions are shown in red. Photographic images from Figure 4r in \cite{Qian1997} are shown on the top row with permission from the publisher.}
    \label{fig:collision}    
\end{figure}

This demonstration of the proposed method simulates the collision of two droplets, where the parameters are chosen to replicate the bouncing collision depicted in Figure 4r of \cite{Qian1997}. Two identical droplets with diameter $D=\qty{360}{\um}$, $\rho_b=\qty{1.225}{\kg\per\cubic\meter}$, $\mu_b=\qty{1.827e-5}{\pascal\second}$, $\rho_a/\rho_b=622$ and $\mu_a/\mu_b=116$ are initialized with a relative velocity $U$ such that $\We=0.0232$ and $\text{Re}_b=28.5$. The initial positions of the droplets have a relative vertical offset $B=\chi/D=0.84$, where $\chi$ is the projection of the separation distance between the droplet centers in the direction normal to $U$, such that $B=0$ corresponds to a head-on collision, and $B=1$ indicates a grazing collision. 
The computational mesh size is $\Delta x=D/23.04$, and the time step size is $\Delta t=\qty{2}{\us}$.
For this test, a threshold value of $\Bar{\mathscr{M}}=0.8$ is used. 
The collision is depicted in a series of snapshots in Figure \ref{fig:collision}, where the interface shapes resulting from ELVIRA and R2P reconstructions are compared. The formation of a micron-thickness subgrid gas film with R2P is indicated by the presence of two-plane interface reconstructions in Figure \ref{subfig:gasfilm}. The formation of this film prevents the droplets from coalescing during the collision, which matches the experimental result. The colliding droplets coalesce when ELVIRA is used because the gas film is not captured. However, the collision outcome, i.e., whether the droplets coalesce or bounce, for an arbitrary choice of parameters is dictated by intermolecular forces that are not captured by the Navier-Stokes equations \citep{mackayGravityApproachCoalescence1963}. As with the liquid film formed during the deformation of the impulsively accelerated droplet, subgrid-scale models could be applied on the gas film cells to recover these non-continuum effects.

\subsection{Buoyant Bubble}
\begin{figure}[hbt]
    \begin{subfigure}[T]{0.5\textwidth}
        \centering\includegraphics[width=\textwidth]{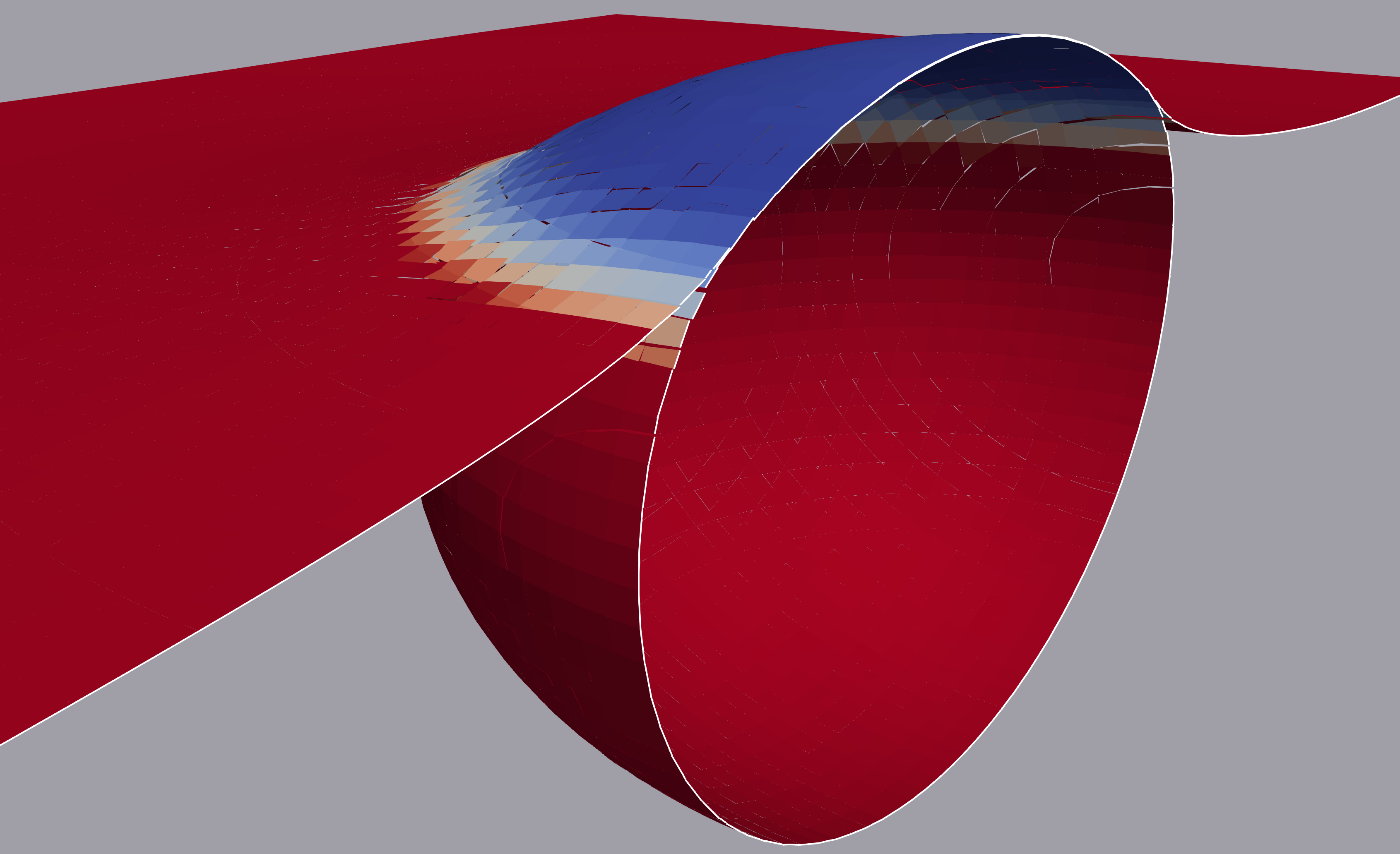}
        \caption{Interface elements colored by film thickness.}
    \end{subfigure}\hfill
    \begin{subfigure}[T]{0.5\textwidth}
        \centering\includegraphics[width=\textwidth]{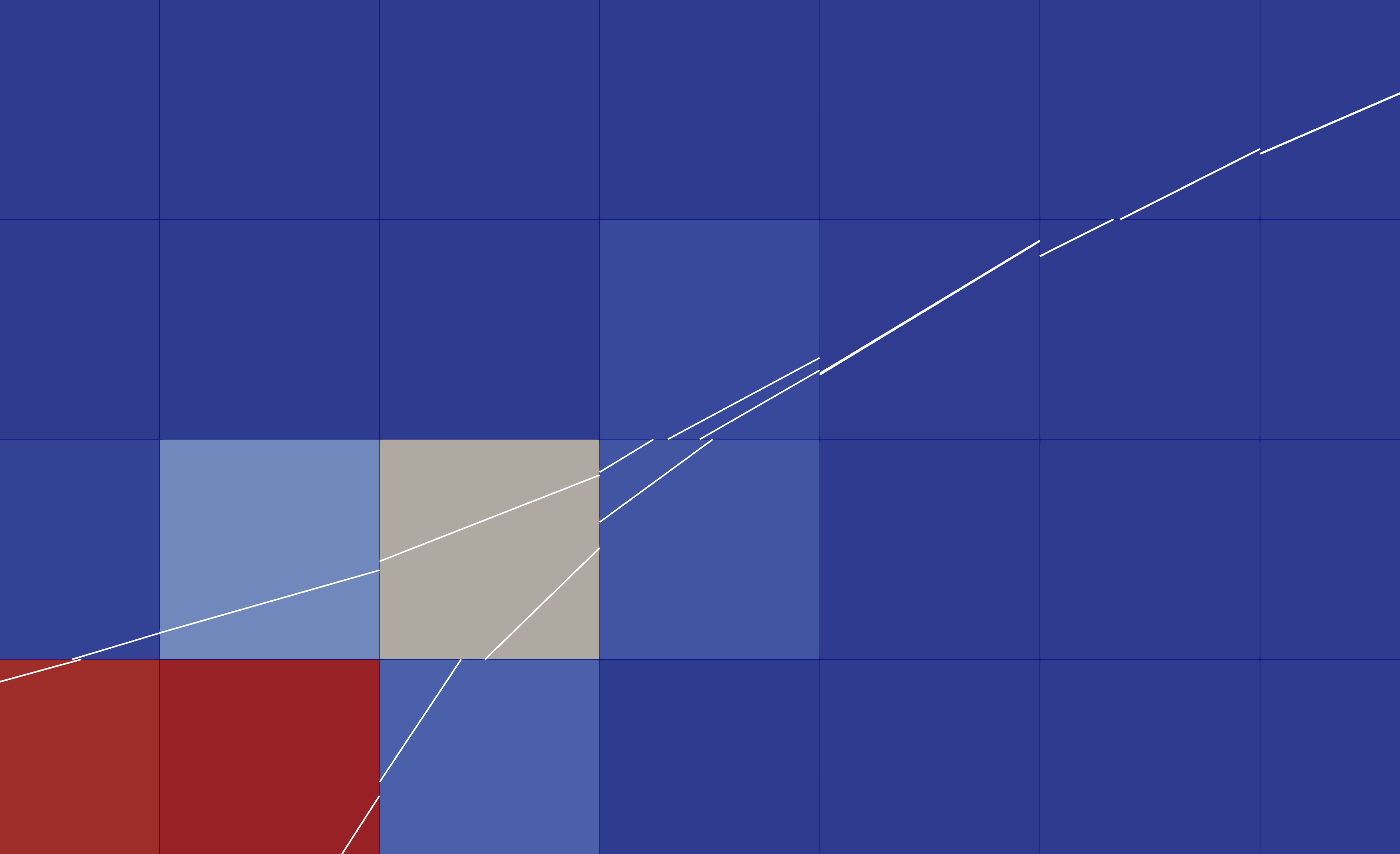}
        \caption{Closeup of the interface. The cells are colored by volume fraction.}
    \end{subfigure}\hfill
    \caption{Interface shape for a buoyant bubble at the surface of a liquid pool.}
    \label{fig:risingbubble}
\end{figure}

This final demonstration simulates a gas bubble of $D=\qty{1}{\mm}$ rising to the surface of a liquid pool due to buoyant forces. Water and air are respectively used as the liquid and gas, with $\rho_a=\qty{1000}{\kg\per\cubic\m}$, $\rho_a/\rho_b=816$ and $\mu_a/\mu_b=63.9$. The Bond number is $\mathrm{Bo} = \Delta\rho gD^2/\sigma=1.96$, and the Morton number is $\mathrm{Mo} = g \mu_a^4 \Delta \rho/\rho_a^2 \sigma^3=\qty{1.31e-7}{}$, where $g=\qty{9.81}{\m\per\square\second}$ is gravitational acceleration and $\Delta\rho=\rho_a-\rho_b$. The initial position of the bubble center is $1.45D$ below the pool surface. The computational mesh size is $\Delta x=D/32$, and the time step size is $\Delta t=\qty{10}{\us}$.
Figure \ref{fig:risingbubble} shows the interface shape when the bubble reaches the pool surface, and a thin liquid film forms between the top of the bubble and the pool surface. This liquid film in the figure is colored by thickness $h$ according to Eq.\ \eqref{eq:thickness-s4}, and the minimum thickness of the film is $h=3.20\times 10^{-3}\Delta x$. This case further illustrates the versatility of R2P in capturing films in flows involving massive differences in length scales between fluid structures, such as bubbly flows.

\section{Conclusions}\label{conclusions}
This work presents R2P, an interface reconstruction method capable of representing liquid-gas interfaces with two planes per computational cell within a geometric VOF framework. The initial guess of the interface centroids and normals is obtained through spherical k-means clustering \citep{Dhillon2001} of the advected interface polygons. This initial guess is refined through a non-linear least squares procedure that minimizes the difference between the transported volume moments and the reconstructed volume moments within the local neighborhood of the interfacial cell while also minimizing the destruction or creation of surface area. A root-finding step during the optimization procedure ensures discrete volume conservation. The pressure jump across subgrid-thickness films due to surface tension is recovered through the use of a ficticious volume fraction field that artificially thickens the film. These routines have been implemented in the Interface Reconstruction Library (IRL) \citep{Chiodi2022,chiodiInterfaceReconstructionLibrary2023} and the two-phase flow solver NGA2 \citep{Desjardins2008,desjardinsNGA22023}.

Tests of R2P with a variety of canonical cases, including the rotation of a slotted disk and the reversible deformation of a cylinder and sphere, demonstrate that R2P seamlessly transitions between one- and two-plane reconstructions, enabling the capturing of thin films at all mesh resolutions. Compared to the one-plane ELVIRA reconstruction, R2P results in lower shape errors at fixed mesh resolution and comparable computational cost at fixed shape error. However, given that the interface reconstruction is a small part of the total cost of a flow solver, R2P's advantages are best observed when utilized within a two-phase Navier-Stokes solver. In the first example, R2P enables the formation of subgrid-thickness liquid bags from impulsively accelerated droplets with initial diameter $D/\Delta x=\order{10}$. The cost increase of R2P from ELVIRA in this test is only 16\% even though ELVIRA entirely fails to form the liquid bag. Similarly, subgrid-thickness gas films can be formed between colliding droplets of comparable diameter. R2P, therefore, enables a potential cost savings of orders of magnitude compared to single-plane PLIC reconstructions that require the full resolution of the micron-scale films. Future work will develop subgrid-scale models to predict droplet formation from the breakup of these films.


\section{Acknowledgements}
Robert Chiodi was supported by the Office of Naval Research through a National Defense Science and Engineering Graduate Research Fellowship. This work was sponsored by the Office of Naval Research (ONR) as part of the Multidisciplinary University Research Initiatives (MURI) Program, under grant number N00014-16-1-2617. The views and conclusions contained herein are those of the authors only and should not be interpreted as representing those of ONR, the U.S. Navy or the U.S. Government. This material is based upon work supported by NSF under CAREER award Grant No. CBET-1351545 and XSEDE computing resources on TACC Stampede 2. This work was supported by the US Department of Energy through the Los Alamos National Laboratory. Los Alamos National Laboratory is operated by Triad National Security, LLC, for the National Nuclear Security Administration of U.S. Department of Energy (Contract No. 89233218CNA000001).

The authors thank Professor Mahdi Esmaily for his comments on a draft of this manuscript.

\appendix
\section{Film Cell Classification}\label{classification}
It is necessary to identify contiguous film regions to which the surface tension treatment described in Section \ref{surface_tension} should be applied. As a film may contain adjacent one-plane and two-plane cells, only applying the proposed surface tension model to two-plane cells would underestimate the surface tension forces across the film. Furthermore, the proposed surface tension strategy requires the identification of the film phase of each cell. This section outlines the classification scheme used to identify the set of liquid film cells $\mathcal{L}$ and gas film cells $\mathcal{G}$ in the domain.

Let $\mathcal{V}$ be the set of liquid-containing cells
\begin{equation}
    \mathcal{V}=\{\Omega_i \mid h_i<1.5\Delta x\},
\end{equation}
where the local film thickness $h$ is calculated as
\begin{equation}
        h_i=\frac{2\min\left(\sum_{\Omega_j \in \mathscr{N}_i}\alpha_j V_j,\sum_{\Omega_j \in \mathscr{N}_i}(1-\alpha_j) V_j\right)}{\sum_{\Omega_j \in \mathscr{N}_i}A_j},
        \label{eq:thickness}
\end{equation}
where $\mathscr{N}_i$ is the set of all cells that share a vertex with cell $\Omega_i$, $V_i$ is the cell volume, and $A_i$ is the surface area of the liquid-gas interface $\Gamma_i$ in the cell. The threshold value of $h_i<1.5\Delta x$ is chosen to avoid overapplication of the film surface tension method and was found to work well for a variety of cases. The factor of 2 is used because a film is bounded by an interface on both sides. The cell neighborhood $\mathscr{N}_i$ is not to be confused with the cell neighborhood $\mathfrak{N}_i$, used in the rest of the subsection, which is the set of cells that share at least one face with $\Omega_i$. Additionally, for each plane $m$ in $\Gamma_i$, where $1 \le m \le N_i$, the normal vector is denoted by $\bm{n}_{i,m}$, and the surface centroid is denoted by $\bm{s}_{i,m}$. The set of liquid film cells $\mathcal{L}$ is the union of interfacial cell sets
\begin{equation}
    \mathcal{L}=\mathcal{L}^2 \cup \mathcal{L}^1,
\end{equation}
where
\begin{equation} 
    \mathcal{L}^2=\{ \Omega_i \in \mathcal{V} \mid (N_i = 2 ) \land ( \left(\bm{s}_{i,2}-\bm{s}_{i,1}\right)\cdot \bm{n}_{i,2} \ge 0) \land (\bm{n}_{i,1}\cdot\bm{n}_{i,2}<0.5)\}
\end{equation}
is the set of two-plane cells containing a liquid film,
and $\mathcal{L}^1=\mathcal{L}^{11} \cup \mathcal{L}^{12} \cup \mathcal{L}^{13}$ is the union of 
\begin{enumerate}
    \item the set of single-plane cells that neighbor a two-plane liquid film cell,
    \begin{equation}
        \mathcal{L}^{11}=\{ \Omega_i \in \mathcal{V} \mid (N_i = 1) \land \left(\exists \Omega_j \in \mathfrak{N}_i\right)\left[N_j \in \mathcal{L}^2 \right]  \},
    \end{equation}
    \item the set of single-plane cells that neighbor another single-plane cell such that a liquid film exists between their respective interfaces,
    \begin{multline}
        \mathcal{L}^{12}=\{ \Omega_i \in V \mid (N_i = 1) \land (\exists \Omega_j \in \mathfrak{N}_i) \\ [(N_j=1 )\land ( \bm{n}_{i,1} \cdot \bm{n}_{j,1} < -0.5 ) \land  ((\bm{s}_{j,1}-\bm{s}_{i,1})\cdot \bm{n}_{j,1} > 0 ) \land ( (\bm{s}_{i,1}-\bm{s}_{j,1})\cdot \bm{n}_{i,1} > 0 ) ]  \},
    \end{multline}
    and
    \item the set of single-plane cells that neighbor a two-plane non-film cell such that a liquid film exists between their respective interfaces
    \begin{multline}
        \mathcal{L}^{13}=\{ \Omega_i \in V \mid (N_i = 1) \land (\exists \Omega_j \in \mathfrak{N}_i) [(N_j=2 )\land(N_j \notin \mathcal{L}^2 ) \\ \land ( \bm{n}_{i,1} \cdot \bm{n}_{j} < -0.5 ) \land  ((\bm{s}_{j}-\bm{s}_{i,1})\cdot \bm{n}_{j} > 0 ) \land ( (\bm{s}_{i,1}-\bm{s}_{j})\cdot \bm{n}_{i,1} > 0 ) ]  \},
    \end{multline}
    where
    \begin{equation}\label{eq:hatnj}
        \bm{n}_j=\frac{\sum_{m=1}^{N_j}w_{j,m}\bm{n}_{j,m}}{\sum_{m=1}^{N_j}w_{j,m}}
    \end{equation}
    and
    \begin{equation}\label{eq:veccj}
        \bm{s}_j=\frac{\sum_{m=1}^{N_j}w_{j,m}\bm{s}_{j,m}}{\sum_{m=1}^{N_j}w_{j,m}}
    \end{equation}
    are the surface-averaged normal and centroid of the neighbor two-plane cell.
\end{enumerate}

Likewise, the set of gas film cells $\mathcal{G}$ is the union of interfacial cell sets
\begin{equation}
    \mathcal{G}=\mathcal{G}^2 \cup \mathcal{G}^1,
\end{equation}
where $\mathcal{G}^2$ is the set of two-plane cells containing a gas film
\begin{equation} 
    \mathcal{G}^2=\{ \Omega_i \in \mathcal{V} \mid (N_i = 2) \land (\left(\bm{s}_{i,2}-\bm{s}_{i,1}\right)\cdot \bm{n}_{i,2} < 0) \land (\bm{n}_{i,1}\cdot\bm{n}_{i,2}<0.5)\},
\end{equation} 
and $\mathcal{G}^1=\mathcal{G}^{11} \cup \mathcal{G}^{12} \cup \mathcal{G}^{13}$ is the union of 
\begin{enumerate}
    \item the set of single-plane cells that neighbor a two-plane gas cell,
    \begin{equation}
        \mathcal{G}^{11}=\{ \Omega_i \in \mathcal{V} \mid (N_i = 1) \land \left(\exists \Omega_j \in \mathfrak{N}_i\right)\left[N_j \in \mathcal{G}^2 \right]  \},
    \end{equation}
    \item the set of single-plane cells that neighbor another single-plane cell such that a gas film exists between their respective interfaces,
    \begin{multline}
        \mathcal{G}^{12}=\{ \Omega_i \in V \mid (N_i = 1 ) \land (\exists \Omega_j \in \mathfrak{N}_i) \\ [(N_j=1 )\land ( \bm{n}_{i,1} \cdot \bm{n}_{j,1} < -0.5 ) \land  ((\bm{s}_{j,1}-\bm{s}_{i,1})\cdot \bm{n}_{j,1} < 0 ) \land ( (\bm{s}_{i,1}-\bm{s}_{j,1})\cdot \bm{n}_{i,1} < 0 ) ]  \},
    \end{multline}
    and
    \item the set of single-plane cells that neighbor a two-plane non-film cell such that a gas film exists between their respective interfaces
    \begin{multline}
        \mathcal{G}^{13}=\{ \Omega_i \in V \mid (N_i = 1) \land (\exists \Omega_j \in \mathfrak{N}_i) [(N_j=2 )\land(N_j \notin \mathcal{G}^2 ) \\ \land ( \bm{n}_{i,1} \cdot \bm{n}_{j} < -0.5 ) \land  ((\bm{s}_{j}-\bm{s}_{i,1})\cdot \bm{n}_{j} < 0 ) \land ( (\bm{s}_{i,1}-\bm{s}_{j})\cdot \bm{n}_{i,1} < 0 ) ]  \},
    \end{multline}
    where $\bm{n}_j$ and $\bm{s}_j$ are defined by Eqs. \ref{eq:hatnj} and \ref{eq:veccj}, respectively.
\end{enumerate}


For sets $\mathcal{L}^{13}$ and $\mathcal{G}^{13}$ it is possible that the dot products $(\bm{s}_{j}-\bm{s}_{i,1})\cdot \bm{n}_{j}$ and $(\bm{s}_{i,1}-\bm{s}_{j})\cdot \bm{n}_{i,1}$ will have opposing signs such that the phase is ambiguous. In this case, the phase of $\Omega_i$ is determined by the phase of $\Omega_j$.

\bibliography{allreferences}

\end{document}